\documentclass[aps,pra,epsf,superscriptaddress,amsmath,amssymb,amsfonts,twocolumn,showpacs]{revtex4-1}

\usepackage[utf8]{inputenc}
\usepackage[T1]{fontenc}
\usepackage[paper=a4paper,left=20mm,right=20mm,top=25mm,bottom=20mm]{geometry}
\usepackage{float}
\usepackage[english]{babel}
\usepackage[normalem]{ulem}
\usepackage{mathtools}
\usepackage{bbold}
\usepackage{graphicx}
\usepackage{multirow}
\usepackage{tabularx}
\usepackage[sgvnames, table]{xcolor}
\usepackage{soul}
\usepackage{comment}
\usepackage[colorlinks=true,allcolors=blue]{hyperref}

\DeclarePairedDelimiter\bra{\langle}{\rvert}
\DeclarePairedDelimiter\ket{\lvert}{\rangle}

\DeclarePairedDelimiterX\braket[2]{\langle}{\rangle}{#1 \delimsize\vert #2}

\begin{document}
	
\title{Counterflow dynamics of two correlated impurities\\ immersed in a bosonic gas}

\author{Friethjof Theel}
\affiliation{Center for Optical Quantum Technologies, University of Hamburg, Department of Physics, Luruper Chaussee 149, D-22761, Hamburg, Germany}
\author{Simeon I. Mistakidis}
\affiliation{ITAMP, Center for Astrophysics $|$ Harvard $\&$ Smithsonian, Cambridge, MA 02138 USA}
\author{Kevin Keiler}
\affiliation{Center for Optical Quantum Technologies, University of Hamburg, Department of Physics, Luruper Chaussee 149, D-22761, Hamburg, Germany}
\author{Peter Schmelcher}
\affiliation{Center for Optical Quantum Technologies, University of Hamburg, Department of Physics, Luruper Chaussee 149, D-22761, Hamburg, Germany}
\affiliation{The Hamburg Centre for Ultrafast Imaging, University of Hamburg, Luruper Chaussee 149, D-22761, Hamburg, Germany}
	
\date{\today}

\begin{abstract}
	The counterflow dynamics of two correlated impurities in a double-well coupled to an one-dimensional bosonic medium is explored. We determine the ground state phase diagram of the system according to the impurity-medium entanglement and the impurities two-body correlations. Specifically, bound impurity structures reminiscent of bipolarons for strong attractive couplings as well as configurations with two clustered or separated impurities in the repulsive case are identified. The interval of existence of these phases depends strongly on the impurity-impurity interactions and external confinement of the medium. Accordingly the impurities dynamical response, triggered by suddenly ramping down the central potential barrier, is affected by the medium's trapping geometry. In particular, for a box-confined medium repulsive impurity-medium couplings lead, due to attractive induced interactions, to the localization of the impurities around the trap center. In contrast, for a harmonically trapped medium the impurities perform a periodic collision and expansion dynamics further interpreted in terms of a two-body effective model. Our findings elucidate the correlation aspects of the collisional physics of impurities which should be accessible in recent cold atom experiments.
\end{abstract}

\maketitle

\section{Introduction}

Ultracold quantum gases provide an exceptional playground for the investigation of fundamental quantum many-body phenomena since they feature an exquisite experimental control \cite{chin2010}. For instance, it is possible to control the shape and dimensionality of the external potential \cite{grimm2000, henderson2009, gaunt2013}, design species selective potentials \cite{haas2007, catani2012, barker2020, barker2020a} and, most importantly, tune the interparticle interactions to an almost arbitrary extend via Feshbach resonances \cite{olshanii1998, kohler2006, chin2010}. A particular research focus has been set on strongly particle imbalanced mixtures, which allow to emulate impurity systems interacting with a bath. The key mechanism is that the bare impurity becomes dressed by the excitations of the bath and, thus, can be considered as a quasi-particle, the so-called polaron \cite{landau1933}. In this regard, several works have been devoted to exemplify the fundamental stationary properties of both Fermi \cite{schirotzek2009, nascimbene2009, kohstall2012, ngampruetikorn2012, massignan2014, schmidt2018a,tajima2021polaron} and Bose polarons \cite{fukuhara2013, jorgensen2016, hu2016, volosniev2017, grusdt2017, mistakidis2019c, ardila2019a, ardila2020, skou2021}, such as their effective mass \cite{nascimbene2009, mistakidis2019c, ardila2020}, energy \cite{jorgensen2016, hu2016} and residue \cite{schirotzek2009, kohstall2012}.

Recently, more attention has been placed on the interplay between several impurities immersed in a quantum gas \cite{grusdt2017a, camacho-guardian2018, huber2019, brauneis2021}. Among others, the coalescence of two bosonic impurities coupled to a harmonically trapped bosonic medium has been predicted \cite{dehkharghani2018} as well as the existence of their induced interactions \cite{charalambous2019, reichert2019, keiler2021}. In the strongly attractive interaction regime, the formation of bipolarons referring to impurity bound states was also unraveled \cite{camacho-guardian2018a, will2021}. Beyond these studies the non-equilibrium dynamics of quasi-particles following an interaction quench \cite{mukherjee2020a, mistakidis2020, bougas2021} has been examined unveiling, in particular, energy redistribution processes, temporal orthogonality catastrophe phenomena and the effective temperature of the impurities \cite{mistakidis2020} e.g. by emulating pump-probe and Ramsey spectroscopy.

Another branch in the field of ultracold quantum gases concerns the collisional aspects of atomic ensembles. Counterflow dynamics can be triggered, e.g., by employing a magnetic field gradient separating two atomic hyperfine states \cite{hamner2011} or releasing an ultracold quantum gas from a double-well potential into a harmonic oscillator \cite{weller2008}. For a single atomic species these protocols result in the oscillation of the formed dark solitons \cite{weller2008} or in the case of a two-component mixture in the spontaneous generation of dark-bright soliton trains \cite{hamner2011}. Another technique to initiate atomic collisions constitutes of two counter-propagating harmonic oscillator potentials \cite{kohler2021} which has been experimentally realized with $^{40}$K and $^{87}$Rb clouds utilizing two optical tweezers \cite{thomas2017, thomas2018}.

In this sense, it is intriguing to explore the counterflow correlated dynamics of impurities in combination with a superimposed superfluid background. A similar question was addressed for fermions~~\cite{tajima2020collisional,kwasniok2020} e.g. showing the formation of shock-waves. Thereby, of immediate interest is the influence of the background on the collisional response and the associated emergent induced interactions between the impurities~\cite{kwasniok2020}. The impact of the bath on the impurity dynamics is expected to depend on the confining potential of the bath, and the impurities coupling strength as well as the interaction between the impurities and the bath particles \cite{theel2021}. To tackle these open questions, herein we consider a minimal model of two bosonic impurities trapped in a double well and immersed in a bosonic bath. The counterflow dynamics between the impurities is induced by suddenly ramping down the potential barrier of their double well and, subsequently, let the system evolve in time for different interaction configurations. 

Specifically, it is shown that already the ground state configurations depend on the chosen impurity-medium and impurity-impurity interaction strength and, importantly, on the type of the underlying trapping potential of the medium. For instance, in the case of a box-confined medium the impurities coalesce for intermediate repulsive impurity-medium couplings independently of the impurity-impurity repulsion. 
On the other hand, for a harmonically trapped bath the impurities separate from each other for strong impurity-impurity and impurity-medium repulsions \cite{dehkharghani2018}. Moreover, we find indications of bipolaron formation~\cite{camacho-guardian2018a} for strong impurity-medium attractions.

The dynamical response of the impurities appears to be strongly affected by the combination of the involved interaction strengths as well as by the external confinement of the bath. More precisely, in the case of a box-confined medium and intermediate repulsive impurity-medium couplings the impurities induced interactions lead to their localization at the trap center after their first collision. However, increasing the impurity-medium interaction strength the impurities experience a periodic collisional response characterized by a damped amplitude: a behavior that is argued to be governed by finite size effects determined by the size of the box potential. Employing a harmonically trapped medium the impurities localize at the trap center for attractive impurity-medium coupling strengths, while for intermediate and strong repulsions they phase separate with the medium~\cite{goold2011, mistakidis2019a} and thus their dressing is suppressed. Considering weakly repulsive impurity-impurity interactions a state transfer manifests from two separated to two coalesced impurities. Importantly, this process is absent in the decoupled case elucidating the role of the coupling with the bath and thus of the interspecies correlations (entanglement). 

A microscopic analysis provides insights into the single-particle excitation processes and the two-body states participating in the dynamics and the aforementioned state transfer. 
To describe the stationary and dynamical properties of the composite impurity-medium system we employ the multi-layer multi-configuration time-dependent Hartree method for atomic mixtures (ML-MCTDHX) \cite{kronke2013, cao2013, cao2017, kohler2019}. This \textit{ab initio} approach allows us to efficiently track the relevant inter- and intraspecies correlations which are anticipated to be enhanced, especially during the dynamics. 
This is in part due to the few-body impurity subsystem as well as the spatial inhomogeneity caused by the external potential.

This work is structured as follows. In section \ref{sec:model} and \ref{sec:method} we present the impurity model under consideration and introduce the ingredients of the variational method, respectively. We proceed in section \ref{sec:GS} with an analysis of the system's ground state and draw a phase diagram with respect to variations of the impurity-impurity and impurity-medium interaction strengths. This analysis is based on the two-body densities quantifying the correlations of the bath particles and the impurities. Next, in section \ref{sec:dyn} and \ref{sec:dyn_ho} the dynamical response of the system following a sudden reduction of the double-well barrier is discussed. In particular, section \ref{sec:dyn} elaborates on the case of a box-confined medium in which we explicate, e.g., the localization of the impurities at intermediate repulsive impurity-medium coupling strengths and the emergence of finite size effects. The case of a harmonically trapped medium is investigated in section \ref{sec:dyn_ho} where the focus is set on the impurities excitation processes and their dependence on the impurity-medium couplings. Our results are summarized in section \ref{sec:conclusion} together with an outlook regarding further research directions. Appendix \ref{ap:box_rel_dist} elaborates on the behavior of the impurities relative distance in their ground state and in Appendix \ref{ap:box_heavy_impurities} the impact of the impurities mass on their collisions is exposed. In Appendix \ref{ap:ramp} we discuss the persistence of the impurities collisional features when a linear ramp is applied to the barrier height of the double-well potential.

\section{Impurity-medium settings}
\label{sec:model}

The system consists of two different bosonic species $B$ and $I$ at ultracold temperatures. In particular, we consider $N_I=2$ impurities of mass $m_I$ and a bosonic bath of $N_B=20$ particles with mass $m_B$. The corresponding Hamiltonian reads
\begin{align}
	\mathcal{\hat{H}} = \mathcal{\hat{H}}_B + \mathcal{\hat{H}}_I + \mathcal{\hat{H}}_{BI},
\label{eq:Hamiltionian}
\end{align}
where $\mathcal{\hat{H}}_\sigma = \sum_{i=1}^{N_\sigma}
\big( - \frac{\hbar^2}{2m_\sigma} \frac{\partial^2}{(\partial x_i^{\sigma})^2} + V_\sigma(x_i^\sigma) + g_{\sigma\sigma} \sum_{i<j} \delta(x_i^\sigma-x_j^\sigma) \big)$ is the interaction Hamiltonian of species $\sigma\in\{B,I\}$. Each component is subject to a different external potential $V_\sigma(x_i^\sigma)$, a scenario that can be achieved via species selective optical potentials \cite{leblanc2007, lercher2011}. 
It is also restricted to one spatial dimension \cite{cazalilla2011} that can be realized experimentally, e.g., by freezing out the transverse degrees of freedom using a strong harmonic confinement \cite{ketterle1996, gorlitz2001}. 

Since we are operating in the ultracold regime it is sufficient to take into account only $s$-wave scattering processes and thus the interaction between two particles of the same species is modeled with a contact interaction potential \cite{olshanii1998} determined by the one-dimensional effective coupling strength parameter $g_{\sigma\sigma}$. Analogously, the coupling between the impurities and the bath is described through a contact interaction potential $\mathcal{\hat{H}}_{BI} = g_{BI} \sum_{i=1}^{N_B}\sum_{j=1}^{N_I}\delta(x_i^B-x_j^I)$, where $g_{BI}$ denotes the impurity-medium interaction strength. Due to the fact that $g_{\sigma\sigma'}$ with $\sigma,\sigma'\in\{B,I\} $ depends, among others, on the three-dimensional $s$-wave scattering length it can be experimentally adjusted, e.g., via Feshbach resonances utilizing either magnetic or optical fields \cite{inouye1998, fedichev1996, kohler2006, chin2010}. Below, we consider a bosonic medium of $^{87}$Rb atoms %with mass $m_B=1$ 
and $^{133}$Cs impurities. 
Thus, the mass ratio is $m_I/m_B=133/87$ \cite{leblanc2007, haas2007, bouton2020, adam2021}.

At $t=0$ the system is prepared in its ground state with a specific combination of interaction strengths ($g_{\sigma\sigma'}$). The impurities are initially confined in a double well $V_I^{\rm{dw}}(x) = \frac{1}{2}m_I\omega_I^2x^2 + \frac{h_I}{w_I\sqrt{2\pi}}\exp(\frac{-x^2}{2w_I^2})$ which is the superposition of a harmonic oscillator potential with frequency $\omega_I$ and a Gaussian of width $w_I$ and height $h_I$ \cite{albiez2005, zollner2008}. For the external potential of the bosonic medium we consider two cases: a box potential of size $L_B=1$ with $V_B(x)=0$ for $-L_B/2<x<L_B/2$ and $V_B(x)=\infty$ elsewhere, and a harmonic oscillator $V_B(x) = \frac{1}{2}m_B\omega_B^2x^2$. In the former scenario we consider $\tilde{x}_{\rm{box}} = L_B/10$ and $\tilde{E}_{\rm{box}} = \frac{\hbar^2}{m_B\tilde{x}_{\rm{box}}^2}$
as characteristic length and energy scales, respectively~\footnote{The characteristic length of the box potential is set such that the resulting interaction scales are comparable with the ones used in the case of a harmonically trapped medium.}. Thus, the time and interaction strength are expressed in units of $\tilde{t}_{\rm{box}} = \frac{m_B\tilde{x}_{\rm{box}}^2}{\hbar}$ and $\tilde{g}_{\rm{box}} = \frac{\hbar^2}{m_B\tilde{x}_{\rm{box}}}$, respectively. However, in the case of a harmonically trapped medium it is more convenient to express the energy in units of the medium's harmonic oscillator $\tilde{E}_{\rm{ho}}=\hbar\tilde{\omega}_{\rm{ho}}$ where $\omega_B/\tilde{\omega}_{\rm{ho}}=1$. It follows that the length, interaction strength and timescales are given in terms of $\tilde{x}_{\rm{ho}}=\sqrt{\frac{\hbar}{m_B\omega_B}}$, $\tilde{g}_{\rm{ho}}=\sqrt{\frac{\hbar^3\omega_B}{m_B}}$ and $\tilde{t}_{\rm{ho}}= \omega_B^{-1}$, respectively. To construct the impurities' double well we employ $\omega_I/\tilde{\omega}_{\rm{box,ho}}=0.6$, $h_I/\tilde{E}_{\rm{box,ho}}\tilde{x}_{\rm{box,ho}}^{-1}=3.0$ and $w_I/\tilde{x}_{\rm{box,ho}}=0.7$.

%As a length scale of the former scenario of a box-confined medium we chose the box size $l$ and, thus,

\begin{figure}[t]
	\centering
	\includegraphics[width=1.0\linewidth]{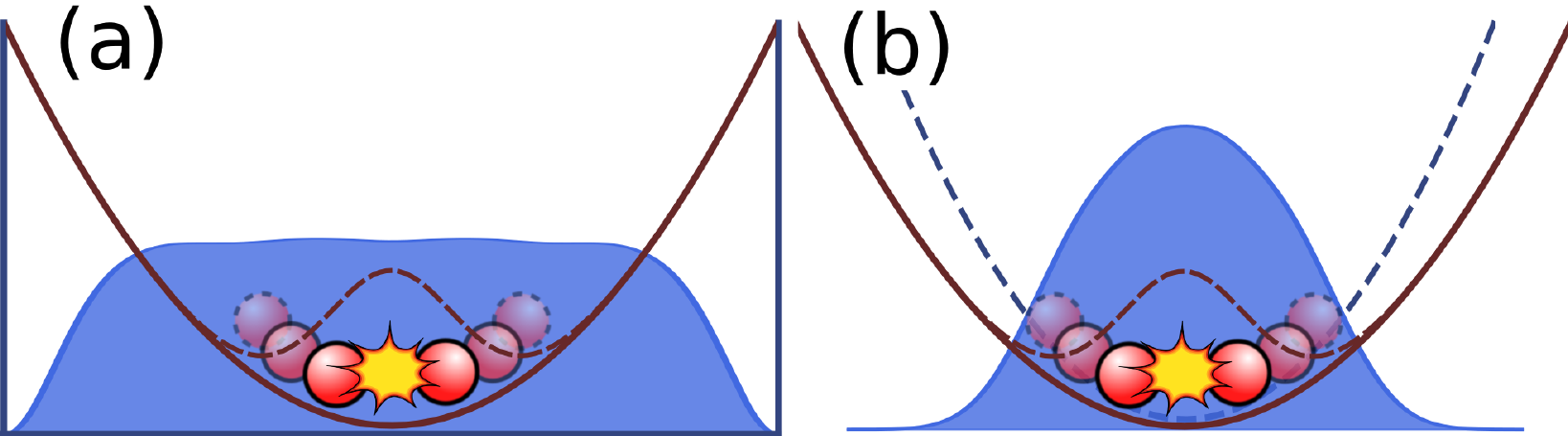}
	\caption{Schematic representation of the considered setup and the quench protocol. Two bosons (red circles) are coupled to a bosonic medium (blue shaded area) which is either confined in (a) a box potential or (b) a harmonic oscillator. First, the two impurities are loaded into the double well and the equilibrium state of the system is obtained for a specific set of inter- and intraspecies interaction parameters ($g_{BB}, g_{II}, g_{BI}$). Subsequently, a counterflow dynamics between the two impurities is induced by suddenly ramping down the barrier of the double-well potential.}
	\label{fig:setup}
\end{figure}

The ground state of the composite system is acquired for a set of values of interaction strengths ($g_{BB}$, $g_{II}$, $g_{BI}$). In the case of a box-confined medium the impurities always exhibit a finite spatial overlap with the bath particles for the considered interaction strengths. 
However, for a harmonically trapped bath the impurity-medium overlap vanishes as long as $g_{BI}>g_{BB}$ (see also the discussion below). Subsequently, the dynamics is triggered by suddenly reducing (at $t=0$) the barrier of the impurities' double-well potential [see Figure \ref{fig:setup}].
Consequently, in the course of the time evolution the impurities collide and experience a harmonic oscillator potential $V_I^{\rm{ho}}(x)$ with frequency $\omega_I$. As we shall argue below, the emerging collisional correlated dynamics depends strongly on the initial phase of the system determined by the interaction parameters ($g_{BB}$, $g_{II}$, $g_{BI}$).

\section{Variational approach and wave function ansatz}
\label{sec:method}

To determine the time-dependent solution of the problem described by the Hamiltonian of Eq.~(\ref{eq:Hamiltionian}) we invoke the ML-MCTDHX method \cite{kronke2013, cao2013, cao2017, kohler2019}. This approach is an \textit{ab-initio} one and optimizes a chosen basis e.g. in terms of the Dirac-Frenkel variational principle~\cite{raab2000}. In particular, the basis set which underlies the many-body wave function $\ket{\Psi^{\rm{MB}}(t)}$ is characterized by a time-dependent and multi-layered structure with individual truncation parameters \footnote{Throughout this work the characterization "many-body" refers to the applied treatment for addressing the properties of the binary mixture. Indeed, the ab-initio ML-MCTDHX method provides access to both intra- and inter-component correlations by considering an expansion of the system's wave function as a superposition of many-particle configurations with variationally optimized single-particle functions.}. Firstly, the many-body wave function is expanded into distinct sets of species functions $\{\ket{\Psi_i^\sigma(t)} \}_{i=1}^{D_{\sigma}}$ with $D_{\sigma}$ denoting the number of the latter for species $\sigma\in\{B,I\}$. Since here we consider a two-component mixture, $\ket{\Psi^{\rm{MB}}(t)}$ is firstly expressed into two such basis sets and, thus, can be written in the form a truncated Schmidt decomposition~\cite{schmidt1907, ekert1995, horodecki2009}
\begin{align}
	\ket{\Psi^{\rm{MB}}(t)} = \sum_{i=1}^{D} \sqrt{\lambda_i(t)} \ket{\Psi_i^B(t)} \ket{\Psi_i^I(t)},
	\label{eq:wavefct_toplayer}
\end{align}
where $D=D_B=D_I$ and $\ket{\Psi_i^\sigma(t)}$ are the so-called natural species functions~\cite{cao2017}. The time-dependent Schmidt coefficients $\lambda_i(t)$ determine the population of the $i$-th natural species function and provide information about the interspecies entanglement~\cite{li2008, horodecki2009}. For instance, in the case that only a single Schmidt coefficient $\lambda_i(t)$ is nonzero, the system is described by a direct product ansatz of species functions indicating the absence of entanglement. On the other hand, the two species are considered to be entangled when more than one Schmidt coefficients are nonzero.

In the next step of the many-body wave function $\ket{\Psi^{\rm{MB}}(t)}$ truncation, each species function is expanded into time-dependent permanents
\begin{align}
	\ket{\Psi_i^\sigma(t)} = \sum_{\vec{n}|N_\sigma} C_{i,\vec{n}}^{\sigma}(t) \ket{\vec{n}(t)}.
\end{align}
Here, each permanent represents one of the $N_\sigma + d_\sigma -1 \choose N_\sigma$ possible configurations to distribute $N_\sigma$ particles on $d_\sigma$ single-particle functions $\ket{\varphi_j^{\sigma}(t)}$. 
A further imposed condition is that the number of occupied single-particle functions for each permanent has to be equal to $N_\sigma$ (indicated by $\vec{n}|N_\sigma$). This expansion enables us to account for intraspecies correlations. Finally, the time-dependent single-particle functions $\ket{\varphi_j^{\sigma}(t)}$ are expanded into a time-independent discrete variable representation \cite{light1985}, which we choose here to consist of 300 grid points in an interval $\{-5, 5\}$ in units of $\tilde{x}_{\rm{box, ho}}$. Additionally, in this work we employ $D=6$ species functions and $d_A=4$, $d_B=6$ single-particle functions for the accurate calculation of the considered systems.

Especially, the multi-layered architecture and the time-dependent basis of the many-body wave function mainly contribute to the high degree of flexibility of the method which enables $\ket{\Psi^{\rm{MB}}(t)}$ to approach the accurate solution for each time instant with a high fidelity even for systems containing a mesoscopic particle number. In this way, the ML-MCTDHX method keeps the number of required wave function coefficients within a computational feasible limit and, at the same time, accounts for the relevant inter- and intraspecies correlations.

%Since the dimension of the Hilbert space grows exponentially with the number of particles, it becomes inevitable to truncate the system's Hilbert space already for a comparatively small number of particles, i.e., of the order of $\sim 10^1$. In the view of this condition the concept of the ML-MCTDHX method is to employ a multi-layered ansatz for the many-body wave function.

\section{Characterization of the ground state}
\label{sec:GS}

In the following we provide an overview of the ground state characteristics of two impurities trapped in a double-well and coupled to a bosonic bath confined either in a box potential or a harmonic oscillator. These ground states will subsequently serve as a starting point for examining the counterflow impurity dynamics immersed in a medium which will be discussed below in Section \ref{sec:dyn}. Unless stated otherwise, the interaction strength between the bath particles is kept fixed to $g_{BB}/\tilde{g}_{\rm{box}}=g_{BB}/\tilde{g}_{\rm{ho}}=0.5$.

\subsection{Main observables of interest}

Let us first introduce the quantities that will be employed for the identification of the ground state phases and the quench dynamics of the two interacting impurities in the cases of a box-confined and a harmonically trapped medium. %\sout{The respective ground state phase diagrams of %the considered composite systems are shown for varying %impurity-impurity ($g_{II}$) and impurity-medium %($g_{BI}$) interaction strengths in Figure %\ref{fig:GS_box}(a) and Figure \ref{fig:GS_ho}(a).} 
The distinction between the emergent ground state configurations is performed with respect to the two-body density distributions of the impurities and the bath particles at $t=0$. The reduced two-body density of two particles of the same species is given by
\begin{align}
\begin{split}
	\rho_{\sigma\sigma}^{(2)}(x_1^\sigma, x_2^\sigma,t) & = \bra{\Psi^{\rm{MB}}(t)} \hat{\Psi}_\sigma^\dagger(x_1^\sigma) \hat{\Psi}_\sigma^\dagger(x_2^\sigma)
	\\& \times \hat{\Psi}_\sigma(x_1^\sigma) \hat{\Psi}_\sigma(x_2^\sigma) \ket{\Psi^{\rm{MB}}(t)},
\end{split}
\end{align}
where $\hat{\Psi}_\sigma^{(\dagger)}(x_1^\sigma)$ denotes the bosonic field operator which annihilates (creates) a particle of species $\sigma\in\{B,I\}$ at position $x_1^\sigma$. In fact, $\rho_{\sigma\sigma}^{(2)}(x_1^\sigma, x_2^\sigma)$ is the probability of finding one particle at $x_1^\sigma$ and, simultaneously, another particle of the same species at $x_2^\sigma$. In the following, we will drop the time parameter since for the ground state $t=0$.

In both considered external confinements of the medium, an increase of the repulsive impurity-medium coupling strength leads to the development of interspecies correlations (entanglement) which eventually impact the ground state configurations \cite{garcia-march2014}. A common measure for quantifying entanglement in a bipartite system is the von Neumann entropy \cite{paskauskas2001, horodecki2009}, defined as
\begin{align}
S^{\rm{vN}} = -\sum_{i=1}^{D} \lambda_i \ln \lambda_i.
\label{eq:SVN}
\end{align}
Recall that $D$ denotes the number of the employed species functions and $\lambda_i$ are the Schmidt coefficients [cf. Eq. (\ref{eq:wavefct_toplayer})]. For a maximally entangled mixture the von Neumann entropy obtains its maximum value, $S_{\rm{max}}^{\rm{vN}} = \ln D$ and $\lambda_i=1/D$. This value corresponds to $S_{\rm{max}}^{\rm{vN}}=1.79$ in our case. In contrast, a vanishing von Neumann entropy indicates a decoupled (i.e. non-entangled) mixture such that the total many-body wave function can be written as a direct product state of the two individual species wave functions.

Moreover, in order to judge the degree of miscibility among the impurity and medium clouds we calculate the interspecies spatial overlap~\cite{jain2011, bandyopadhyay2017} which is quantified through
\begin{align}
\Lambda_{BI} = \frac{[\int \rm{d}x\rho_B^{(1)}(x) \rho_I^{(1)}(x) ]^2}{\int \rm{d}x[\rho_B^{(1)}(x)]^2 \int \rm{d}x[\rho_I^{(1)}(x)]^2}. 
\end{align}
Here, $\rho_{\sigma}^{(1)}(x) = \bra{\Psi^{\rm{MB}}} \hat{\Psi}_\sigma^\dagger(x)\hat{\Psi}_\sigma(x) \ket{\Psi^{\rm{MB}}}$ being the one-body density of $\sigma\in\{B,I\}$ species~\cite{mueller2006}.

\begin{figure*}[t]
	\centering
	\includegraphics[width=\linewidth]{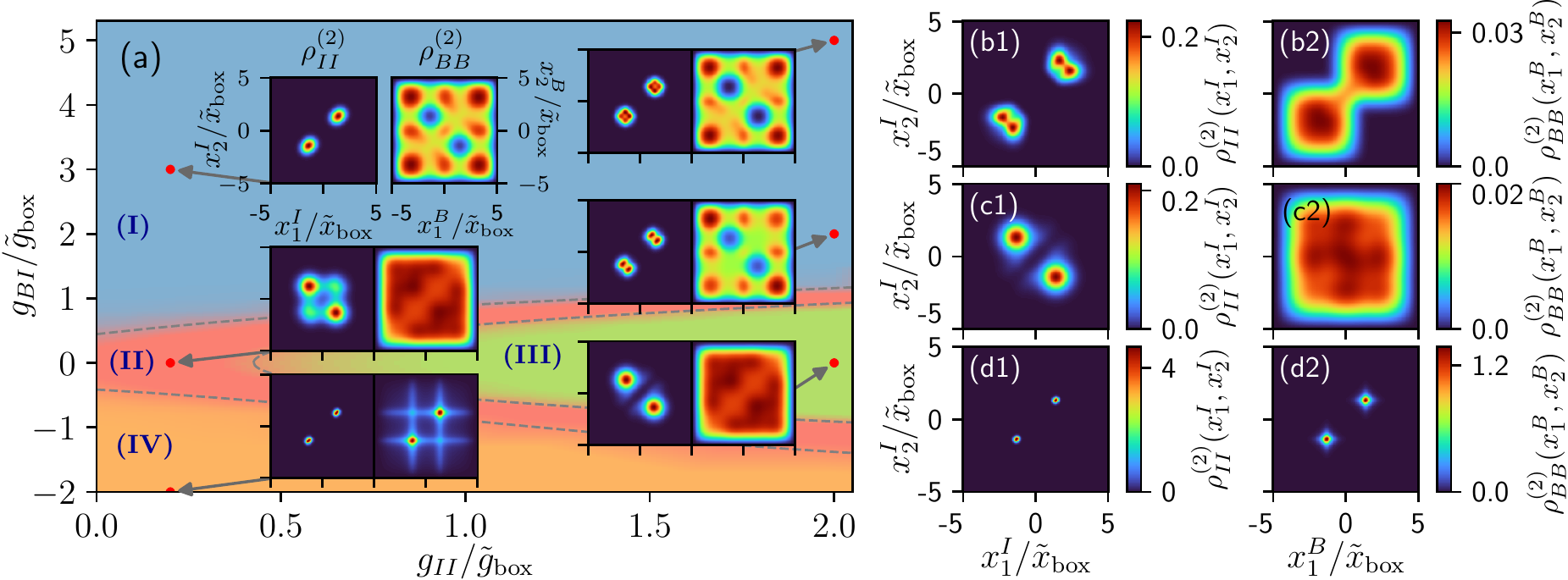}
	\caption{(a) Phase diagram of the impurities-bath ground state with respect to the impurity-medium ($g_{BI}$) and impurity-impurity ($g_{II}$) interaction strengths for constant interactions of the medium ($g_{BB}/\tilde{g}_{\rm{box}}=0.5$). The impurities are confined in a double well and the medium is trapped in a box potential. The respective ground state configurations are characterized in terms of the two-body densities of the bath $\rho_{BB}^{(2)}(x_1^B, x_2^B)$ and the impurities $\rho_{II}^{(2)}(x_1^I, x_2^I)$. All crossovers among the involved phases are smooth. For each inset the range of the color map is optimally chosen and maximally extends from 0 to 2 [0.2] for $\rho_{II}^{(2)}(x_1^I, x_2^I)$ [$\rho_{BB}^{(2)}(x_1^B, x_2^B)$]. In panels (b1)-(d2) the ground state two-body densities of the bath and the impurities are presented in terms of a weakly interacting bath ($g_{BB}/\tilde{g}_{\rm{box}}=0.1$) and strongly interacting impurities ($g_{II}/\tilde{g}_{\rm{box}}=2.0$). The interspecies interaction strength from top to bottom row is $g_{BI}/\tilde{g}_{\rm{box}}=5.0$, $0.2$, $-2.0$, respectively.}
	\label{fig:GS_box}
\end{figure*}

\subsection{Decoupled case $g_{BI}=0$}

Before addressing the ground state properties of the coupled mixture, we focus on the simpler scenario where the bath and the impurities are decoupled from each other ($g_{BI}=0$) and thus they can be treated individually. 
Accordingly, the impurity-bath entanglement is vanishing, i.e. $S^{\rm{vN}}=0$.
Then, the system reduces to two interacting bosons in a double well \cite{zollner2006, zollner2006a, murphy2007} with the bath being homogeneous or harmonically trapped [cf. Figure \ref{fig:GS_box}(a) and Figure \ref{fig:GS_ho}(a) for $g_{BI}=0$]. Here, we distinguish between weakly and strongly coupled impurities. In the former case, the two impurities are delocalized over the two sites of the double well, see the dominant population of the off-diagonal compared to the diagonal elements of $\rho_{II}^{(2)}(x_1^I, x_2^I)$ [cf. insets of Figure \ref{fig:GS_box}(a) and Figure \ref{fig:GS_ho}(a) corresponding to regime (II)]. 

On the other hand, for larger impurity-impurity repulsion $g_{II}$ the density maxima along the diagonal vanish and only density peaks at the off-diagonal remain [cf. inset of Figure \ref{fig:GS_box}(a) belonging to regime (III)]. This configuration of $\rho_{II}^{(2)}(x_1^I, x_2^I)$ is described by the conditional probability of finding one impurity at the left and one impurity at the right site of the double well or vice versa. In this sense, the impurities tend to separate from each other and are anti-correlated within the same site of the double well. Thus, they reside in a Mott-type state. Notice that for a decoupled mixture the above described impurity configurations corresponding to regime (II) and (III) occur independently of the particular trapping geometry of the bath [cf. $g_{BI}=0$ in Figures \ref{fig:GS_box}(a) and \ref{fig:GS_ho}(a)]. Thereby, the medium extends almost homogeneously over the box potential as also reflected by the shape of its two-body density [see inset of Figure \ref{fig:GS_box}(a)], while in the harmonically trapped scenario it exhibits a Gaussian profile [cf. inset of Figure \ref{fig:GS_ho}(a)].

\subsection{Finite interspecies interactions with the medium confined in a box potential}
\label{sec:gs_box_finite_gBI}

Having analyzed the spatial configurations of the interacting impurities for a suppressed impurity-bath coupling we then discuss the ground state properties of the composite system when the bath is confined in a box potential and the interspecies interaction strength $g_{BI}$ becomes finite. The respective ground state phase-diagram is mapped out and presented in Figure~\ref{fig:GS_box}(a) based on the underlying two-body configurations identified in $\rho_{BB}^{(2)}(x_1^B, x_2^B)$ and $\rho_{II}^{(2)}(x_1^I, x_2^I)$.
Overall, we find that upon varying $g_{BI}$ and $g_{II}$, the system deforms smoothly across the different phases, which are analyzed in detail in the following, see also~\footnote{To discern among the different phases we rely on the integrated two-body density of the impurities $\int_{-\infty}^{0} dx_{1}^Idx_{2}^I\rho_{II}^{(2)}(x_1^I, x_2^I)$. It immediately distinguishes the cases of two separated (coalesced) impurities where it acquires the value 0.0 (0.5). Furthermore, it allows to also identify intermediate regimes where both the diagonal and the off-diagonal elements of $\rho_{II}^{(2)}(x_1^I, x_2^I)$ are populated. Notice that for a box confined medium we additionally weight the integrated density by $\mathrm{sgn}(g_{BI})$ in order to distinguish between regimes (I) and (III) appearing at strongly repulsive and attractive $g_{BI}$, respectively.}.

In the case of strong repulsive $g_{BI}$, the mixture enters regime (I) [see Figure \ref{fig:GS_box}(a)] \footnote{Throughout this work we refer to weakly (strongly) interacting impurities when $g_{II}<g_{BB}$ ($g_{II}>g_{BB}$). Similarly, we use the term weak (intermediate, strong) impurity-medium interaction strength if $g_{BI}<g_{BB}$ ($g_{BI}>g_{BB}$, $g_{BI}\gg g_{BB}$).}.
Here, the impurities two-body density exhibits two peaks along its diagonal meaning that the two impurities occupy simultaneously a single site of the double-well. Such a behavior is referred to as the coalescence of the impurities and has been observed also for the case where the impurities and bath particles are both harmonically confined \cite{dehkharghani2018}. 
Intuitively, we explain this behavior as follows. One impurity lying at a specific site of the double well repels the bath particles and, thereby, creates an effective hole which attracts the other impurity~\cite{pflanzer2009, pflanzer2010}.
%In turn, due to the impurity-medium repulsion the bath particles avoid to stay close with the impurities which cluster at a single site. Thus, the probability for two bath particles to lie at the opposite double-well sites is highly suppressed which is manifested by the holes in the off-diagonal of $\rho_{BB}^{(2)}(x_1^B, x_2^B)$ [see inset in regime (I) of Figure~\ref{fig:GS_box}(a)]. However, a situation in which two bath particles reside, e.g., at the right double-well site while both impurities occupy the left site is still conceivable. 
%This can be seen by the non-vanishing density at the diagonal of $\rho_{BB}^{(2)}(x_1^B, x_2^B)$ in the inset of Figure~\ref{fig:GS_box}(a) which denotes the probability of finding two bath particles at the same location.
On the other hand, the impurities impact accordingly the bath. 
This backaction manifests, for instance, in the off-diagonal parts of the medium's two-body density which exhibits strongly suppressed spatial regions at the location of the impurities [see inset in regime (I) of Figure~\ref{fig:GS_box}(a)]. 
Indeed, the probability to find two bath particles at positions corresponding to opposite double-well sites is vanishing. 
This is due to the fact that the impurities lie both either at the left or at the right double-well site as it becomes apparent from their reduced two-body density. 
However, a configuration where two bath particles reside simultaneously at the same double-well site is still conceivable, assumed that the impurities are at the opposite site, thereby, avoiding the bath particles [note the non-vanishing density at the diagonal of $\rho_{BB}^{(2)}(x_1^B, x_2^B)$ in the inset of Figure~\ref{fig:GS_box}(a)].

Increasing the impurity-medium interaction strength within regime (I) for a fixed $g_{II}/\tilde{g}_{\rm{box}} \in [0, 2]$ we observe two prominent features appearing in terms of $\rho_{BB}^{(2)}(x_1^B, x_2^B)$. Firstly, the two-body density holes at the off-diagonal of $\rho_{BB}^{(2)}(x_1^B, x_2^B)$ become more pronounced for increasing $g_{BI}$ and, secondly, for a $g_{BI}/\tilde{g}_{\rm{box}} \gtrsim 2.5$ two bath particles are correlated at the most right and most left or at opposite sites of the bath cloud [see the outermost density peaks at the diagonal and off-diagonal elements of $\rho_{BB}^{(2)}(x_1^B, x_2^B)$ in the inset of Figure~\ref{fig:GS_box}(a)]. From this latter behavior we can conclude that the bath particles exhibit two-body long distance correlations. Moreover, we note that in case of strong impurity-impurity repulsions, e.g. for $g_{II}/\tilde{g}_{\rm{box}}=2.0$, the two-body state of the impurities begins to fermionize and the diagonal peaks of $\rho_{II}^{(2)}(x_1^I, x_2^I)$ broaden and, eventually, fragment \cite{liu2015} [see corresponding inset of Figure \ref{fig:GS_box}(a)].

Similarly to the coalescence of the impurities in the repulsive case [regime (I)] also in the attractive scenario the two impurities simultaneously occupy either the left or right site of the double well, as it can be deduced from their diagonal and highly localized two-body density configuration $\rho_{II}^{(2)}(x_1^I, x_2^I)$ [see the inset of Figure~\ref{fig:GS_box}(a)]. In both the repulsive and the attractive cases, the bath mediates an induced attractive interaction between the impurities such that the latter coalesce and tend to occupy the same double-well site. 
For a more detailed discussion regarding the presence of the attractive induced interactions between the impurities via their relative distance, see Appendix~\ref{ap:box_rel_dist}. Furthermore, due to the attractive interactions the bath particles localize in the vicinity of the impurities such that also the two-body density of the medium exhibits two dominant peaks along the diagonal [Figure \ref{fig:GS_box}(a)]. Also, we remark that regime (II) in Figure \ref{fig:GS_box}(a) includes ground states corresponding to delocalized impurities, i.e., where the diagonal and off-diagonal elements of $\rho_{II}^{(2)}(x_1^I, x_2^I)$ are simultaneously populated. However, with varying $g_{II}$ and $g_{BI}$ the particular density peaks are deformed compared to the depicted insets of regime (II) in Figure \ref{fig:GS_box}(a). For instance, in the case of $g_{II}/\tilde{g}_{\rm{box}}=2.0$ and $g_{BI}/\tilde{g}_{\rm{box}}=1.0$ corresponding to regime (II) in Figure~\ref{fig:GS_box}(a) the two-body density of the impurities exhibits peaks at its off-diagonal elements [similar to regime (III)] and fragmented density humps at its diagonal [as observed in regime (I)].

\begin{figure}[t]
	\centering
	\includegraphics[width=1.0\linewidth]{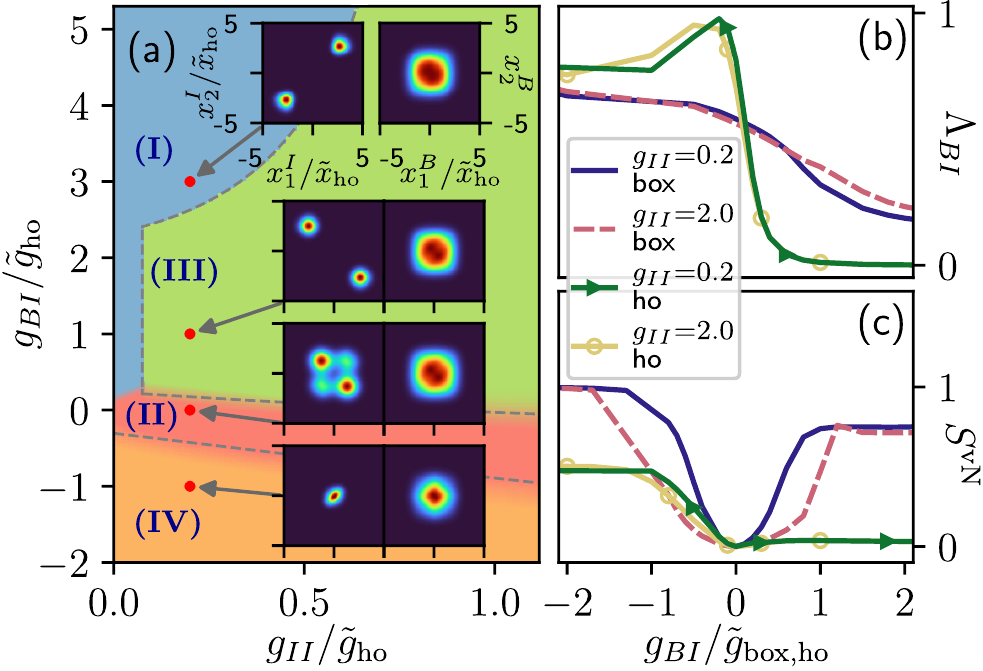}
	\caption{(a) Ground state phase diagram of the impurities inside a harmonically trapped medium. The crossover from region (I) to (III) is abrupt marked by the dashed line, while all others are smooth. (b) Spatial overlap $\Lambda_{BI}$ and (c) von Neumann entropy $S^{\rm{vN}}$ between the bath and the impurities for varying impurity-medium interaction strength and different fixed impurity-impurity couplings (in units of $\tilde{g}_{\rm{box,ho}}$) as well as for distinct external potentials of the bath (see legend). The interaction strength among the bath particles is fixed to $g_{BB}/\tilde{g}_{\rm{box}}=g_{BB}/\tilde{g}_{\rm{ho}}=0.5$.}
	\label{fig:GS_ho}
\end{figure}

Moreover, increasing (decreasing) the interspecies interaction to large repulsive (attractive) values leads to a noticeable growth of the von Neumann entropy. 
Namely, the impurities become highly entangled with the bath [cf. Figure~\ref{fig:GS_ho}(b)]. 
At the same time the impurities and the bath share a finite spatial overlap with each other for all the considered values of $g_{BI}$ and $g_{II}$ [Figure \ref{fig:GS_ho}(c)]. Therefore, since these two features constitute a basic requirement for the formation of quasi-particles, e.g. as discussed in Refs. \cite{rath2013, mistakidis2019a,mistakidis2019c, mistakidis2020}, in principle, the impurities can be dressed by the excitations of the bath and, thus, form Bose polarons.

Additionally, we investigated the ground state of the system for a weakly interacting bath, e.g. for $g_{BB}/\tilde{g}_{\rm{box}}=0.1$. As we will argue, the enhanced compressibility of the bath can alter the ground state configurations and this is evident, among other observables, in terms of the two-body density. In Figures~\ref{fig:GS_box}(b1)-(d2) we exemplary present the two-body densities of the impurities and the bath particles for $g_{BB}/\tilde{g}_{\rm{box}}=0.1$ and $g_{II}/\tilde{g}_{\rm{box}}=2.0$. Here, the impurities two-body densities exhibit a qualitatively similar structure with the ones corresponding to a moderately interacting bath [Figure~\ref{fig:GS_box}(a)]. For instance, a localization at the sites of the double-well is observed for strongly attractive $g_{BI}$ [Figure~\ref{fig:GS_box}(d1)] as well as an anti-correlated behavior for weak impurity-medium couplings [Figure~\ref{fig:GS_box}(c1)] and the coalescence of the impurities for strong $g_{BI}$ [Figure~\ref{fig:GS_box}(b1)]. Similarly, the two-body density of the weakly interacting medium resembles the one of a moderately interacting bath in the cases of weak attractive and repulsive as well as strong attractive $g_{BI}$ [Figures~\ref{fig:GS_box}(c2) and (d2)]. However, for strong repulsive $g_{BI}$ the medium's two-body density is modified for weak $g_{BB}$, i.e., the off-diagonal parts of $\rho_{BB}^{(2)}(x_1^B, x_2^B)$ are depopulated and only the diagonal ones are occupied [Figure~\ref{fig:GS_box}(b2)]. We attribute this behavior to the increased compressibility of the bath which suppresses correlations between two bath particles residing at longer distances e.g. the opposite edges of the cloud as observed in Figure~\ref{fig:GS_box}(a). Additionally, this behavior is accompanied by strong anti-correlations between the impurities and the bath particles, i.e., the respective two-body density $\rho_{BI}^{(2)}(x_1^B, x_2^I)$ exhibits only peaks at its off-diagonal (not shown).

Furthermore, we have found that in the corresponding ground state phase diagram of a weakly interacting medium ($g_{BB}/\tilde{g}_{\rm{box}}=0.1$) with $N_B=20$, the regimes (II) and (III) shrink as compared to the $g_{BB}/\tilde{g}_{\rm{box}}=0.5$ case shown in Figure \ref{fig:GS_box}(a). 
Specifically, their phase boundaries are shifted towards the line corresponding to $g_{BI}=0$.
A similar, but less pronounced, shift of the phase boundaries is observed when increasing the number of bath particles to $N_B=30$ but keeping $g_{BB}$ fixed. 
Summarizing, both decreasing $g_{BB}$ or increasing $N_B$ while considering fixed all other parameters leads to an enhancement of the magnitude of the attractive induced interactions between the impurities when $g_{BI}$ is switched on towards finite attractive or repulsive values.

\subsection{Harmonically trapped medium}

We then proceed to analyze in more detail the system consisting of a harmonically trapped medium. This change of the external confinement reduces the mobility of the bath particles which are then naturally bounded by the harmonic oscillator around the trap center. 
The respective phases presented in Figure \ref{fig:GS_ho}(a) feature smooth crossovers among them besides the one between the regimes (I) and (III) which is abrupt. 
To testify the "smoothness" of the underlying crossover regions we track, as in the box-confined scenario, the behavior of the impurities two-body densities (cf. \cite{Note3}). 

An increasing impurity-medium repulsion such that $g_{BI}>g_{BB}$ leads to a phase separation between the impurities and the bath particles as it is captured by the diminishing spatial overlap depicted in Figure \ref{fig:GS_ho}(b). In this case the impurities are no longer dressed by the excitations of the bath and, thus, the quasi-particle notion is essentially lost \cite{goold2011, mistakidis2019a, mistakidis2020}. Thereby, we distinguish between two cases according to the impurity-impurity interaction strength. In the case of weak $g_{II}$ and strong $g_{BI}$ corresponding to regime (I) in Figure \ref{fig:GS_ho}(a) the impurities coalesce in a similar manner as described above (see also Ref. \cite{dehkharghani2018}). However, as the impurity-impurity interaction strength becomes large enough or the impurity-medium repulsion sufficiently small, regime (III) is accessed in which the impurities spatially separate. This is identified by the exclusive population of the off-diagonals of $\rho_{II}^{(2)}(x_1^I, x_2^I)$ [cf. corresponding inset of Figure \ref{fig:GS_ho}(a)]. Turning to strongly attractive impurity-medium interaction strengths assigned as regime (IV) in Figure \ref{fig:GS_ho}(a) a localization of the impurities in the barrier of the double-well is observed, see in particular the elongation of $\rho_{II}^{(2)}(x_1^I, x_2^I)$ along its diagonal. This property is related to the formation of a bipolaron, referring to a dimer bound state consisting of two polarons \cite{camacho-guardian2018a, mistakidis2020, keiler2021}. We base our argument of bipolaron formation on the following observations which have also been used in Refs~. \cite{camacho-guardian2018a} to expose the existence of such states in three-dimensions.
The continuous decrease of the so-called bipolaron energy $E_{\mathrm{BP}}=E_2-E_1+E_0$ has been verified for increasing impurity-medium attraction, where $E_i$ denotes the total energy of the bosonic gas containing $i=0,1,2$ impurities.
In the same manner, also the size of the dimer state quantified in our case by $1/\sqrt{\langle \hat{r}_{II}^2 \rangle}$ increases for larger $g_{BI}$ (not shown here).

Concluding, let us mention in passing that similarly to the case of a box-confined medium, the main requirements for the formation of Bose polarons are also fulfilled in the presence of a harmonic trap. The only exception consists of the region of phase-separation among the impurities and the medium at $g_{BI}>g_{BB}$. Thus, the impurities response, to be presented below, can be interpreted as the counterflow correlated dynamics of two quasi-particles, here Bose polarons.

% species mean-field: regime with drifting apart not observed

% enlarge the box: drifting regime not observed

\section{Collisional many-body dynamics for a box confined medium}
\label{sec:dyn}

%In section \ref{sec:GS} we discussed the ground state properties of two impurities confined in a double well potential and immersed in a bosonic medium with respect to varying impurity-impurity and impurity-medium interaction strengths

Next, we investigate the time evolution of the composite system upon suddenly ramping down the central barrier of the impurities' double-well potential such that they are henceforth externally confined in a harmonic oscillator and, thus, their counterflow dynamics is triggered. In a decoupled mixture ($g_{BI}=0$) this quench results in an undamped periodic impurities motion where they collide and subsequently expand repeatedly. Turning to finite impurity-medium couplings the response is substantially altered and depends strongly on the trapping potential of the medium, as we will argue below. 
%We start with the investigation of the dynamics when the %bath is confined in a box potential and, afterwards, %discuss the case of a harmonically trapped medium.
%Throughout this section the bath is confined in a box %potential. 

\subsection{Response through the time-evolution of the density}
\label{sec:dyn_box}

\begin{figure}[t]
	\centering
	\includegraphics[width=1.0\linewidth]{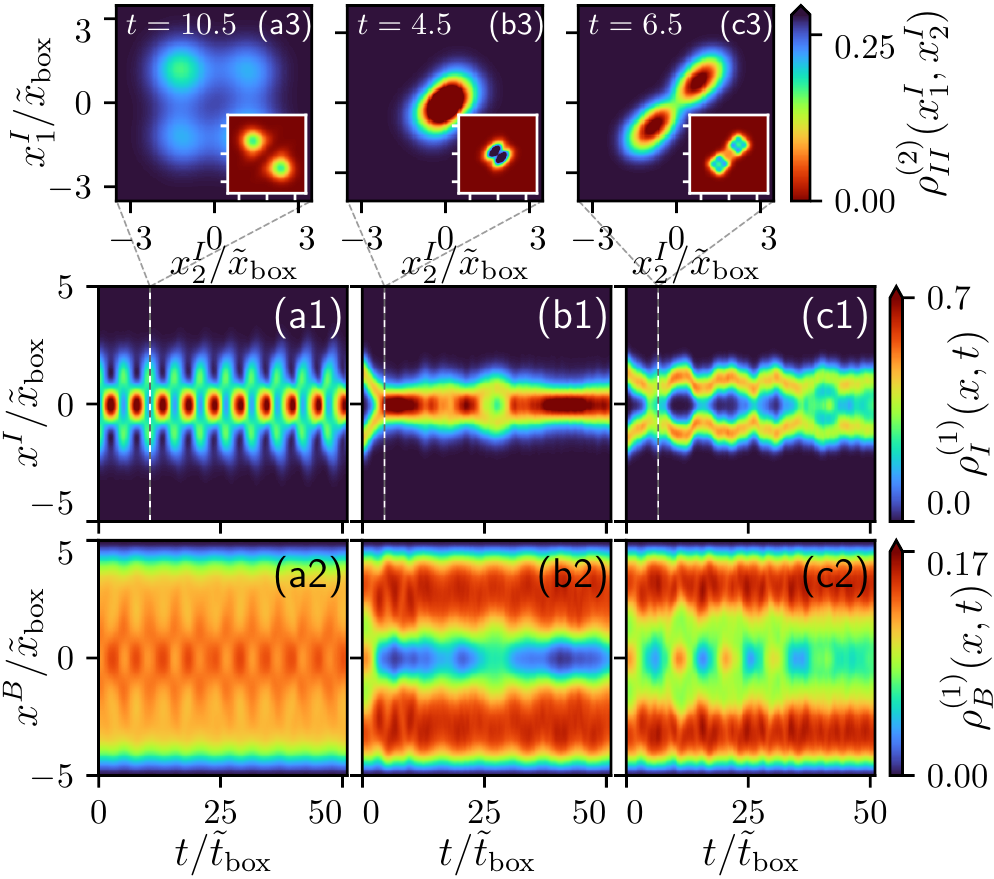}
	\caption{Time evolution of the one-body density of (a1)-(c1) the harmonically trapped impurities and (a2)-(c2) the bath particles confined in a box potential. Each column represents the dynamics for a fixed impurity-medium interaction strength which is from left to right $g_{BI}/\tilde{g}_{\rm{box}}=-0.2, 1.5, 5.0$. In all cases, the impurity-impurity coupling is $g_{II}/\tilde{g}_{\rm{box}}=0.2$. (a3)-(c3) Snapshots of the impurities two-body density at different time instants (see legends). The insets depict the two-body densities at the same time instants and scales, but for two strongly interacting impurities, i.e. $g_{II}/\tilde{g}_{\rm{box}}=2.0$. At those time instants, the one-body densities for two weakly and strongly interacting impurities reveal similar features, thus, allowing the comparison on the two-body density level.}
	\label{fig:box_gpop_dm2}
\end{figure}

We monitor the counterflow dynamics of two impurities coupled via $g_{BI}$ to a bosonic medium trapped in a box potential. After ramping down the potential barrier of the double well, the impurities are left to evolve in the resulting harmonic trap. As a first step, we categorize the emergent dynamical response regimes by inspecting the one-body densities $\rho_{I}^{(1)}(x,t)$ and $\rho_{B}^{(1)}(x,t)$ depicted in Figures \ref{fig:box_gpop_dm2}(a1)-(c1) and (a2)-(c2), respectively. The impurity-impurity coupling is kept fixed $g_{II}/\tilde{g}_{\rm{box}}=0.2$ and only the impurity-medium interaction is varied. We are able to identify four distinct dynamical response regimes taking place at strong attractive ($g_{BI}/\tilde{g}_{\rm{box}}<-0.8$), weak attractive and repulsive ($-0.8\lesssim g_{BI}/\tilde{g}_{\rm{box}}<0.8$), intermediate repulsive ($0.8\lesssim g_{BI}/\tilde{g}_{\rm{box}}\lesssim 2.0$) and strongly repulsive ($2<g_{BI}/\tilde{g}_{\rm{box}}$) values of $g_{BI}$. 
These dynamical regimes are, of course, inherently related to the corresponding phases unraveled in the ground state of the system, see Figure~\ref{fig:GS_box}(a). Note that the behavior of the one-body densities in the respective regions do not qualitatively alter for varying $g_{II}$ from small to large repulsive values at least in the range of $0\leq g_{II}/\tilde{g}_{\rm{box}}\leq 2.0$ considered herein. Only by inspecting higher-body observables, such as the reduced two-body density, reveals significant alterations of the impurities response regarding variations of $g_{II}$.

\subsection{Dynamics for weakly attractive and repulsive impurity-medium couplings}

In the case of either weakly attractive or repulsive $g_{BI}$ the impurities one-body densities feature a periodic motion consisting of a collision and an expansion of their cloud [cf. Figure \ref{fig:box_gpop_dm2}(a1) for $g_{BI}/\tilde{g}_{\rm{box}}=-0.2$]. As a consequence, the bath is only weakly perturbed from its initial homogeneous profile showing small amplitude distortions at the instantaneous location of the impurities [see Figure \ref{fig:box_gpop_dm2}(a2)]. This response emerges when considering initial configurations corresponding to the interaction regimes (II) and (III) discussed in Figure \ref{fig:GS_box}(a). Interestingly, the time-evolution of the one-body density does not depend strongly on variations of the impurity-impurity interaction strength~\footnote{We have verified that varying from $g_{II}/\tilde{g}_{\rm{box}}=0.0$ to 2.0 the overall shape of the one-body density of both species does not qualitatively alter for fixed $g_{BI}$. However, quantitative deviation are in play, for example, for strong $g_{II}$ for which the periodic motion consisting of the impurities collision and expansion is more pronounced than for weak $g_{II}$.}. Therefore, one has to rely on two-body observables, such as the two-body density $\rho_{II}^{(2)}(x_1^I, x_2^I)$, e.g. presented in Figure \ref{fig:box_gpop_dm2}(a3) for $g_{BI}/\tilde{g}_{\rm{box}}=-0.2$ and $g_{II}/\tilde{g}_{\rm{box}}=0.2$ at $t/\tilde{t}_{\rm{box}}=10.5$ and in the respective inset for two strongly interacting impurities with $g_{II}/\tilde{g}_{\rm{box}}=2.0$. In the former case, the two weakly interacting impurities are initially and throughout the evolution delocalized over both sites of the double well since both the diagonal and the off-diagonal elements of $\rho_{II}^{(2)}(x_1^I, x_2^I)$ are non-vanishing. 

However, for strongly interacting impurities we find that in the course of the evolution a pronounced correlation hole occurs \cite{mistakidis2020}, i.e., solely the off-diagonal of $\rho_{II}^{(2)}(x_1^I, x_2^I)$ is populated [cf. inset of Figure \ref{fig:box_gpop_dm2}(a3)]. In other words, due to the strong repulsion the impurities reside at spatially opposite positions and avoid each other during the dynamical evolution. The respective dynamics of the impurities is characterized by the periodic expansion and contraction of their cloud around the trap center while avoiding to reside at the same location, see the correlation hole of $\rho_{II}^{(2)}(x_1^I, x_2^I)$. However, independently of $g_{II}$ the impurities remain within the medium, thus, forming a polaron due to the finite $g_{BI}$.

\subsection{Time-evolution for intermediate interspecies repulsions}

Increasing the impurity-medium repulsion to intermediate values ($g_{BI}/\tilde{g} \lesssim2.0$) and, thereby, entering regime (I) in Figure~\ref{fig:GS_box}(a) a comparatively altered response is realized. Indeed, once the two impurities collide at the trap center, they remain localized~\footnote{A similar localization behavior of the impurities is also observed for the dynamical evolution at intermediate and strong attractive $g_{BI}$ (not shown here). However, in this case also the medium localizes around the trap center.}, see Figure \ref{fig:box_gpop_dm2}(b1) for $g_{BI}/\tilde{g}_{\rm{box}}=1.5$. Consequently, because of the repulsive character of the employed $g_{BI}$, the bath is pushed towards the edges of the box with each density branch undergoing weak amplitude oscillations due to its reflection from the walls of the confining box [Figure \ref{fig:box_gpop_dm2}(b2)]. Simultaneously, the bosonic medium becomes highly depleted, meaning that higher-lying natural orbitals (being the eigenvalues of the reduced one-body density matrix $\rho_{B}^{(1)}(x,x', t)$) are macroscopically populated. 
Accordingly, the bosonic gas is correlated and deviates from a perfect BEC.
This behavior is in contrast to the case of $g_{BI}/\tilde{g}_{\mathrm{box}} < 0.8$ where the first orbital is dominantly occupied.

However, the corresponding two-body density $\rho_{II}^{(2)}(x_1^I, x_2^I)$ when the impurities collide at the trap center at $t/\tilde{t}_{\rm{box}}=4.5$ is shown in Figure \ref{fig:box_gpop_dm2}(b3). The elongated shape of $\rho_{II}^{(2)}(x_1^I, x_2^I)$ along the diagonal indicates the presence of an attractive interaction between the impurities induced by the coupling with the bath \cite{dehkharghani2018, mistakidis2019a, mistakidis2020}. Apparently the strength of this induced attraction is larger than for $g_{BI}/\tilde{g}_{\rm{box}}=0.2$. Moreover, by comparing this case to the one of strongly interacting impurities ($g_{II}/\tilde{g}_{\rm{box}}=2.0$), we find once again significant differences only on the behavior of their two-body density. At $t=0$, the shape of $\rho_{II}^{(2)}(x_1^I, x_2^I)$ corresponds to two coalesced impurities whose density peaks lay on its diagonal and are fragmented [cf. inset of Figure \ref{fig:GS_box}(a)]. In the course of the dynamics the impurities collide at the center where they remain in the course of the evolution [see inset of Figure \ref{fig:box_gpop_dm2}(b3)]. Thereby, the strong repulsion between the impurities hinders a population at the exact diagonal of the two-body density. Notice their small spatial overlap with the bath hinting towards their suppressed dressing~\footnote{However, we remark that for an adequate description of the quasi-particle notion more elaborated measures need to be considered such as the residue, which goes beyond the present scope.}.

\subsection{Dynamical response for strong impurity-medium repulsions}

For even stronger impurity-medium repulsions ($g_{BI}/\tilde{g}>2.0$), the impurities once they collide around $x=0$ they drift apart from each other and collide again [see Figure \ref{fig:box_gpop_dm2}(c1) for $g_{BI}/\tilde{g}_{\rm{box}}=5.0$]. This behavior is repeated in the course of time with a damped collision amplitude. Thereby, the population on the diagonal of $\rho_{II}^{(2)}(x_1^I, x_2^I)$ becomes narrower as compared to the case of intermediate repulsive $g_{BI}$, indicating the presence of an even stronger strength of induced interactions [cf. Figure \ref{fig:box_gpop_dm2}(c3)]. Again two strongly repulsive impurities fragment along the diagonal of $\rho_{II}^{(2)}(x_1^I, x_2^I)$ [see the inset of Figure \ref{fig:box_gpop_dm2}(c3)] with the two respective density fragments exhibiting each four faint maxima which persist until an evolution time of $t/\tilde{t}_{\rm{box}}=8$. During the evolution the one-body density of the bath allocates at the edges of the box when the impurities collide and reoccupies the trap center when the latter drift apart, see also below for a more detailed discussion.

\begin{figure}[t]
	\centering
	\includegraphics[width=1.0\linewidth]{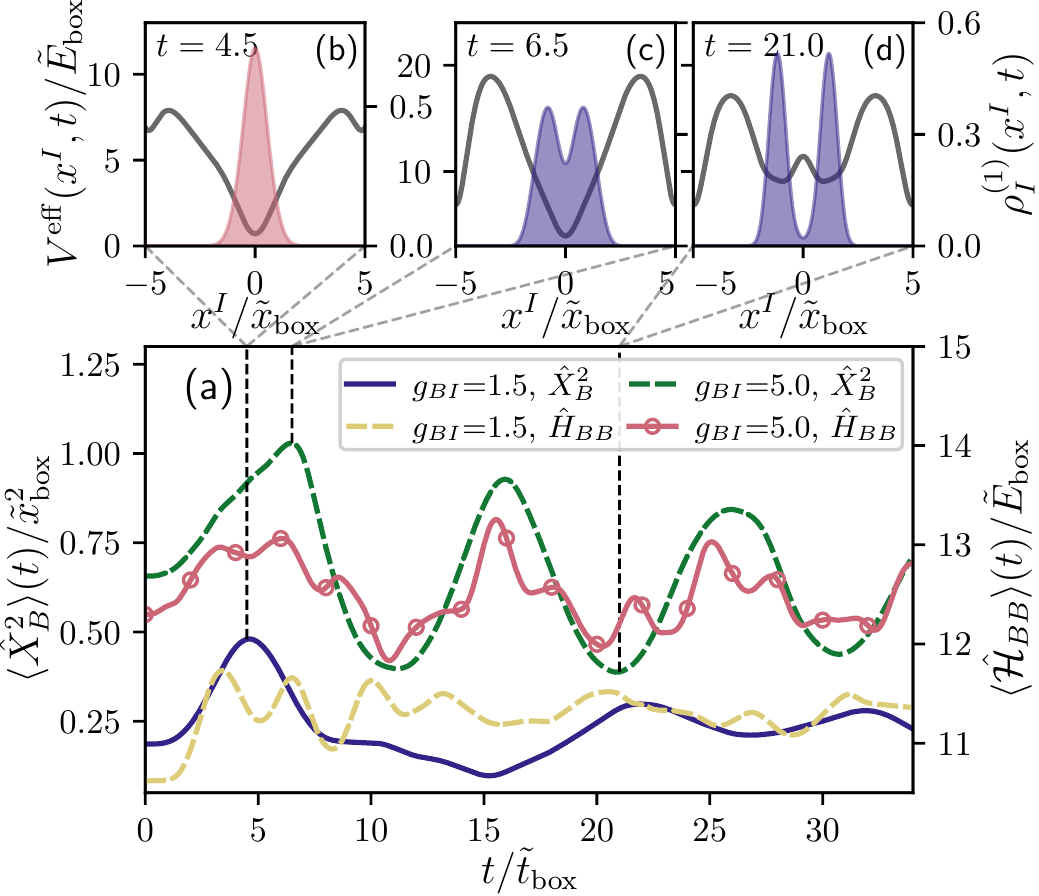}
	\caption{(a) Dynamics of the width of the medium cloud (captured by the spatial variance $\langle \hat{X}_B^2\rangle(t)$) and internal interaction energy of the bath particles $\langle \hat{\mathcal{H}}_{BB}\rangle$ for different impurity-medium couplings (see legend) and fixed impurity-impurity interaction strength $g_{II}/\tilde{g}_{\rm{box}}=0.2$. In (b)-(d) profiles of the impurities one-body density $\rho_I^{(1)}(x^I,t)$ are shown together with their effective potential (gray lines) for specific time instants (see legends). In panel (b) $g_{II}/\tilde{g}_{\rm{box}}=1.5$ while in (c) and (d) $g_{II}/\tilde{g}_{\rm{box}}=5.0$. The time and interaction are expressed in units of $\tilde{t}_{\rm{box}}$ and $\tilde{g}_{\rm{box}}$, respectively.}
	\label{fig:box_x2_effpot}
\end{figure}

We attribute the impurities' density splitting for large $g_{BI}$, even though the induced attractive interaction is higher in this case than for intermediate repulsive couplings, to the finite size of the considered box potential. In order to elucidate the underlying mechanism we show in Figure \ref{fig:box_x2_effpot}(a) the time-dependent spatial variance of the medium $\langle \hat{X}_B^2\rangle(t)$ which serves as a measure for the instantaneous spatial extension of the medium's cloud \cite{fukuhara2013, ronzheimer2013}. At $t=0$ and for sufficiently strong repulsive $g_{BI}$ the impurities reside both either at the right or at the left site of the double well while the bath particles avoid these pairs [cf. regime (I) in Figure \ref{fig:GS_box}(a)], leading for larger $g_{BI}$ to an increased $\langle \hat{X}_B^2\rangle(t=0)$ [see Figure \ref{fig:box_x2_effpot}(a)].

Subsequently, after ramping down the barrier of the double well, the impurities collide around $x=0$ enforcing the medium to depopulate the trap center, a process that results in the increase of $\langle \hat{X}_B^2\rangle(t)$. To facilitate further our discussion, we provide specific profiles of the impurities' one-body density and their effective potential \cite{theel2020, mistakidis2019c, mistakidis2020}. The effective potential is constructed from the superposition of the impurities' (post-quench) harmonic oscillator and the one-body density of the bath weighted by the impurity-medium coupling strength. It reads,
\begin{align}
	V^{\rm{eff}}(x^I,t) = V_I^{\rm{dw}}(x^I) + N_Bg_{BI}\rho_B^{(1)}(x^I, t).
	\label{eq:eff_pot}
\end{align}
A maximum of $\langle \hat{X}_B^2\rangle(t)$ is reached, i.e. the spatial extend of the medium is largest, when the impurities allocate at the trap center [see Figures \ref{fig:box_x2_effpot}(b) and (c)]. Thereby, the impurities transfer energy to the medium leading to an increased interaction energy between the particles of the latter, i.e. $\langle \hat{\mathcal{H}}_{BB}\rangle = \langle g_{BB} \sum_{i<j} \delta(x^B_i - x^B_j) \rangle$, is maximized [cf. Figure~\ref{fig:box_x2_effpot}(a)]. After reaching a maximum of $\langle \hat{\mathcal{H}}_{BB}\rangle$ the bath reoccupies the trap center as indicated by the reduction of $\langle \hat{X}_B^2\rangle(t)$ and the impurities' density splits again, a behavior that is repeated in the course of time. As argued below, this dynamical response can be attributed to finite size effects stemming from the size of the medium's box potential.

\subsection{Impact of the barrier height, atom number and box size on the impurity dynamics}

In order to check the robustness of the observed dynamical response regimes of the impurities against the system parameters we have additionally varied the height of the double well $h_I$, the size of the box potential $L_B$ and the number of bath particles $N_B$. For small $g_{BI}$ and fixed $g_{II}/\tilde{g}_{\rm{box}}=0.2$, increasing the height of the double well from $h_I/\tilde{E}_{\rm{box}}\tilde{x}_{\rm{box}}^{-1}=2$ to 7 leads to a crossover of the impurities, i.e. from a superfluid to a Mott-insulating phase. The former phase corresponds to a two-body density $\rho_{II}^{(2)}(x_1^I, x_2^I)$ where in all quarters prominent density peaks are present. 
In the latter case only the off-diagonal of $\rho_{II}^{(2)}(x_1^I, x_2^I)$ is populated [compare with the ground state configurations in regime (III) of Figure \ref{fig:GS_box}(a)]. Further increasing the impurity-medium coupling to intermediate repulsive values the impurities coalesce again independently of the barrier height of the double-well. Thereby, the impurities' ground state corresponds to the one of regime (I) depicted in Figure \ref{fig:GS_box}(a). Regarding the dynamical response of the above-described system, we did not find a qualitatively different behavior when $h_I$ is varied but rather a shifting of the identified regimes towards a larger value of $g_{BI}$.

In Section \ref{sec:gs_box_finite_gBI} it was mentioned that the ground state phase diagram is altered with respect to variations of $g_{BB}$ and $N_B$. 
As such, also the dynamical response of the system is affected upon tuning these parameters by means that the specific dynamical features are realized for smaller values of $g_{BI}$. 
This is attributed to the fact that the impurities feature an enhanced magnitude of attractive induced interactions for either a decreased $g_{BB}$ (e.g. $g_{BB}/\tilde{g}_{\rm{box}}=0.1$ and especially in the interval $|g_{BI}|<g_{BB}$) or an increased $N_B$ (for instance at $N_B=30$). Consequently, they localize around the trap center after their first collision for a smaller $g_{BI}$ than for $g_{BB}/\tilde{g}_{\rm{box}}=0.5$ or a larger $N_B$ such as $N_B=30$. This effect is arguably more prominent in the former scenario and during the dynamics manifests by the enhanced impurities localization at the trap center.

An opposite behavior is observed for varying only the size of the medium's box potential $L_B$. In this case, an increasing $L_B$ broadens the medium such that the impurities are less affected by the presence of bath particles which, eventually, for intermediate repulsive $g_{BI}$ increases the relative distance between the impurities for a larger $L_B$ (not shown here). However, when the box size is of the order of the distance of the double well minima, the bath particles localize at the trap center (between the two double-well sites) for intermediate repulsive impurity-medium interaction strengths. Additionally, for strong repulsive $g_{BI}$ an increasing box size leads to the localization of the impurities around $x=0$ and a dynamical response similar to the one observed in Figure~\ref{fig:box_gpop_dm2}(c1) is absent. This holds also when we simultaneously increase the number of bath particles and the size of the box potential while keeping the ratio $N_B/L_B$ fixed~\footnote{Since in this scenario the size of the impurities double-well potential is not altered, the absence of the dynamical response corresponding to Figure~\ref{fig:box_gpop_dm2}(c1) provides further evidence for the involvement of finite size effects in this latter case.}.

\section{Impurities dynamics for the harmonically trapped bath}
\label{sec:dyn_ho}

Next, we examine the counterflow dynamics (induced by the same quench protocol) of the two impurities coupled to a harmonically confined bath. Since in this case the medium tends to localize at the trap center a phase separation between the two species is facilitated for intermediate to strong repulsive impurity-medium interaction strengths which eventually prohibits a subsequent dynamical mixing of the species. Indeed, we find that for values larger than $g_{BI}/\tilde{g}_{\rm{ho}}=0.6>g_{BB}/\tilde{g}_{\rm{ho}}$, corresponding to a vanishing spatial overlap at $t=0$ [cf. Figure \ref{fig:GS_ho}(b)], the initial phase separation between the impurities and the medium persists also in the course of the propagation. However, for $g_{BI}/\tilde{g}<0.6$ an intriguing response is observed.

\begin{figure}[tt]
	\centering
	\includegraphics[width=1.0\linewidth]{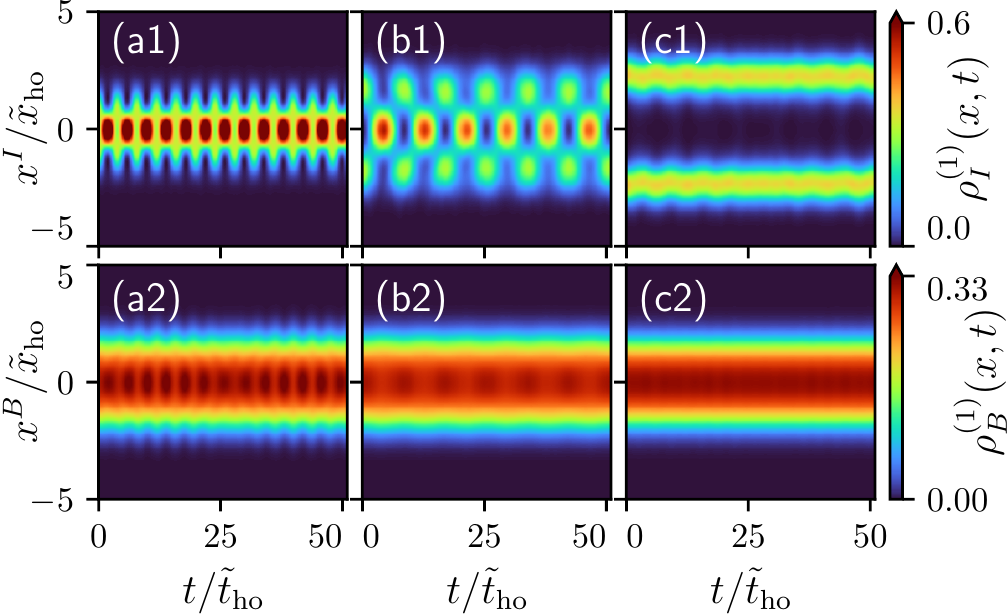}
	\caption{Spatiotemporal evolution of the one-body densities for two weakly interacting impurities ($g_{II}/\tilde{g}_{\rm{ho}}=0.2$) and a weakly interacting bath ($g_{BB}/\tilde{g}_{\rm{ho}}=0.5$) trapped in a harmonic oscillator. The dynamics is induced by ramping down the barrier of the double well in which the impurities initially reside. In each column a different impurity-medium interaction strength is considered, which is from left to right $g_{BI}/\tilde{g}_{\rm{ho}}=-0.2,0.2,0.6$.}
	\label{fig:ho_gpop}
\end{figure}

\subsection{Collision features in terms of the one-body density}

Figure \ref{fig:ho_gpop} illustrates the time evolution of $\rho_{B}^{(1)}(x, t)$ and $\rho_{I}^{(1)}(x, t)$ for weak impurity-impurity couplings, i.e. $g_{II}/\tilde{g}_{\rm{ho}}=0.2$, and for varying impurity-medium interaction strength. As it can be exemplary inferred from Figures \ref{fig:ho_gpop}(a1) and (b1) for weak attractive or repulsive $g_{BI}$ the impurities perform a periodic motion within the harmonic trap and induce only small deformations to the bath density associated with sound-wave emission of the latter [Figures \ref{fig:ho_gpop}(a2) and (b2)]. In the former case, a persisting breathing dynamics of the initially localized impurities takes place being somewhat similar to the one which has been previously discussed for a medium confined in a box potential [Fig. \ref{fig:box_gpop_dm2}(a1)]. In the latter scenario corresponding to Figure~\ref{fig:ho_gpop}(b2), the originally spatially separated impurities collide around $x=0$ and then split in a periodic fashion (see also the discussion below). 

Moreover, we show the respective density evolution in the phase-separation regime, i.e. for $g_{BI}/\tilde{g}_{\rm{ho}}=0.6>g_{BB}/\tilde{g}_{\rm{ho}}$ [Figures \ref{fig:ho_gpop}(c1) and (c2)]. It can be readily deduced that here the impurities are already initially phase separated with the bath and remain in this state also in the course of the evolution while performing small amplitude oscillations due to their collisions with the bath edges~\footnote{We remark that such a diminishing response for these values of $g_{BI}$ is observed besides the one-body density level also for the corresponding two-body densities which preserve their initial diagonal or off-diagonal shape depending on the species (not shown here).}. In particular, for $g_{BI}/\tilde{g}_{\rm{ho}}>0.6$ the impurities remain in the coalescence regime if the initial chosen values for $g_{BI}$ and $g_{II}$ coincide with regime (I) [Figure \ref{fig:GS_ho}(a)] and are spatially separated if the values for $g_{BI}$ and $g_{II}$ correspond to regime (III) [Figure \ref{fig:GS_ho}(a)]. Furthermore, for very strong attractive $g_{BI}$ the impurities and the medium localize together at the trap center where they remain throughout the time-evolution (not shown), see also Ref. \cite{mistakidis2020b} for a similar dynamics. We finally remark that as in the case of a box-confined medium increasing the number of bath atoms, e.g. to $N_B=30$, does not lead to significant alterations of the observed dynamical response.

\begin{figure}[t]
	\centering
	\includegraphics[width=1.0\linewidth]{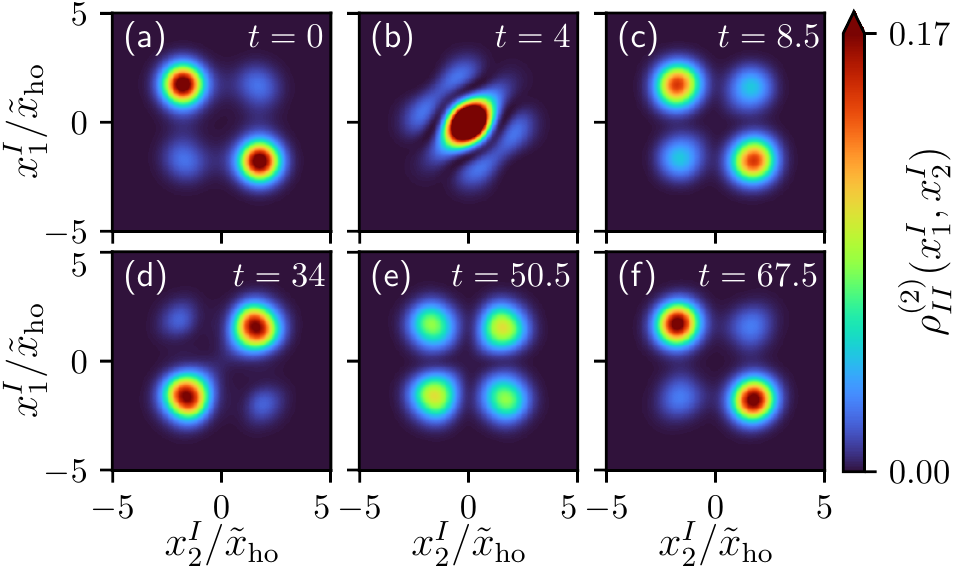}
	\caption{Dynamical evolution of the impurities two-body density $\rho_{II}^{(2)}(x_1^I, x_2^I)$ for $g_{BI}/\tilde{g}_{\rm{ho}}=0.2$ and $g_{II}/\tilde{g}_{\rm{ho}}=0.2$ at specific time-instants (see legends) in units of $\tilde{t}_{\rm{ho}}$. Following the quench the impurities perform a periodic motion consisting of their collision at the trap center and a subsequent expansion. Panel (b) corresponds to a collision event of the impurities, while the other panels refer to time-instants at which the impurities expand [cf. Figure~\ref{fig:ho_gpop}(b1)].
	\label{fig:ho_dm2}}
	\end{figure}

\subsection{Two-body density evolution for weak impurity-medium repulsions}

Let us now focus on the dynamical properties of two impurities which are weakly repulsively coupled to the medium, e.g., via $g_{BI}/\tilde{g}_{\rm{ho}}=0.2$ and interacting among each other with $g_{II}/\tilde{g}_{\rm{ho}}=0.2$ [Figure \ref{fig:ho_gpop}(b1)]. 
Specifically, we are interested in the dynamical evolution of the impurities' reduced two-body density $\rho_{II}^{(2)}(x_1^I, x_2^I)$ depicted in Figure~\ref{fig:ho_dm2}. At $t=0$, corresponding to the ground state in which the two impurities are confined in a double well, their two-body density exhibits two dominant density peaks across its off-diagonal and two suppressed peaks at its diagonal elements [see Figure~\ref{fig:ho_dm2}(a)]. Considering the two diagonal peaks as sufficiently small, this two-body configuration can be interpreted as the probability for the impurities to occupy opposite double-well sites. Following the quench, the impurities collide at the trap center and their cloud starts to contract and expand with a frequency corresponding to the periodic motion of the one-body density [Figure~\ref{fig:ho_gpop}(b1)]. In the course of this periodic motion the two-body configuration alters from a two-body superposition state where both diagonals and off-diagonal elements are populated [Figure~\ref{fig:ho_dm2}(c)] upon expansion of the cloud to a diagonal structure when featuring contraction [see Figure~\ref{fig:ho_dm2}(d)] and vice versa [Figures~\ref{fig:ho_dm2}(f)]. Notice that this dynamical response is inherently related to a process which is hidden on the one-body level [cf. Figure~\ref{fig:ho_gpop}(b1)].

\begin{figure}[t]
	\centering
	\includegraphics[width=1.0\linewidth]{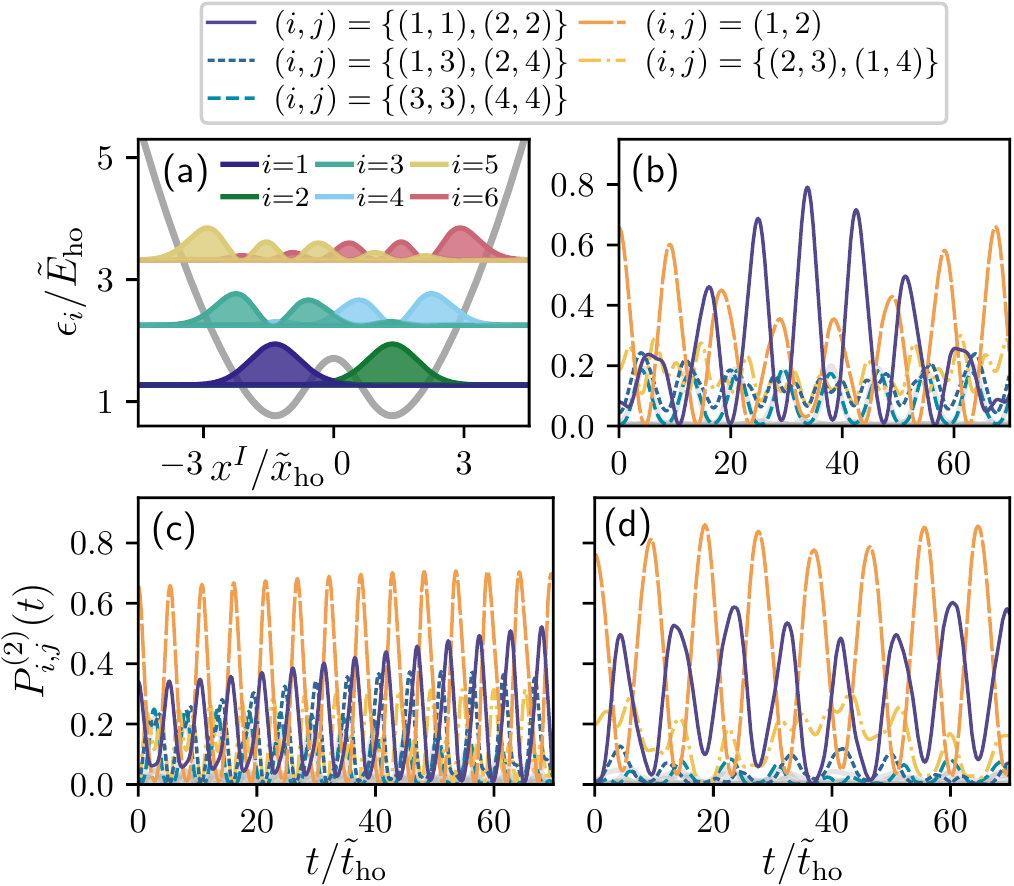}
	\caption{(a) Wannier states corresponding to the first six energetically lowest eigenfunctions of the one-body Hamiltonian consisting of the double-well potential. (b)-(d) Temporal evolution of the two-body probabilities $P_{i,j}^{(2)}(t)$ for the impurities to simultaneously occupy the $i$-th and $j$-th Wannier state for different sets of interaction strengths ($g_{BI}$, $g_{II}$), which are in panel (b) (0.2, 0.2), (c) (0.0, 0.2) and (d) (0.2, 2.0), expressed in units of $\tilde{g}_{\rm{ho}}$. Probabilities involving either the fifth or the sixth Wannier states are suppressed having at most an amplitude of 0.05 and are shown in gray. The dynamics is induced by ramping down the barrier of the impurities double-well potential.}
	\label{fig:ho_occup}
\end{figure}

\subsection{Single-particle dynamical excitation processes}

To obtain insights into the underlying microscopic processes in the course of the impurity dynamics we project the many-body wave function onto basis functions consisting of the generalized Wannier functions ${\phi_i^I(x^I)}$ of the initially considered (pre-quenched) double-well potential \cite{kivelson1982, kivelson1982a}. In this way we can retrieve the probabilities for the impurities to occupy certain localized states of this basis and distinguish between the left and right double-well sites. The Wannier functions are constructed as a superposition from the six energetically lowest eigenfunctions of the one-body Hamiltonian $\hat{H}^{(1),\rm{dw}}= - \frac{\hbar^2}{2m_I} \frac{\partial^2}{(\partial x^I)^2} + V^{\rm{dw}}_I(x^I)$ and are provided in Figure \ref{fig:ho_occup}(a) together with their associated eigenenergies $\epsilon_i$. In particular, the Wannier state corresponding to $i=1$ ($i=2$) is the energetically lowest one at the left (right) site. Analogously, $i=3,5$ ($i=4,6$) signify the first and second excited Wannier states at the left (right) site. Note that, even though for the analysis a basis of a double-well potential is utilized, the impurities' dynamics still takes place within a harmonic oscillator. The respective two-body probabilities for the impurities to simultaneously occupy the $i$-th and $j$-th Wannier state are given by
\begin{align}
	P_{i,j}^{(2)}(t) = \langle \Psi^{\rm{MB}}(t) | \mathbb{1}_B \otimes \ket{\phi_i^I} \bra{\phi_i^I} \otimes \ket{\phi_j^I} \bra{\phi_j^I} \Psi^{\rm{MB}}(t) \rangle.
\end{align}
Here $\mathbb{1}_B$ is the unit operator defined in the subspace of the bath and $\ket{\phi_i^I} \bra{\phi_i^I}$ are the one-body projectors of the $i$-th Wannier state acting on a single impurity. The quality of the basis is tested by summing up all probabilities $P_{i,j}^{(2)}(t)$ for each time instant and verifying that $\sum_{i,j} P_{i,j}^{(2)}(t) > 0.97$ holds until $t/\tilde{t}_{\rm{ho}}=70$. 

\begin{figure}[t]
	\centering
	\includegraphics[width=1.0\linewidth]{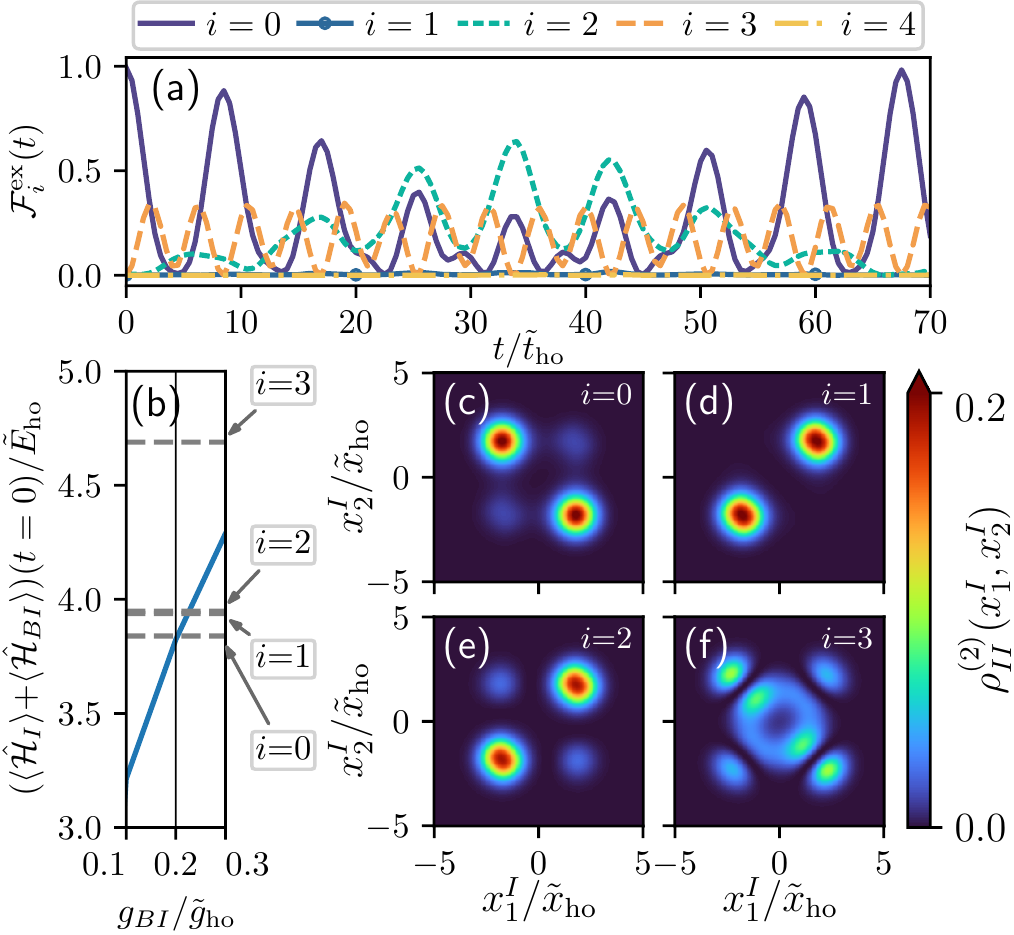}
	\caption{(a) Time evolution of the fidelity between the many-body wave function for $g_{BI}/\tilde{g}_{\rm{ho}}=0.2$ and $g_{II}/\tilde{g}_{\rm{ho}}=0.2$ and the excited states of the effective two-body Hamiltonian $\hat{H}^{(2), \rm{eff}}$ (see also Eq.~(\ref{eq:eff_pot})). (b) Sum of the impurity-medium interaction energy $\langle\hat{\mathcal{H}}_{BI}\rangle$ and the energy of the impurities $\langle\hat{\mathcal{H}}_{I}\rangle$ in the many-body approach for the ground state as a function of the impurity-medium coupling (blue line) depicted together with the eigenenergies of $\hat{H}^{(2),\rm{eff}}$ (grey horizontal dashed lines). The two-body density of the ground state and the first three excited eigenstates are provided in (c)-(f).}
	\label{fig:ho_ex}
\end{figure}

The above-described two-body probabilities for $g_{BI}/\tilde{g}_{\rm{ho}}=0.2$ and $g_{II}/\tilde{g}_{\rm{ho}}=0.2$ are presented in Figure \ref{fig:ho_occup}(b). All probabilities show an oscillatory behavior stemming from the periodic collision and expansion of the impurities [cf. Figure \ref{fig:ho_gpop}(b)]. Beyond this rapid motion, a decay and revival of $P_{1,2}^{(2)}(t)$ takes place at longer time scales, where $P_{1,2}^{(2)}(t)$ corresponds to the probability of finding one impurity in the energetically lowest left site Wannier state while the other one occupies the right site Wannier state. 
On the other hand, when the envelope of $P_{1,2}^{(2)}(t)$ reaches a minimum the probability of finding two impurities both in the left $P_{1,1}^{(2)}(t)$ (right, $P_{2,2}^{(2)}(t)$) Wannier state is maximized as demonstrated in Figure~\ref{fig:ho_gpop}(b). This observation implies that a single-particle intraband excitation process takes place. Moreover, also energetically higher-lying Wannier states contribute to the ground state configuration of the impurities as well as to their dynamical response. For instance, the second and third (first and fourth) Wannier states contribute with $P_{2,3}(0)=P_{1,4}(0)=19\,\%$ to the ground state configuration. 
Therefore, the initial state is a superposition of different single-particle states. 
This is to be contrasted with the discussion below [see Figure~\ref{fig:ho_ex}] where a two-body basis is employed accounting in a more natural way for effects stemming from impurity-impurity interactions and the coupling to the bath.

For comparison the case of two weakly interacting impurities $g_{II}/\tilde{g}_{\rm{ho}}=0.2$ which are decoupled from the bath ($g_{BI}=0$) is showcased in Figure \ref{fig:ho_occup}(c). 
It can be readily seen that the impurities do not perform a state transfer similar to the one depicted in Figure~\ref{fig:ho_occup}(b) but rather retain their delocalized configuration. 
In particular, at time-instants corresponding to an expansion of the impurities cloud the two-body probabilities associated with the energetically lowest Wannier states lying at opposite and the same sites significantly contribute to the impurities many-body wave function, thus confirming the former statement. Otherwise, the impurities response is characterized by excitations to energetically higher-lying states.
Next, we inspect the case of strongly interacting impurities ($g_{II}/\tilde{g}_{\rm{ho}}=2.0$) which are weakly coupled to the bath ($g_{BI}/\tilde{g}=0.2$) [Figure \ref{fig:ho_occup}(d)]. The dynamics begins with initially separated impurities, viz. $P_{1,2}^{(2)}(t)$ obtains a maximum at $t=0$, and continues with the collision of the impurities at the trap center where they both dominantly populate the same energetically lowest left or right Wannier state [cf. $P_{1,1}^{(2)}(t)=P_{2,2}^{(2)}(t)$]. Subsequently, the strong impurity-impurity repulsion enforces the impurities to occupy again opposite double-well sites. This scheme repeats itself during the evolution and, in particular, lasts until $T/\tilde{t}_{\rm{ho}}=200$. 
Again a state transfer process occurs as it can be seen from the competition of $P_{1,2}^{(2)}(t)$ and $P_{1,1}^{(2)}(t)$. However, in this case the transfer is less transparent and not as dominant as for $g_{II}/\tilde{g}_{\rm{ho}}=0.2$, implying that an increasing $g_{II}/\tilde{g}_{\rm{ho}}$ results in the suppression of this process.

Therefore, the intraband excitation process observed for $g_{BI}/\tilde{g}_{\rm{ho}}=0.2$ and $g_{II}/\tilde{g}_{\rm{ho}}=0.2$ proves to be sensitive to the impurity-impurity interaction strength and, most importantly, requires a finite impurity-medium coupling. In this manner, we can conclude that this state transfer of the impurities is induced by the presence of the bath. Moreover, we have verified the absence of this mechanism for a species mean-field ansatz ($D=1$ in Eq.~(\ref{eq:wavefct_toplayer})), i.e. when the entanglement is not taken into account. Thus, we can deduce that many-body effects and, in particular, the impurity-medium entanglement play a crucial role for the realization of such processes.

\subsection{Effective two-body impurity mechanisms}

To further understand the participating excitation processes we consider a projection of the many-body wave function onto a two-body basis set. In particular, we choose for this investigation the ground state and the first four energetically lowest excited states of an effective Hamiltonian. This effective Hamiltonian $\hat{H}^{(2),\rm{eff}}$ describes two weakly interacting impurities ($g_{II}/\tilde{g}_{\rm{ho}}=0.2$) trapped in the effective potential defined in Eq.~(\ref{eq:eff_pot}) with $g_{BI}/\tilde{g}_{\rm{ho}}=0.2$. In this manner, we take the backaction induced by the medium into account. In Figure~\ref{fig:ho_ex}(c)-(f) we present the impurities two-body density for the ground state as well as the first three excited states $|\Phi_i^I\rangle$ of $\hat{H}^{(2),\rm{eff}}$. As such, we associate the ground state ($i=0$) with the two-body state $(|LR\rangle + |RL\rangle)/\sqrt{2}$ where $|L\rangle$ ($|R\rangle$) represents a single-particle state corresponding to the left (right) site of the double well. Analogously, we refer to the first ($i=1$) and second ($i=2$) excited states as the configurations $(|LL\rangle - |RR\rangle)/\sqrt{2}$ and $(|LL\rangle + |RR\rangle)/\sqrt{2}$, respectively. The corresponding eigenenergies are shown in Figure~\ref{fig:ho_ex}(b). Note that the first ($i=1$) and second ($i=2$) excited eigenstates are approximately degenerate. In order to support the validity of this two-body approach for two impurities coupled to a larger medium, we additionally provide the sum of the impurity energy and the interaction energy ($\langle\hat{\mathcal{H}}_I\rangle + \langle\hat{\mathcal{H}}_{BI}\rangle$) at $t=0$ as predicted within the many-body approach, namely when the impurities are still trapped in a double-well potential. Since this energy matches at $g_{BI}/\tilde{g}_{\rm{ho}}=0.2$ the ground state energy of the effective approach [cf. intersection of $\langle\hat{\mathcal{H}}_I\rangle + \langle\hat{\mathcal{H}}_{BI}\rangle$ with $i=0$ in Figure~\ref{fig:ho_ex}(b)] we conclude that the effective potential adequately accounts for the presence of the medium at $t=0$.

As a next step, we calculate the fidelity of the two-body eigenstates $|\Phi_i^I\rangle$ with the time-dependent species functions of the impurities being coupled to the medium. In this way, the probabilities of the contributing two-body configurations are revealed. Therefore, we estimate the absolute square of the projection of $\sum_{j=1}^{D} |\Psi_j^{B}(t)\rangle \otimes | \Phi_i^I \rangle$ on the many-body wave function $|\Psi^{\rm{MB}}(t)\rangle$ defined in Eq. (\ref{eq:wavefct_toplayer}), which reads as
\begin{align}
	\mathcal{F}_i^{\rm{ex}}(t) = \left| \sum_{j=1}^{D} \sqrt{\lambda_j(t)} \langle \Phi_i^{I}| \Psi_j^I(t)\rangle \right|^2.
\end{align}
The dynamics of the fidelity with respect to the ground state and the first four excited eigenstates of $\hat{H}^{(2),\rm{eff}}$ is provided in Figure \ref{fig:ho_ex}(a). Analogously with the analysis regarding the Wannier states, we observe besides a fast periodic motion a slower decay and revival of the ground state $|\Phi_0^I\rangle$ associated with two separated impurities. This behavior is accompanied by a complementary increase of the second excited state $|\Phi_2^I\rangle = (|LL\rangle + |RR\rangle)/\sqrt{2}$ associated with the coalescence of the impurities. Since the first excited state, corresponding to the anti-symmetric configuration $(|LL\rangle - |RR\rangle)/\sqrt{2}$, is strongly suppressed we conclude that the impurities undergo the two-body state transfer from $(|LR\rangle + |RL\rangle)/\sqrt{2}$ to $(|LL\rangle + |RR\rangle)/\sqrt{2}$. Moreover, we note that during the impurities' collision an appreciable amount of higher excited states need to be taken into account as indicated, for instance, by the non-negligible occupation of the third exited state $|\Phi_3^I\rangle$.

Concluding we have explicated the microscopic mechanisms on both the one- and the two-body level taking place during the collision dynamics of interacting impurities coupled to a harmonically confined bath. For instance, the single-particle intraband excitation process appears to be sensitive with respect to the impurity-impurity interaction strength and requires a finite coupling to the bath. 
Moreover, an analysis with respect to a two-body basis deciphered the transitions among particular two-body configurations.

\section{Conclusions and Outlook}
\label{sec:conclusion}

We have investigated the ground state and correlated dynamics of two interacting bosonic impurities confined in a double well and immersed in a bosonic medium. The latter either experiences a box potential or it is confined in a harmonic trap. We establish the phase diagram of the ground state for varying impurity-impurity and impurity-medium coupling strengths. Thereby, the emergent ground states have been characterized with the aid of the two-body densities and impurity-medium entanglement. An analysis of the impact of different trapping geometries on the formation of these phases has been performed. For instance, we explicate that the coalescence of the impurities at strong (repulsive or attractive) impurity-medium interaction strengths is preserved for different impurity-impurity repulsions when the bath is in a box. However, in the case of a harmonically trapped bath the impurities separate from each other for strong impurity-impurity repulsion residing in a Mott-type configuration. Moreover, in the latter scenario, we observe at strong impurity-medium attractions indications for the formation of a bipolaron.

Focusing on a specific interaction-dependent ground state configuration we trigger the dynamics by suddenly ramping down the potential barrier of the impurities' double well. Firstly, the dynamical response regimes of the impurities coupled to a box confined medium are unraveled with respect to their associated one- and two-body densities. In particular, for intermediate impurity-medium repulsions a localization of the impurities at the trap center after the original collision is realized. The impurities' two-body density features an elongated shape along the diagonal for weak impurity-impurity repulsion which suggests the presence of attractive induced interactions mediated by the bath. This induced localization of the impurities persists also when the coupling strength between the impurities is further increased. This observation together with the existence of a spatial overlap for finite impurity-medium interaction strengths support the formation of quasi-particles, i.e. two interacting polarons. We have attested the robustness of the above phenomena with respect to variations of the number of bath particles and the size of the box potential. However, for strong repulsions finite size effects of the medium's box potential come into play and govern the dynamical response of the system. Essentially, after the impurities' collision at the trap center they drift apart and then the medium reoccupies the center.

By considering a harmonically confined bath the impurities' response is qualitatively altered. Due to the spatial localization of the bath at the trap center the impurities and the medium undergo a phase separation already for intermediate impurity-medium repulsions as it was also observed on the ground state level. The response becomes especially intriguing for weak impurity-medium couplings where the impurities are able to perform a breathing motion within the bath. Specifically, for weak impurity-medium repulsions we observe a state transfer of the impurities starting with two spatially separated ones located at different double-well sites and evolving into a coalesced configuration, i.e. the impurities cluster. Interestingly, this state transfer process does not emerge for strongly interacting impurities and, most importantly requires a finite impurity-medium interaction strength, viz. it is induced by the coupling to the bath. Moreover, it is shown that this mechanism can be well understood in terms of a single-particle Wannier basis of the double-well. Additionally, we reveal the participating two-body states in this process using an analysis in terms of a two-body basis which consists of the low-lying excited states of a corresponding effective two-body Hamiltonian.

There are several possible extensions of our results. 
An immediate one will be to investigate the collision features of the impurities immersed in a spatially extended bosonic gas with the aim to unveil their possible damping mechanisms and appreciate the corresponding drag force. 
In another context, it would be worth including a spin degree of freedom for the impurities. Here, the dynamics of the emergent spin-spin correlations is of interest especially when the impurities localize around the trap center. Moreover, it would be intriguing to consider two impurities with different masses, e.g. a light and heavy one, coupled to a background. This way it would be feasible to investigate the influence of the mass on the emergent collisional aspects of the impurities and their induced interactions as well as trigger specific population transfer channels by considering a Rabi-coupling term.

\begin{acknowledgements}
This work has been funded by the Deutsche Forschungsgemeinschaft (DFG, Germany Research Foundation) --- SFB 925 --- project 170620586. S.I.M. gratefully acknowledges financial support from the NSF through a grant for ITAMP at Harvard University. The authors acknowledge fruitful discussions with G. M. Koutentakis.

\end{acknowledgements}

\appendix

\section{Bath confined in a box potential: relative distance between the impurities}
\label{ap:box_rel_dist}

\begin{figure}[t]
	\centering
	\includegraphics[width=0.9\linewidth]{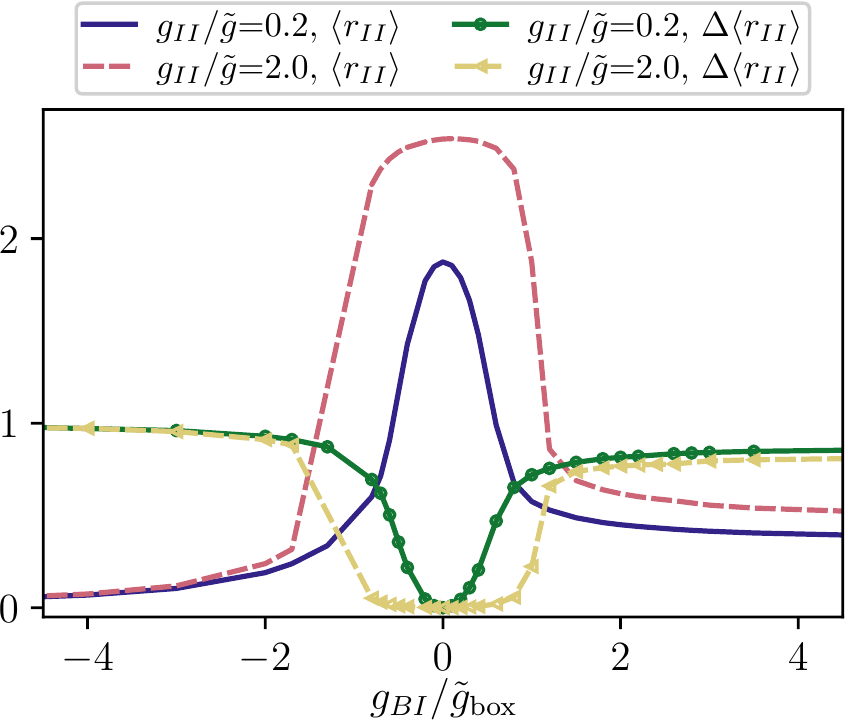}
	\caption{Ground state relative distance between the impurities $\langle r_{II}\rangle$ (in units of $\tilde{x}_{\rm{ho}}$) with respect to the impurity-medium interaction strength $g_{BI}$ for different impurity-impurity couplings (see legend). The relative difference $\Delta\langle r_{II}\rangle$ between the impurity-impurity distance predicted within the many-body approach ($\langle r_{II}\rangle$) and the effective Hamiltonian of Eq.~(\ref{eq:eff_pot}) with $V_I^{\rm{dw}}(x^I)=0$ ($\langle r_{II}^{\rm{eff}}\rangle$) is also provided (see main text).}
	\label{fig:box_rel_dist}
\end{figure}

In the following, we examine the impurities relative distance $\langle r_{II}\rangle$ which can serve as an indicator for the existence of their induced interactions mediated by the bath~\cite{mistakidis2019b, mistakidis2020, mukherjee2020a}. 
This quantity, which is accessible through {\it in-situ} spin-resolved
single-shot measurements on the impurities state~\cite{bergschneider2018spin}, reads
\begin{align}
\langle r_{II} \rangle(t) = \frac{1}{N_I(N_I-1)}\int {\rm{d}} x_1^I {\rm{d}} x_2^I \left|x_1^I - x_2^I \right| \rho_{II}^{(2)}(x_1^I, x_2^I).
\label{eq:rel_dist}
\end{align}
The impurities relative distance is presented in Figure~\ref{fig:box_rel_dist} for the ground state of the system ($t=0$) for varying impurity-medium interaction strength and different impurity-impurity couplings. 
In the case of two weakly interacting impurities as well as for two strongly interacting ones we observe that with an increasing absolute value of $g_{BI}$ the relative distance between the impurities reduces. This behavior implies an induced attraction mediated by the coupling to the bath. 
Furthermore, it is evident that for larger impurity-impurity interactions $\langle r_{II}\rangle$ is enhanced when compared to the one of weakly interacting impurities. 
This behavior is caused by the increased intraspecies impurities direct repulsion compensating their induced attraction.

To further justify the presence of induced interactions, we compare the resulting impurities relative distance as obtained using the 
%by evaluating Eq.~(\ref{eq:Hamiltionian}) in terms of a 
many-body approach ($\langle r_{II}\rangle$) with the one ($\langle r_{II}^{\rm{eff}}\rangle$)
%relative distance obtained 
predicted within the effective two-body Hamiltonian $\bar{H}^{(2),\rm{eff}}$.
The latter was introduced in section~\ref{sec:dyn_ho} for a harmonically confined medium. 
Notice that here this effective Hamiltonian describes the interplay between two interacting particles confined in an effective potential constructed by the one-body density of a box-trapped medium [see Eq. (\ref{eq:eff_pot})]. As such, the entanglement between the impurities and the medium is neglected while effects stemming from the backaction to the bath are taken into account. By comparing $\langle r_{II}\rangle$ between these two methods we can determine whether the decrease of $\langle r_{II}\rangle$ for increasing $|g_{BI}|$ [see Figure~\ref{fig:box_rel_dist}] originates from an entanglement-assisted induced interaction or it is due to an alteration of the effective potential. 

Inspecting $\Delta\langle r_{II}\rangle = (\langle r_{II}^{\rm{eff}}\rangle - \langle r_{II}\rangle)/\langle r_{II}^{\rm{eff}}\rangle$ depicted in Figure~\ref{fig:box_rel_dist} a large deviation among $\langle r_{II}\rangle$ and $\langle r_{II}^{\rm{eff}}\rangle$ becomes evident for increasing $|g_{BI}|$. 
This confirms the presence of attractive induced interactions between the impurities. 
The fact that $\Delta\langle r_{II}\rangle$ is finite can be traced back to the shape of 
%of A hint for the underlying reason can be found by %inspecting 
the impurities two-body densities as obtained within the aforementioned approaches for large $g_{BI}$ [corresponding to regime I in Figure~\ref{fig:GS_box}(a)]. While in the many-body scenario 
%case where interspecies correlations are taken into %account 
the impurities coalesce (namely only the diagonal of $\rho_{II}^{(2)}(x_1^I, x_2^I)$ is occupied), in the effective potential case both the diagonal and the off-diagonal elements of $\rho_{II}^{(2)}(x_1^I, x_2^I)$ are equally populated. 
This naturally leads to a larger relative distance. Concluding, the comparison with the effective model underlies the importance of considering correlations in the system and reveals the presence of attractive induced interactions between the impurities, see more details in Refs.~\cite{chen2018,mistakidis2020}.

\section{Collisional properties of two heavy impurities}
\label{ap:box_heavy_impurities}

\begin{figure}[t]
	\centering
	\includegraphics[width=1.0\linewidth]{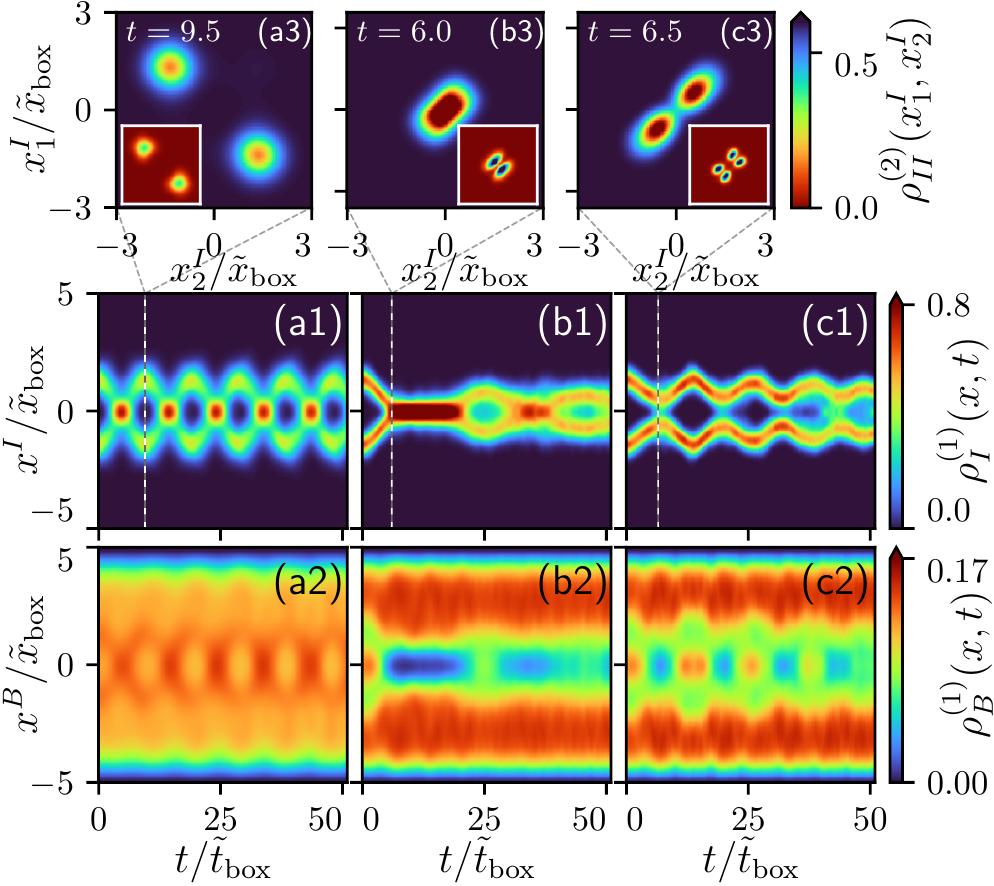}
	\caption{Time evolution of the one-body density of (a1)-(c1) two heavy impurities and (a2)-(c2) a box-confined bosonic bath. Each column represents the dynamics for a fixed impurity-medium interaction strength which is from left to right $g_{BI}/\tilde{g}_{\rm{box}}=-0.2, 1.5, 5.0$. The impurity-impurity coupling remains constant being $g_{II}/\tilde{g}_{\rm{box}}=0.2$. (a3)-(c3) Snapshots of the impurities two-body density. The insets show the two-body densities at the same time instants, but for two strongly interacting impurities, i.e. $g_{II}/\tilde{g}_{\rm{box}}=2.0$.}
	\label{fig:box_massI_gpop_dm2}
\end{figure}
Throughout this work we have considered a system consisting of $^{87}$Rb bath particles and $^{133}$Cs impurities corresponding to a mass ratio of $m_I=133/87m_B$. 
Below, we employ a mixture characterized by a mass ratio $m_I'=5m_B$ in order to elaborate on the dynamical response of two heavy impurities~\cite{mukherjee2020a} immersed in a bath confined in a box potential~\footnote{Additionally, the trap frequency of the harmonic confinement has been adjusted according to $\omega_I'=\omega_I\sqrt{\frac{m_I}{m_I'}}$ in order to preserve the initial shape of the double-well potential.}. The time evolution of the corresponding one- and two-body densities of the impurities and the bath particles are depicted in Figures \ref{fig:box_massI_gpop_dm2}(a1)-(c1) and (a2)-(c2) for interaction parameters as the ones employed in the main text [Figure \ref{fig:box_gpop_dm2}]. In particular, we assume two weakly interacting impurities ($g_{II}/\tilde{g}_{\rm{box}}=0.2$) and vary the impurity-medium interaction strength $g_{BI}$. Regarding the one-body density evolution we do observe a qualitatively similar behavior as compared to the case of lighter impurities. Only in the case of weak attractive $g_{BI}$ [Figure \ref{fig:box_massI_gpop_dm2}(a1)] the heavy impurities perform a more pronounced breathing oscillation with a larger oscillation period. Inspecting a two-body density snapshot reveals that the impurities are spatially separated from each other and oscillate along the off-diagonal of $\rho_{II}^{(2)}(x_1^I, x_2^I)$ [cf. Figure \ref{fig:box_massI_gpop_dm2}(a3)]. Therefore, they are not delocalized as their lighter counterparts [Figure \ref{fig:box_gpop_dm2}(a3)], and this behavior persists for two strongly interacting impurities [cf. inset of Figure \ref{fig:box_massI_gpop_dm2}(a3)].

Additionally, for intermediate impurity-medium repulsions, i.e. $g_{BI}/\tilde{g}_{\rm{box}}=1.5$, the one-body densities of the impurities and the medium [Figures \ref{fig:box_massI_gpop_dm2}(b1) and (b2)] as well as the impurities two-body density [Figure \ref{fig:box_massI_gpop_dm2}(b3) and its inset] do not reveal a qualitatively different response with respect to the case of lighter impurities [cf. Figures \ref{fig:box_gpop_dm2}(b1)-(b3)]. 
Indeed the impurities remain very close throughout the time-evolution exhibiting a more pronounced localization trend around the trap center as compared to lighter ones. 
The same holds also for the case of strong impurity-medium interactions [see Figures \ref{fig:box_massI_gpop_dm2}(c1)-(c3) for $g_{BI}/\tilde{g}_{\rm{box}}=5.0$] where the impurities feature multiple collisions with a dissipative amplitude. Only, the fragmentation in terms of the diagonal of the two-body density for $g_{II}/\tilde{g}_{\rm{box}}=2.0$ is more prominent in the case of heavy impurities than for lighter ones [cf. inset of Figure \ref{fig:box_massI_gpop_dm2}(c3)] and becomes visible even on the one-body density level (not shown here).

\begin{figure}[t]
	\centering
	\includegraphics[width=0.8\linewidth]{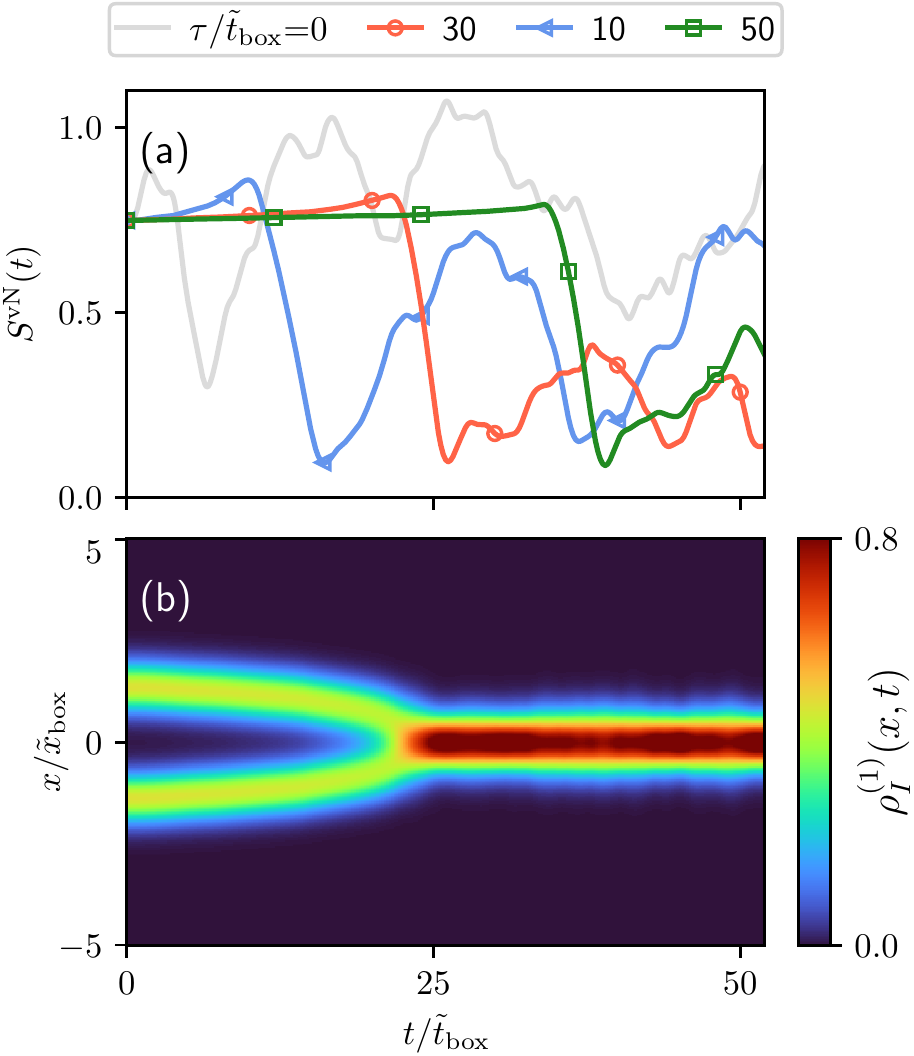}
	\caption{(a) Time-evolution of the von Neumann entropy upon linearly ramping down the barrier height $h_I$ of the double well for different ramp times $\tau$ (see legend). The impurities are weakly interacting with $g_{II}/\tilde{g}_{\mathrm{box}}=0.2$ and are coupled to a box-confined medium with an interspecies coupling $g_{BI}/\tilde{g}_{\mathrm{box}}=1.5$. An increasing ramp time maintains an almost constant magnitude of entanglement until the impurities collision. (b) The corresponding one-body density evolution of the impurities for $\tau/\tilde{t}_{\mathrm{box}}=30$. As it can be seen the linear ramp delays the impurities first collision event when compared to Fig.~\ref{fig:box_gpop_dm2} ($b_1$) referring to the corresponding quench dynamics.}
	\label{fig:ramp}
\end{figure}

\section{Impurity dynamics after a linear ramp of their double-well potential}
\label{ap:ramp}

In the following we address the robustness of the impurities dynamical response when ramping down their barrier height $h_I$ in a time-dependent manner and not suddenly as in the main text. 
Specifically, we apply the following linear protocol $\tilde{h}_I(t) = h_I - h_It/\tau$ if $0 \leq t \leq \tau$ whilst $\tilde{h}_I(t)=0$ as long as $\tau < t$. The ramp time $\tau$ is defined as the one at which the barrier height vanishes, i.e. when $\tilde{h}_I(\tau)=0$. For our purposes, we restrict our study to finite values of $\tau$ which deviate from the instantaneous quench but also do not refer to an adiabatic linear ramping.

To visualize the impact of the time-dependent protocol on the impurities collision process we present in Figure~\ref{fig:ramp} the time-evolution of the von Neumann entanglement entropy [Eq.~(\ref{eq:SVN})] and the impurities one body density. 
We follow a linear ramp of the impurities barrier height characterized by a finite ramp time $\tau/\tilde{t}_{\mathrm{box}}=30$. Notice that the latter appreciably deviates from the quench scenario ($\tau/\tilde{t}_{\mathrm{box}}=0$). 
Moreover, we exemplarily invoke the system where the medium is confined in a box potential while the relevant interactions are $g_{BB}/\tilde{g}_{\mathrm{box}}=0.5$, $g_{II}/\tilde{g}_{\mathrm{box}}=0.2$ and $g_{BI}/\tilde{g}_{\mathrm{box}}=1.5$. 
Recall that in the main text it has been shown that in this regime the impurities localize at the trap center after their first collision event, see also Figure~\ref{fig:box_gpop_dm2}(b1). 
As it can be readily seen [Figure~\ref{fig:ramp}(b)], the impurities dynamical response in terms of $\rho_I^{(1)}(x,t)$ remains qualitatively unchanged as compared to the quench [Figure~\ref{fig:box_gpop_dm2}(b1)]. 
The most prominent difference regards the timescale at which the impurities collide and subsequently localize at the trap center. As expected the initial collision can be delayed for an increasing ramp time. Turning to the evolution of the respective von Neumann entropies for varying ramp times we observe an interesting behavior. At short timescales the impurities and the medium are highly entangled [Figure~\ref{fig:ramp}(a)]. The magnitude of the entanglement is maintained in the course of the evolution until the impurities collide and then localize at the trap center where it suddenly decreases. 
%sudden decrease of the entanglement occurs. 
This drop of the entanglement depends strongly on the ramp time and in particular it takes place faster for smaller $\tau$ since in this case the collision event is accelerated.

Concluding, we remark that the linear protocol affects the remaining response regimes, occurring for other interspecies interaction strengths, in a similar vein. 
Namely the main features as described in Section~\ref{sec:dyn} do not substantially alter but rather the underlying timescales change. For instance, considering weak attractive or repulsive $g_{BI}$ referring to the impurities breathing motion [Figure~\ref{fig:box_gpop_dm2}(a1)], we find a decreasing tendency of the breathing amplitude and frequency for larger $\tau$ (not shown). This is attributed to the fact that for increasing $\tau$ the collision of the impurities is less violent thus producing a less pronounced breathing. Analogous effects are observed for a harmonically trapped medium where, for instance, also in this case the amplitude and frequency of the underlying breathing motion depend strongly on the ramp time $\tau$.

\bibliography{literature.bib}

\begin{thebibliography}{101}%
\makeatletter
\providecommand \@ifxundefined [1]{%
 \@ifx{#1\undefined}
}%
\providecommand \@ifnum [1]{%
 \ifnum #1\expandafter \@firstoftwo
 \else \expandafter \@secondoftwo
 \fi
}%
\providecommand \@ifx [1]{%
 \ifx #1\expandafter \@firstoftwo
 \else \expandafter \@secondoftwo
 \fi
}%
\providecommand \natexlab [1]{#1}%
\providecommand \enquote  [1]{``#1''}%
\providecommand \bibnamefont  [1]{#1}%
\providecommand \bibfnamefont [1]{#1}%
\providecommand \citenamefont [1]{#1}%
\providecommand \href@noop [0]{\@secondoftwo}%
\providecommand \href [0]{\begingroup \@sanitize@url \@href}%
\providecommand \@href[1]{\@@startlink{#1}\@@href}%
\providecommand \@@href[1]{\endgroup#1\@@endlink}%
\providecommand \@sanitize@url [0]{\catcode `\\12\catcode `\$12\catcode
  `\&12\catcode `\#12\catcode `\^12\catcode `\_12\catcode `\%12\relax}%
\providecommand \@@startlink[1]{}%
\providecommand \@@endlink[0]{}%
\providecommand \url  [0]{\begingroup\@sanitize@url \@url }%
\providecommand \@url [1]{\endgroup\@href {#1}{\urlprefix }}%
\providecommand \urlprefix  [0]{URL }%
\providecommand \Eprint [0]{\href }%
\providecommand \doibase [0]{http://dx.doi.org/}%
\providecommand \selectlanguage [0]{\@gobble}%
\providecommand \bibinfo  [0]{\@secondoftwo}%
\providecommand \bibfield  [0]{\@secondoftwo}%
\providecommand \translation [1]{[#1]}%
\providecommand \BibitemOpen [0]{}%
\providecommand \bibitemStop [0]{}%
\providecommand \bibitemNoStop [0]{.\EOS\space}%
\providecommand \EOS [0]{\spacefactor3000\relax}%
\providecommand \BibitemShut  [1]{\csname bibitem#1\endcsname}%
\let\auto@bib@innerbib\@empty
%</preamble>
\bibitem [{\citenamefont {Chin}\ \emph {et~al.}(2010)\citenamefont {Chin},
  \citenamefont {Grimm}, \citenamefont {Julienne},\ and\ \citenamefont
  {Tiesinga}}]{chin2010}%
  \BibitemOpen
  \bibfield  {author} {\bibinfo {author} {\bibfnamefont {C.}~\bibnamefont
  {Chin}}, \bibinfo {author} {\bibfnamefont {R.}~\bibnamefont {Grimm}},
  \bibinfo {author} {\bibfnamefont {P.}~\bibnamefont {Julienne}}, \ and\
  \bibinfo {author} {\bibfnamefont {E.}~\bibnamefont {Tiesinga}},\ }\href
  {\doibase 10.1103/RevModPhys.82.1225} {\bibfield  {journal} {\bibinfo
  {journal} {Rev. Mod. Phys.}\ }\textbf {\bibinfo {volume} {82}},\ \bibinfo
  {pages} {1225} (\bibinfo {year} {2010})}\BibitemShut {NoStop}%
\bibitem [{\citenamefont {Grimm}\ \emph {et~al.}(2000)\citenamefont {Grimm},
  \citenamefont {Weidem{\"u}ller},\ and\ \citenamefont
  {Ovchinnikov}}]{grimm2000}%
  \BibitemOpen
  \bibfield  {author} {\bibinfo {author} {\bibfnamefont {R.}~\bibnamefont
  {Grimm}}, \bibinfo {author} {\bibfnamefont {M.}~\bibnamefont
  {Weidem{\"u}ller}}, \ and\ \bibinfo {author} {\bibfnamefont {Y.~B.}\
  \bibnamefont {Ovchinnikov}},\ }\href {\doibase 10.1016/S1049-250X(08)60186-X}
  {\bibfield  {journal} {\bibinfo  {journal} {Adv. At. Mol. Opt. Phys.}\
  }\textbf {\bibinfo {volume} {42}},\ \bibinfo {pages} {95} (\bibinfo {year}
  {2000})}\BibitemShut {NoStop}%
\bibitem [{\citenamefont {Henderson}\ \emph {et~al.}(2009)\citenamefont
  {Henderson}, \citenamefont {Ryu}, \citenamefont {MacCormick},\ and\
  \citenamefont {Boshier}}]{henderson2009}%
  \BibitemOpen
  \bibfield  {author} {\bibinfo {author} {\bibfnamefont {K.}~\bibnamefont
  {Henderson}}, \bibinfo {author} {\bibfnamefont {C.}~\bibnamefont {Ryu}},
  \bibinfo {author} {\bibfnamefont {C.}~\bibnamefont {MacCormick}}, \ and\
  \bibinfo {author} {\bibfnamefont {M.~G.}\ \bibnamefont {Boshier}},\ }\href
  {\doibase 10.1088/1367-2630/11/4/043030} {\bibfield  {journal} {\bibinfo
  {journal} {New J. Phys.}\ }\textbf {\bibinfo {volume} {11}},\ \bibinfo
  {pages} {043030} (\bibinfo {year} {2009})}\BibitemShut {NoStop}%
\bibitem [{\citenamefont {Gaunt}\ \emph {et~al.}(2013)\citenamefont {Gaunt},
  \citenamefont {Schmidutz}, \citenamefont {Gotlibovych}, \citenamefont
  {Smith},\ and\ \citenamefont {Hadzibabic}}]{gaunt2013}%
  \BibitemOpen
  \bibfield  {author} {\bibinfo {author} {\bibfnamefont {A.~L.}\ \bibnamefont
  {Gaunt}}, \bibinfo {author} {\bibfnamefont {T.~F.}\ \bibnamefont
  {Schmidutz}}, \bibinfo {author} {\bibfnamefont {I.}~\bibnamefont
  {Gotlibovych}}, \bibinfo {author} {\bibfnamefont {R.~P.}\ \bibnamefont
  {Smith}}, \ and\ \bibinfo {author} {\bibfnamefont {Z.}~\bibnamefont
  {Hadzibabic}},\ }\href {\doibase 10.1103/PhysRevLett.110.200406} {\bibfield
  {journal} {\bibinfo  {journal} {Phys. Rev. Lett.}\ }\textbf {\bibinfo
  {volume} {110}},\ \bibinfo {pages} {200406} (\bibinfo {year}
  {2013})}\BibitemShut {NoStop}%
\bibitem [{\citenamefont {Haas}\ \emph {et~al.}(2007)\citenamefont {Haas},
  \citenamefont {Leung}, \citenamefont {Frese}, \citenamefont {Haubrich},
  \citenamefont {John}, \citenamefont {Weber}, \citenamefont {Rauschenbeutel},\
  and\ \citenamefont {Meschede}}]{haas2007}%
  \BibitemOpen
  \bibfield  {author} {\bibinfo {author} {\bibfnamefont {M.}~\bibnamefont
  {Haas}}, \bibinfo {author} {\bibfnamefont {V.}~\bibnamefont {Leung}},
  \bibinfo {author} {\bibfnamefont {D.}~\bibnamefont {Frese}}, \bibinfo
  {author} {\bibfnamefont {D.}~\bibnamefont {Haubrich}}, \bibinfo {author}
  {\bibfnamefont {S.}~\bibnamefont {John}}, \bibinfo {author} {\bibfnamefont
  {C.}~\bibnamefont {Weber}}, \bibinfo {author} {\bibfnamefont
  {A.}~\bibnamefont {Rauschenbeutel}}, \ and\ \bibinfo {author} {\bibfnamefont
  {D.}~\bibnamefont {Meschede}},\ }\href {\doibase 10.1088/1367-2630/9/5/147}
  {\bibfield  {journal} {\bibinfo  {journal} {New J. Phys.}\ }\textbf {\bibinfo
  {volume} {9}},\ \bibinfo {pages} {147} (\bibinfo {year} {2007})}\BibitemShut
  {NoStop}%
\bibitem [{\citenamefont {Catani}\ \emph {et~al.}(2012)\citenamefont {Catani},
  \citenamefont {Lamporesi}, \citenamefont {Naik}, \citenamefont {Gring},
  \citenamefont {Inguscio}, \citenamefont {Minardi}, \citenamefont {Kantian},\
  and\ \citenamefont {Giamarchi}}]{catani2012}%
  \BibitemOpen
  \bibfield  {author} {\bibinfo {author} {\bibfnamefont {J.}~\bibnamefont
  {Catani}}, \bibinfo {author} {\bibfnamefont {G.}~\bibnamefont {Lamporesi}},
  \bibinfo {author} {\bibfnamefont {D.}~\bibnamefont {Naik}}, \bibinfo {author}
  {\bibfnamefont {M.}~\bibnamefont {Gring}}, \bibinfo {author} {\bibfnamefont
  {M.}~\bibnamefont {Inguscio}}, \bibinfo {author} {\bibfnamefont
  {F.}~\bibnamefont {Minardi}}, \bibinfo {author} {\bibfnamefont
  {A.}~\bibnamefont {Kantian}}, \ and\ \bibinfo {author} {\bibfnamefont
  {T.}~\bibnamefont {Giamarchi}},\ }\href {\doibase 10.1103/PhysRevA.85.023623}
  {\bibfield  {journal} {\bibinfo  {journal} {Phys. Rev. A}\ }\textbf {\bibinfo
  {volume} {85}},\ \bibinfo {pages} {023623} (\bibinfo {year}
  {2012})}\BibitemShut {NoStop}%
\bibitem [{\citenamefont {Barker}\ \emph
  {et~al.}(2020{\natexlab{a}})\citenamefont {Barker}, \citenamefont {Sunami},
  \citenamefont {Garrick}, \citenamefont {Beregi}, \citenamefont {Luksch},
  \citenamefont {Bentine},\ and\ \citenamefont {Foot}}]{barker2020}%
  \BibitemOpen
  \bibfield  {author} {\bibinfo {author} {\bibfnamefont {A.~J.}\ \bibnamefont
  {Barker}}, \bibinfo {author} {\bibfnamefont {S.}~\bibnamefont {Sunami}},
  \bibinfo {author} {\bibfnamefont {D.}~\bibnamefont {Garrick}}, \bibinfo
  {author} {\bibfnamefont {A.}~\bibnamefont {Beregi}}, \bibinfo {author}
  {\bibfnamefont {K.}~\bibnamefont {Luksch}}, \bibinfo {author} {\bibfnamefont
  {E.}~\bibnamefont {Bentine}}, \ and\ \bibinfo {author} {\bibfnamefont
  {C.~J.}\ \bibnamefont {Foot}},\ }\href {\doibase 10.1088/1367-2630/abbced}
  {\bibfield  {journal} {\bibinfo  {journal} {New J. Phys.}\ }\textbf {\bibinfo
  {volume} {22}},\ \bibinfo {pages} {103040} (\bibinfo {year}
  {2020}{\natexlab{a}})}\BibitemShut {NoStop}%
\bibitem [{\citenamefont {Barker}\ \emph
  {et~al.}(2020{\natexlab{b}})\citenamefont {Barker}, \citenamefont {Sunami},
  \citenamefont {Garrick}, \citenamefont {Beregi}, \citenamefont {Luksch},
  \citenamefont {Bentine},\ and\ \citenamefont {Foot}}]{barker2020a}%
  \BibitemOpen
  \bibfield  {author} {\bibinfo {author} {\bibfnamefont {A.~J.}\ \bibnamefont
  {Barker}}, \bibinfo {author} {\bibfnamefont {S.}~\bibnamefont {Sunami}},
  \bibinfo {author} {\bibfnamefont {D.}~\bibnamefont {Garrick}}, \bibinfo
  {author} {\bibfnamefont {A.}~\bibnamefont {Beregi}}, \bibinfo {author}
  {\bibfnamefont {K.}~\bibnamefont {Luksch}}, \bibinfo {author} {\bibfnamefont
  {E.}~\bibnamefont {Bentine}}, \ and\ \bibinfo {author} {\bibfnamefont
  {C.~J.}\ \bibnamefont {Foot}},\ }\href {\doibase 10.1088/1361-6455/ab9152}
  {\bibfield  {journal} {\bibinfo  {journal} {J. Phys. B: At. Mol. Opt. Phys.}\
  }\textbf {\bibinfo {volume} {53}},\ \bibinfo {pages} {155001} (\bibinfo
  {year} {2020}{\natexlab{b}})}\BibitemShut {NoStop}%
\bibitem [{\citenamefont {Olshanii}(1998)}]{olshanii1998}%
  \BibitemOpen
  \bibfield  {author} {\bibinfo {author} {\bibfnamefont {M.}~\bibnamefont
  {Olshanii}},\ }\href {\doibase 10.1103/PhysRevLett.81.938} {\bibfield
  {journal} {\bibinfo  {journal} {Phys. Rev. Lett.}\ }\textbf {\bibinfo
  {volume} {81}},\ \bibinfo {pages} {938} (\bibinfo {year} {1998})}\BibitemShut
  {NoStop}%
\bibitem [{\citenamefont {K{\"o}hler}\ \emph {et~al.}(2006)\citenamefont
  {K{\"o}hler}, \citenamefont {G{\'o}ral},\ and\ \citenamefont
  {Julienne}}]{kohler2006}%
  \BibitemOpen
  \bibfield  {author} {\bibinfo {author} {\bibfnamefont {T.}~\bibnamefont
  {K{\"o}hler}}, \bibinfo {author} {\bibfnamefont {K.}~\bibnamefont
  {G{\'o}ral}}, \ and\ \bibinfo {author} {\bibfnamefont {P.~S.}\ \bibnamefont
  {Julienne}},\ }\href {\doibase 10.1103/RevModPhys.78.1311} {\bibfield
  {journal} {\bibinfo  {journal} {Rev. Mod. Phys.}\ }\textbf {\bibinfo {volume}
  {78}},\ \bibinfo {pages} {1311} (\bibinfo {year} {2006})}\BibitemShut
  {NoStop}%
\bibitem [{\citenamefont {Landau}(1933)}]{landau1933}%
  \BibitemOpen
  \bibfield  {author} {\bibinfo {author} {\bibfnamefont {L.~D.}\ \bibnamefont
  {Landau}},\ }\href@noop {} {\bibfield  {journal} {\bibinfo  {journal} {Phys.
  Z. Sowjetunion}\ }\textbf {\bibinfo {volume} {3}},\ \bibinfo {pages} {644}
  (\bibinfo {year} {1933})}\BibitemShut {NoStop}%
\bibitem [{\citenamefont {Schirotzek}\ \emph {et~al.}(2009)\citenamefont
  {Schirotzek}, \citenamefont {Wu}, \citenamefont {Sommer},\ and\ \citenamefont
  {Zwierlein}}]{schirotzek2009}%
  \BibitemOpen
  \bibfield  {author} {\bibinfo {author} {\bibfnamefont {A.}~\bibnamefont
  {Schirotzek}}, \bibinfo {author} {\bibfnamefont {C.-H.}\ \bibnamefont {Wu}},
  \bibinfo {author} {\bibfnamefont {A.}~\bibnamefont {Sommer}}, \ and\ \bibinfo
  {author} {\bibfnamefont {M.~W.}\ \bibnamefont {Zwierlein}},\ }\href {\doibase
  10.1103/PhysRevLett.102.230402} {\bibfield  {journal} {\bibinfo  {journal}
  {Phys. Rev. Lett.}\ }\textbf {\bibinfo {volume} {102}},\ \bibinfo {pages}
  {230402} (\bibinfo {year} {2009})}\BibitemShut {NoStop}%
\bibitem [{\citenamefont {Nascimb{\`e}ne}\ \emph {et~al.}(2009)\citenamefont
  {Nascimb{\`e}ne}, \citenamefont {Navon}, \citenamefont {Jiang}, \citenamefont
  {Tarruell}, \citenamefont {Teichmann}, \citenamefont {McKeever},
  \citenamefont {Chevy},\ and\ \citenamefont {Salomon}}]{nascimbene2009}%
  \BibitemOpen
  \bibfield  {author} {\bibinfo {author} {\bibfnamefont {S.}~\bibnamefont
  {Nascimb{\`e}ne}}, \bibinfo {author} {\bibfnamefont {N.}~\bibnamefont
  {Navon}}, \bibinfo {author} {\bibfnamefont {K.~J.}\ \bibnamefont {Jiang}},
  \bibinfo {author} {\bibfnamefont {L.}~\bibnamefont {Tarruell}}, \bibinfo
  {author} {\bibfnamefont {M.}~\bibnamefont {Teichmann}}, \bibinfo {author}
  {\bibfnamefont {J.}~\bibnamefont {McKeever}}, \bibinfo {author}
  {\bibfnamefont {F.}~\bibnamefont {Chevy}}, \ and\ \bibinfo {author}
  {\bibfnamefont {C.}~\bibnamefont {Salomon}},\ }\href {\doibase
  10.1103/PhysRevLett.103.170402} {\bibfield  {journal} {\bibinfo  {journal}
  {Phys. Rev. Lett.}\ }\textbf {\bibinfo {volume} {103}},\ \bibinfo {pages}
  {170402} (\bibinfo {year} {2009})}\BibitemShut {NoStop}%
\bibitem [{\citenamefont {Kohstall}\ \emph {et~al.}(2012)\citenamefont
  {Kohstall}, \citenamefont {Zaccanti}, \citenamefont {Jag}, \citenamefont
  {Trenkwalder}, \citenamefont {Massignan}, \citenamefont {Bruun},
  \citenamefont {Schreck},\ and\ \citenamefont {Grimm}}]{kohstall2012}%
  \BibitemOpen
  \bibfield  {author} {\bibinfo {author} {\bibfnamefont {C.}~\bibnamefont
  {Kohstall}}, \bibinfo {author} {\bibfnamefont {M.}~\bibnamefont {Zaccanti}},
  \bibinfo {author} {\bibfnamefont {M.}~\bibnamefont {Jag}}, \bibinfo {author}
  {\bibfnamefont {A.}~\bibnamefont {Trenkwalder}}, \bibinfo {author}
  {\bibfnamefont {P.}~\bibnamefont {Massignan}}, \bibinfo {author}
  {\bibfnamefont {G.~M.}\ \bibnamefont {Bruun}}, \bibinfo {author}
  {\bibfnamefont {F.}~\bibnamefont {Schreck}}, \ and\ \bibinfo {author}
  {\bibfnamefont {R.}~\bibnamefont {Grimm}},\ }\href {\doibase
  10.1038/nature11065} {\bibfield  {journal} {\bibinfo  {journal} {Nature}\
  }\textbf {\bibinfo {volume} {485}},\ \bibinfo {pages} {615} (\bibinfo {year}
  {2012})}\BibitemShut {NoStop}%
\bibitem [{\citenamefont {Ngampruetikorn}\ \emph {et~al.}(2012)\citenamefont
  {Ngampruetikorn}, \citenamefont {Levinsen},\ and\ \citenamefont
  {Parish}}]{ngampruetikorn2012}%
  \BibitemOpen
  \bibfield  {author} {\bibinfo {author} {\bibfnamefont {V.}~\bibnamefont
  {Ngampruetikorn}}, \bibinfo {author} {\bibfnamefont {J.}~\bibnamefont
  {Levinsen}}, \ and\ \bibinfo {author} {\bibfnamefont {M.~M.}\ \bibnamefont
  {Parish}},\ }\href {\doibase 10.1209/0295-5075/98/30005} {\bibfield
  {journal} {\bibinfo  {journal} {Europhys. Lett.}\ }\textbf {\bibinfo {volume}
  {98}},\ \bibinfo {pages} {30005} (\bibinfo {year} {2012})}\BibitemShut
  {NoStop}%
\bibitem [{\citenamefont {Massignan}\ \emph {et~al.}(2014)\citenamefont
  {Massignan}, \citenamefont {Zaccanti},\ and\ \citenamefont
  {Bruun}}]{massignan2014}%
  \BibitemOpen
  \bibfield  {author} {\bibinfo {author} {\bibfnamefont {P.}~\bibnamefont
  {Massignan}}, \bibinfo {author} {\bibfnamefont {M.}~\bibnamefont {Zaccanti}},
  \ and\ \bibinfo {author} {\bibfnamefont {G.~M.}\ \bibnamefont {Bruun}},\
  }\href {\doibase 10.1088/0034-4885/77/3/034401} {\bibfield  {journal}
  {\bibinfo  {journal} {Rep. Prog. Phys.}\ }\textbf {\bibinfo {volume} {77}},\
  \bibinfo {pages} {034401} (\bibinfo {year} {2014})}\BibitemShut {NoStop}%
\bibitem [{\citenamefont {Schmidt}\ \emph {et~al.}(2018)\citenamefont
  {Schmidt}, \citenamefont {Knap}, \citenamefont {Ivanov}, \citenamefont {You},
  \citenamefont {Cetina},\ and\ \citenamefont {Demler}}]{schmidt2018a}%
  \BibitemOpen
  \bibfield  {author} {\bibinfo {author} {\bibfnamefont {R.}~\bibnamefont
  {Schmidt}}, \bibinfo {author} {\bibfnamefont {M.}~\bibnamefont {Knap}},
  \bibinfo {author} {\bibfnamefont {D.~A.}\ \bibnamefont {Ivanov}}, \bibinfo
  {author} {\bibfnamefont {J.-S.}\ \bibnamefont {You}}, \bibinfo {author}
  {\bibfnamefont {M.}~\bibnamefont {Cetina}}, \ and\ \bibinfo {author}
  {\bibfnamefont {E.}~\bibnamefont {Demler}},\ }\href {\doibase
  10.1088/1361-6633/aa9593} {\bibfield  {journal} {\bibinfo  {journal} {Rep.
  Prog. Phys.}\ }\textbf {\bibinfo {volume} {81}},\ \bibinfo {pages} {024401}
  (\bibinfo {year} {2018})}\BibitemShut {NoStop}%
\bibitem [{\citenamefont {Tajima}\ \emph {et~al.}(2021)\citenamefont {Tajima},
  \citenamefont {Takahashi}, \citenamefont {Mistakidis}, \citenamefont
  {Nakano},\ and\ \citenamefont {Iida}}]{tajima2021polaron}%
  \BibitemOpen
  \bibfield  {author} {\bibinfo {author} {\bibfnamefont {H.}~\bibnamefont
  {Tajima}}, \bibinfo {author} {\bibfnamefont {J.}~\bibnamefont {Takahashi}},
  \bibinfo {author} {\bibfnamefont {S.~I.}\ \bibnamefont {Mistakidis}},
  \bibinfo {author} {\bibfnamefont {E.}~\bibnamefont {Nakano}}, \ and\ \bibinfo
  {author} {\bibfnamefont {K.}~\bibnamefont {Iida}},\ }\href {\doibase
  10.3390/atoms9010018} {\bibfield  {journal} {\bibinfo  {journal} {Atoms}\
  }\textbf {\bibinfo {volume} {9}},\ \bibinfo {pages} {18} (\bibinfo {year}
  {2021})}\BibitemShut {NoStop}%
\bibitem [{\citenamefont {Fukuhara}\ \emph {et~al.}(2013)\citenamefont
  {Fukuhara}, \citenamefont {Kantian}, \citenamefont {Endres}, \citenamefont
  {Cheneau}, \citenamefont {Schau{\ss}}, \citenamefont {Hild}, \citenamefont
  {Bellem}, \citenamefont {Schollw{\"o}ck}, \citenamefont {Giamarchi},
  \citenamefont {Gross}, \citenamefont {Bloch},\ and\ \citenamefont
  {Kuhr}}]{fukuhara2013}%
  \BibitemOpen
  \bibfield  {author} {\bibinfo {author} {\bibfnamefont {T.}~\bibnamefont
  {Fukuhara}}, \bibinfo {author} {\bibfnamefont {A.}~\bibnamefont {Kantian}},
  \bibinfo {author} {\bibfnamefont {M.}~\bibnamefont {Endres}}, \bibinfo
  {author} {\bibfnamefont {M.}~\bibnamefont {Cheneau}}, \bibinfo {author}
  {\bibfnamefont {P.}~\bibnamefont {Schau{\ss}}}, \bibinfo {author}
  {\bibfnamefont {S.}~\bibnamefont {Hild}}, \bibinfo {author} {\bibfnamefont
  {D.}~\bibnamefont {Bellem}}, \bibinfo {author} {\bibfnamefont
  {U.}~\bibnamefont {Schollw{\"o}ck}}, \bibinfo {author} {\bibfnamefont
  {T.}~\bibnamefont {Giamarchi}}, \bibinfo {author} {\bibfnamefont
  {C.}~\bibnamefont {Gross}}, \bibinfo {author} {\bibfnamefont
  {I.}~\bibnamefont {Bloch}}, \ and\ \bibinfo {author} {\bibfnamefont
  {S.}~\bibnamefont {Kuhr}},\ }\href {\doibase 10.1038/nphys2561} {\bibfield
  {journal} {\bibinfo  {journal} {Nat. Phys.}\ }\textbf {\bibinfo {volume}
  {9}},\ \bibinfo {pages} {235} (\bibinfo {year} {2013})}\BibitemShut {NoStop}%
\bibitem [{\citenamefont {J{\o}rgensen}\ \emph {et~al.}(2016)\citenamefont
  {J{\o}rgensen}, \citenamefont {Wacker}, \citenamefont {Skalmstang},
  \citenamefont {Parish}, \citenamefont {Levinsen}, \citenamefont
  {Christensen}, \citenamefont {Bruun},\ and\ \citenamefont
  {Arlt}}]{jorgensen2016}%
  \BibitemOpen
  \bibfield  {author} {\bibinfo {author} {\bibfnamefont {N.~B.}\ \bibnamefont
  {J{\o}rgensen}}, \bibinfo {author} {\bibfnamefont {L.}~\bibnamefont
  {Wacker}}, \bibinfo {author} {\bibfnamefont {K.~T.}\ \bibnamefont
  {Skalmstang}}, \bibinfo {author} {\bibfnamefont {M.~M.}\ \bibnamefont
  {Parish}}, \bibinfo {author} {\bibfnamefont {J.}~\bibnamefont {Levinsen}},
  \bibinfo {author} {\bibfnamefont {R.~S.}\ \bibnamefont {Christensen}},
  \bibinfo {author} {\bibfnamefont {G.~M.}\ \bibnamefont {Bruun}}, \ and\
  \bibinfo {author} {\bibfnamefont {J.~J.}\ \bibnamefont {Arlt}},\ }\href
  {\doibase 10.1103/PhysRevLett.117.055302} {\bibfield  {journal} {\bibinfo
  {journal} {Phys. Rev. Lett.}\ }\textbf {\bibinfo {volume} {117}},\ \bibinfo
  {pages} {055302} (\bibinfo {year} {2016})}\BibitemShut {NoStop}%
\bibitem [{\citenamefont {Hu}\ \emph {et~al.}(2016)\citenamefont {Hu},
  \citenamefont {{Van de Graaff}}, \citenamefont {Kedar}, \citenamefont
  {Corson}, \citenamefont {Cornell},\ and\ \citenamefont {Jin}}]{hu2016}%
  \BibitemOpen
  \bibfield  {author} {\bibinfo {author} {\bibfnamefont {M.-G.}\ \bibnamefont
  {Hu}}, \bibinfo {author} {\bibfnamefont {M.~J.}\ \bibnamefont {{Van de
  Graaff}}}, \bibinfo {author} {\bibfnamefont {D.}~\bibnamefont {Kedar}},
  \bibinfo {author} {\bibfnamefont {J.~P.}\ \bibnamefont {Corson}}, \bibinfo
  {author} {\bibfnamefont {E.~A.}\ \bibnamefont {Cornell}}, \ and\ \bibinfo
  {author} {\bibfnamefont {D.~S.}\ \bibnamefont {Jin}},\ }\href {\doibase
  10.1103/PhysRevLett.117.055301} {\bibfield  {journal} {\bibinfo  {journal}
  {Phys. Rev. Lett.}\ }\textbf {\bibinfo {volume} {117}},\ \bibinfo {pages}
  {055301} (\bibinfo {year} {2016})}\BibitemShut {NoStop}%
\bibitem [{\citenamefont {Volosniev}\ and\ \citenamefont
  {Hammer}(2017)}]{volosniev2017}%
  \BibitemOpen
  \bibfield  {author} {\bibinfo {author} {\bibfnamefont {A.~G.}\ \bibnamefont
  {Volosniev}}\ and\ \bibinfo {author} {\bibfnamefont {H.-W.}\ \bibnamefont
  {Hammer}},\ }\href {\doibase 10.1103/PhysRevA.96.031601} {\bibfield
  {journal} {\bibinfo  {journal} {Phys. Rev. A}\ }\textbf {\bibinfo {volume}
  {96}},\ \bibinfo {pages} {031601(R)} (\bibinfo {year} {2017})}\BibitemShut
  {NoStop}%
\bibitem [{\citenamefont {Grusdt}\ \emph
  {et~al.}(2017{\natexlab{a}})\citenamefont {Grusdt}, \citenamefont {Schmidt},
  \citenamefont {Shchadilova},\ and\ \citenamefont {Demler}}]{grusdt2017}%
  \BibitemOpen
  \bibfield  {author} {\bibinfo {author} {\bibfnamefont {F.}~\bibnamefont
  {Grusdt}}, \bibinfo {author} {\bibfnamefont {R.}~\bibnamefont {Schmidt}},
  \bibinfo {author} {\bibfnamefont {Y.~E.}\ \bibnamefont {Shchadilova}}, \ and\
  \bibinfo {author} {\bibfnamefont {E.}~\bibnamefont {Demler}},\ }\href
  {\doibase 10.1103/PhysRevA.96.013607} {\bibfield  {journal} {\bibinfo
  {journal} {Phys. Rev. A}\ }\textbf {\bibinfo {volume} {96}},\ \bibinfo
  {pages} {013607} (\bibinfo {year} {2017}{\natexlab{a}})}\BibitemShut
  {NoStop}%
\bibitem [{\citenamefont {Mistakidis}\ \emph
  {et~al.}(2019{\natexlab{a}})\citenamefont {Mistakidis}, \citenamefont
  {Katsimiga}, \citenamefont {Koutentakis}, \citenamefont {Busch},\ and\
  \citenamefont {Schmelcher}}]{mistakidis2019c}%
  \BibitemOpen
  \bibfield  {author} {\bibinfo {author} {\bibfnamefont {S.~I.}\ \bibnamefont
  {Mistakidis}}, \bibinfo {author} {\bibfnamefont {G.~C.}\ \bibnamefont
  {Katsimiga}}, \bibinfo {author} {\bibfnamefont {G.~M.}\ \bibnamefont
  {Koutentakis}}, \bibinfo {author} {\bibfnamefont {T.}~\bibnamefont {Busch}},
  \ and\ \bibinfo {author} {\bibfnamefont {P.}~\bibnamefont {Schmelcher}},\
  }\href {\doibase 10.1103/PhysRevLett.122.183001} {\bibfield  {journal}
  {\bibinfo  {journal} {Phys. Rev. Lett.}\ }\textbf {\bibinfo {volume} {122}},\
  \bibinfo {pages} {183001} (\bibinfo {year} {2019}{\natexlab{a}})}\BibitemShut
  {NoStop}%
\bibitem [{\citenamefont {Pe{\~n}a~Ardila}\ \emph {et~al.}(2019)\citenamefont
  {Pe{\~n}a~Ardila}, \citenamefont {J{\"o}rgensen}, \citenamefont {Pohl},
  \citenamefont {Giorgini}, \citenamefont {Bruun},\ and\ \citenamefont
  {Arlt}}]{ardila2019a}%
  \BibitemOpen
  \bibfield  {author} {\bibinfo {author} {\bibfnamefont {L.~A.}\ \bibnamefont
  {Pe{\~n}a~Ardila}}, \bibinfo {author} {\bibfnamefont {N.~B.}\ \bibnamefont
  {J{\"o}rgensen}}, \bibinfo {author} {\bibfnamefont {T.}~\bibnamefont {Pohl}},
  \bibinfo {author} {\bibfnamefont {S.}~\bibnamefont {Giorgini}}, \bibinfo
  {author} {\bibfnamefont {G.~M.}\ \bibnamefont {Bruun}}, \ and\ \bibinfo
  {author} {\bibfnamefont {J.~J.}\ \bibnamefont {Arlt}},\ }\href {\doibase
  10.1103/PhysRevA.99.063607} {\bibfield  {journal} {\bibinfo  {journal} {Phys.
  Rev. A}\ }\textbf {\bibinfo {volume} {99}},\ \bibinfo {pages} {063607}
  (\bibinfo {year} {2019})}\BibitemShut {NoStop}%
\bibitem [{\citenamefont {Pe{\~n}a~Ardila}\ \emph {et~al.}(2020)\citenamefont
  {Pe{\~n}a~Ardila}, \citenamefont {Astrakharchik},\ and\ \citenamefont
  {Giorgini}}]{ardila2020}%
  \BibitemOpen
  \bibfield  {author} {\bibinfo {author} {\bibfnamefont {L.~A.}\ \bibnamefont
  {Pe{\~n}a~Ardila}}, \bibinfo {author} {\bibfnamefont {G.~E.}\ \bibnamefont
  {Astrakharchik}}, \ and\ \bibinfo {author} {\bibfnamefont {S.}~\bibnamefont
  {Giorgini}},\ }\href {\doibase 10.1103/PhysRevResearch.2.023405} {\bibfield
  {journal} {\bibinfo  {journal} {Phys. Rev. Research}\ }\textbf {\bibinfo
  {volume} {2}},\ \bibinfo {pages} {023405} (\bibinfo {year}
  {2020})}\BibitemShut {NoStop}%
\bibitem [{\citenamefont {Skou}\ \emph {et~al.}(2021)\citenamefont {Skou},
  \citenamefont {Skov}, \citenamefont {J{\o}rgensen}, \citenamefont {Nielsen},
  \citenamefont {{Camacho-Guardian}}, \citenamefont {Pohl}, \citenamefont
  {Bruun},\ and\ \citenamefont {Arlt}}]{skou2021}%
  \BibitemOpen
  \bibfield  {author} {\bibinfo {author} {\bibfnamefont {M.~G.}\ \bibnamefont
  {Skou}}, \bibinfo {author} {\bibfnamefont {T.~G.}\ \bibnamefont {Skov}},
  \bibinfo {author} {\bibfnamefont {N.~B.}\ \bibnamefont {J{\o}rgensen}},
  \bibinfo {author} {\bibfnamefont {K.~K.}\ \bibnamefont {Nielsen}}, \bibinfo
  {author} {\bibfnamefont {A.}~\bibnamefont {{Camacho-Guardian}}}, \bibinfo
  {author} {\bibfnamefont {T.}~\bibnamefont {Pohl}}, \bibinfo {author}
  {\bibfnamefont {G.~M.}\ \bibnamefont {Bruun}}, \ and\ \bibinfo {author}
  {\bibfnamefont {J.~J.}\ \bibnamefont {Arlt}},\ }\href {\doibase
  10.1038/s41567-021-01184-5} {\bibfield  {journal} {\bibinfo  {journal} {Nat.
  Phys.}\ }\textbf {\bibinfo {volume} {17}},\ \bibinfo {pages} {731} (\bibinfo
  {year} {2021})}\BibitemShut {NoStop}%
\bibitem [{\citenamefont {Grusdt}\ \emph
  {et~al.}(2017{\natexlab{b}})\citenamefont {Grusdt}, \citenamefont
  {Astrakharchik},\ and\ \citenamefont {Demler}}]{grusdt2017a}%
  \BibitemOpen
  \bibfield  {author} {\bibinfo {author} {\bibfnamefont {F.}~\bibnamefont
  {Grusdt}}, \bibinfo {author} {\bibfnamefont {G.~E.}\ \bibnamefont
  {Astrakharchik}}, \ and\ \bibinfo {author} {\bibfnamefont {E.}~\bibnamefont
  {Demler}},\ }\href {\doibase 10.1088/1367-2630/aa8a2e} {\bibfield  {journal}
  {\bibinfo  {journal} {New J. Phys.}\ }\textbf {\bibinfo {volume} {19}},\
  \bibinfo {pages} {103035} (\bibinfo {year} {2017}{\natexlab{b}})}\BibitemShut
  {NoStop}%
\bibitem [{\citenamefont {{Camacho-Guardian}}\ and\ \citenamefont
  {Bruun}(2018)}]{camacho-guardian2018}%
  \BibitemOpen
  \bibfield  {author} {\bibinfo {author} {\bibfnamefont {A.}~\bibnamefont
  {{Camacho-Guardian}}}\ and\ \bibinfo {author} {\bibfnamefont {G.~M.}\
  \bibnamefont {Bruun}},\ }\href {\doibase 10.1103/PhysRevX.8.031042}
  {\bibfield  {journal} {\bibinfo  {journal} {Phys. Rev. X}\ }\textbf {\bibinfo
  {volume} {8}},\ \bibinfo {pages} {031042} (\bibinfo {year}
  {2018})}\BibitemShut {NoStop}%
\bibitem [{\citenamefont {Huber}\ \emph {et~al.}(2019)\citenamefont {Huber},
  \citenamefont {Hammer},\ and\ \citenamefont {Volosniev}}]{huber2019}%
  \BibitemOpen
  \bibfield  {author} {\bibinfo {author} {\bibfnamefont {D.}~\bibnamefont
  {Huber}}, \bibinfo {author} {\bibfnamefont {H.-W.}\ \bibnamefont {Hammer}}, \
  and\ \bibinfo {author} {\bibfnamefont {A.~G.}\ \bibnamefont {Volosniev}},\
  }\href {\doibase 10.1103/PhysRevResearch.1.033177} {\bibfield  {journal}
  {\bibinfo  {journal} {Phys. Rev. Research}\ }\textbf {\bibinfo {volume}
  {1}},\ \bibinfo {pages} {033177} (\bibinfo {year} {2019})}\BibitemShut
  {NoStop}%
\bibitem [{\citenamefont {Brauneis}\ \emph {et~al.}(2021)\citenamefont
  {Brauneis}, \citenamefont {Hammer}, \citenamefont {Lemeshko},\ and\
  \citenamefont {Volosniev}}]{brauneis2021}%
  \BibitemOpen
  \bibfield  {author} {\bibinfo {author} {\bibfnamefont {F.}~\bibnamefont
  {Brauneis}}, \bibinfo {author} {\bibfnamefont {H.-W.}\ \bibnamefont
  {Hammer}}, \bibinfo {author} {\bibfnamefont {M.}~\bibnamefont {Lemeshko}}, \
  and\ \bibinfo {author} {\bibfnamefont {A.}~\bibnamefont {Volosniev}},\ }\href
  {\doibase 10.21468/SciPostPhys.11.1.008} {\bibfield  {journal} {\bibinfo
  {journal} {SciPost Phys.}\ }\textbf {\bibinfo {volume} {11}},\ \bibinfo
  {pages} {008} (\bibinfo {year} {2021})}\BibitemShut {NoStop}%
\bibitem [{\citenamefont {Dehkharghani}\ \emph {et~al.}(2018)\citenamefont
  {Dehkharghani}, \citenamefont {Volosniev},\ and\ \citenamefont
  {Zinner}}]{dehkharghani2018}%
  \BibitemOpen
  \bibfield  {author} {\bibinfo {author} {\bibfnamefont {A.~S.}\ \bibnamefont
  {Dehkharghani}}, \bibinfo {author} {\bibfnamefont {A.~G.}\ \bibnamefont
  {Volosniev}}, \ and\ \bibinfo {author} {\bibfnamefont {N.~T.}\ \bibnamefont
  {Zinner}},\ }\href {\doibase 10.1103/PhysRevLett.121.080405} {\bibfield
  {journal} {\bibinfo  {journal} {Phys. Rev. Lett.}\ }\textbf {\bibinfo
  {volume} {121}},\ \bibinfo {pages} {080405} (\bibinfo {year}
  {2018})}\BibitemShut {NoStop}%
\bibitem [{\citenamefont {Charalambous}\ \emph {et~al.}(2019)\citenamefont
  {Charalambous}, \citenamefont {{Garcia-March}}, \citenamefont {Lampo},
  \citenamefont {Mehboud},\ and\ \citenamefont
  {Lewenstein}}]{charalambous2019}%
  \BibitemOpen
  \bibfield  {author} {\bibinfo {author} {\bibfnamefont {C.}~\bibnamefont
  {Charalambous}}, \bibinfo {author} {\bibfnamefont {M.~A.}\ \bibnamefont
  {{Garcia-March}}}, \bibinfo {author} {\bibfnamefont {A.}~\bibnamefont
  {Lampo}}, \bibinfo {author} {\bibfnamefont {M.}~\bibnamefont {Mehboud}}, \
  and\ \bibinfo {author} {\bibfnamefont {M.}~\bibnamefont {Lewenstein}},\
  }\href {\doibase 10.21468/SciPostPhys.6.1.010} {\bibfield  {journal}
  {\bibinfo  {journal} {SciPost Phys.}\ }\textbf {\bibinfo {volume} {6}},\
  \bibinfo {pages} {010} (\bibinfo {year} {2019})}\BibitemShut {NoStop}%
\bibitem [{\citenamefont {Reichert}\ \emph {et~al.}(2019)\citenamefont
  {Reichert}, \citenamefont {Ristivojevic},\ and\ \citenamefont
  {Petkovi{\'c}}}]{reichert2019}%
  \BibitemOpen
  \bibfield  {author} {\bibinfo {author} {\bibfnamefont {B.}~\bibnamefont
  {Reichert}}, \bibinfo {author} {\bibfnamefont {Z.}~\bibnamefont
  {Ristivojevic}}, \ and\ \bibinfo {author} {\bibfnamefont {A.}~\bibnamefont
  {Petkovi{\'c}}},\ }\href {\doibase 10.1088/1367-2630/ab1b8e} {\bibfield
  {journal} {\bibinfo  {journal} {New J. Phys.}\ }\textbf {\bibinfo {volume}
  {21}},\ \bibinfo {pages} {053024} (\bibinfo {year} {2019})}\BibitemShut
  {NoStop}%
\bibitem [{\citenamefont {Keiler}\ \emph {et~al.}(2021)\citenamefont {Keiler},
  \citenamefont {Mistakidis},\ and\ \citenamefont {Schmelcher}}]{keiler2021}%
  \BibitemOpen
  \bibfield  {author} {\bibinfo {author} {\bibfnamefont {K.}~\bibnamefont
  {Keiler}}, \bibinfo {author} {\bibfnamefont {S.~I.}\ \bibnamefont
  {Mistakidis}}, \ and\ \bibinfo {author} {\bibfnamefont {P.}~\bibnamefont
  {Schmelcher}},\ }\href {\doibase 10.1103/PhysRevA.104.L031301} {\bibfield
  {journal} {\bibinfo  {journal} {Phys. Rev. A}\ }\textbf {\bibinfo {volume}
  {104}},\ \bibinfo {pages} {L031301} (\bibinfo {year} {2021})}\BibitemShut
  {NoStop}%
\bibitem [{\citenamefont {{Camacho-Guardian}}\ \emph
  {et~al.}(2018)\citenamefont {{Camacho-Guardian}}, \citenamefont
  {Pe{\~n}a~Ardila}, \citenamefont {Pohl},\ and\ \citenamefont
  {Bruun}}]{camacho-guardian2018a}%
  \BibitemOpen
  \bibfield  {author} {\bibinfo {author} {\bibfnamefont {A.}~\bibnamefont
  {{Camacho-Guardian}}}, \bibinfo {author} {\bibfnamefont {L.~A.}\ \bibnamefont
  {Pe{\~n}a~Ardila}}, \bibinfo {author} {\bibfnamefont {T.}~\bibnamefont
  {Pohl}}, \ and\ \bibinfo {author} {\bibfnamefont {G.~M.}\ \bibnamefont
  {Bruun}},\ }\href {\doibase 10.1103/PhysRevLett.121.013401} {\bibfield
  {journal} {\bibinfo  {journal} {Phys. Rev. Lett.}\ }\textbf {\bibinfo
  {volume} {121}},\ \bibinfo {pages} {013401} (\bibinfo {year}
  {2018})}\BibitemShut {NoStop}%
\bibitem [{\citenamefont {Will}\ \emph {et~al.}(2021)\citenamefont {Will},
  \citenamefont {Astrakharchik},\ and\ \citenamefont
  {Fleischhauer}}]{will2021}%
  \BibitemOpen
  \bibfield  {author} {\bibinfo {author} {\bibfnamefont {M.}~\bibnamefont
  {Will}}, \bibinfo {author} {\bibfnamefont {G.~E.}\ \bibnamefont
  {Astrakharchik}}, \ and\ \bibinfo {author} {\bibfnamefont {M.}~\bibnamefont
  {Fleischhauer}},\ }\href {\doibase 10.1103/PhysRevLett.127.103401} {\bibfield
   {journal} {\bibinfo  {journal} {Phys. Rev. Lett.}\ }\textbf {\bibinfo
  {volume} {127}},\ \bibinfo {pages} {103401} (\bibinfo {year}
  {2021})}\BibitemShut {NoStop}%
\bibitem [{\citenamefont {Mukherjee}\ \emph {et~al.}(2020)\citenamefont
  {Mukherjee}, \citenamefont {Mistakidis}, \citenamefont {Majumder},\ and\
  \citenamefont {Schmelcher}}]{mukherjee2020a}%
  \BibitemOpen
  \bibfield  {author} {\bibinfo {author} {\bibfnamefont {K.}~\bibnamefont
  {Mukherjee}}, \bibinfo {author} {\bibfnamefont {S.~I.}\ \bibnamefont
  {Mistakidis}}, \bibinfo {author} {\bibfnamefont {S.}~\bibnamefont
  {Majumder}}, \ and\ \bibinfo {author} {\bibfnamefont {P.}~\bibnamefont
  {Schmelcher}},\ }\href {\doibase 10.1103/PhysRevA.102.053317} {\bibfield
  {journal} {\bibinfo  {journal} {Phys. Rev. A}\ }\textbf {\bibinfo {volume}
  {102}},\ \bibinfo {pages} {053317} (\bibinfo {year} {2020})}\BibitemShut
  {NoStop}%
\bibitem [{\citenamefont {Mistakidis}\ \emph
  {et~al.}(2020{\natexlab{a}})\citenamefont {Mistakidis}, \citenamefont
  {Koutentakis}, \citenamefont {Katsimiga}, \citenamefont {Busch},\ and\
  \citenamefont {Schmelcher}}]{mistakidis2020}%
  \BibitemOpen
  \bibfield  {author} {\bibinfo {author} {\bibfnamefont {S.~I.}\ \bibnamefont
  {Mistakidis}}, \bibinfo {author} {\bibfnamefont {G.~M.}\ \bibnamefont
  {Koutentakis}}, \bibinfo {author} {\bibfnamefont {G.~C.}\ \bibnamefont
  {Katsimiga}}, \bibinfo {author} {\bibfnamefont {T.}~\bibnamefont {Busch}}, \
  and\ \bibinfo {author} {\bibfnamefont {P.}~\bibnamefont {Schmelcher}},\
  }\href {\doibase 10.1088/1367-2630/ab7599} {\bibfield  {journal} {\bibinfo
  {journal} {New J. Phys.}\ }\textbf {\bibinfo {volume} {22}},\ \bibinfo
  {pages} {043007} (\bibinfo {year} {2020}{\natexlab{a}})}\BibitemShut
  {NoStop}%
\bibitem [{\citenamefont {Bougas}\ \emph {et~al.}(2021)\citenamefont {Bougas},
  \citenamefont {Mistakidis},\ and\ \citenamefont {Schmelcher}}]{bougas2021}%
  \BibitemOpen
  \bibfield  {author} {\bibinfo {author} {\bibfnamefont {G.}~\bibnamefont
  {Bougas}}, \bibinfo {author} {\bibfnamefont {S.~I.}\ \bibnamefont
  {Mistakidis}}, \ and\ \bibinfo {author} {\bibfnamefont {P.}~\bibnamefont
  {Schmelcher}},\ }\href {\doibase 10.1103/PhysRevA.103.023313} {\bibfield
  {journal} {\bibinfo  {journal} {Phys. Rev. A}\ }\textbf {\bibinfo {volume}
  {103}},\ \bibinfo {pages} {023313} (\bibinfo {year} {2021})}\BibitemShut
  {NoStop}%
\bibitem [{\citenamefont {Hamner}\ \emph {et~al.}(2011)\citenamefont {Hamner},
  \citenamefont {Chang}, \citenamefont {Engels},\ and\ \citenamefont
  {Hoefer}}]{hamner2011}%
  \BibitemOpen
  \bibfield  {author} {\bibinfo {author} {\bibfnamefont {C.}~\bibnamefont
  {Hamner}}, \bibinfo {author} {\bibfnamefont {J.~J.}\ \bibnamefont {Chang}},
  \bibinfo {author} {\bibfnamefont {P.}~\bibnamefont {Engels}}, \ and\ \bibinfo
  {author} {\bibfnamefont {M.~A.}\ \bibnamefont {Hoefer}},\ }\href {\doibase
  10.1103/PhysRevLett.106.065302} {\bibfield  {journal} {\bibinfo  {journal}
  {Phys. Rev. Lett.}\ }\textbf {\bibinfo {volume} {106}},\ \bibinfo {pages}
  {065302} (\bibinfo {year} {2011})}\BibitemShut {NoStop}%
\bibitem [{\citenamefont {Weller}\ \emph {et~al.}(2008)\citenamefont {Weller},
  \citenamefont {Ronzheimer}, \citenamefont {Gross}, \citenamefont {Esteve},
  \citenamefont {Oberthaler}, \citenamefont {Frantzeskakis}, \citenamefont
  {Theocharis},\ and\ \citenamefont {Kevrekidis}}]{weller2008}%
  \BibitemOpen
  \bibfield  {author} {\bibinfo {author} {\bibfnamefont {A.}~\bibnamefont
  {Weller}}, \bibinfo {author} {\bibfnamefont {J.~P.}\ \bibnamefont
  {Ronzheimer}}, \bibinfo {author} {\bibfnamefont {C.}~\bibnamefont {Gross}},
  \bibinfo {author} {\bibfnamefont {J.}~\bibnamefont {Esteve}}, \bibinfo
  {author} {\bibfnamefont {M.~K.}\ \bibnamefont {Oberthaler}}, \bibinfo
  {author} {\bibfnamefont {D.~J.}\ \bibnamefont {Frantzeskakis}}, \bibinfo
  {author} {\bibfnamefont {G.}~\bibnamefont {Theocharis}}, \ and\ \bibinfo
  {author} {\bibfnamefont {P.~G.}\ \bibnamefont {Kevrekidis}},\ }\href
  {\doibase 10.1103/PhysRevLett.101.130401} {\bibfield  {journal} {\bibinfo
  {journal} {Phys. Rev. Lett.}\ }\textbf {\bibinfo {volume} {101}},\ \bibinfo
  {pages} {130401} (\bibinfo {year} {2008})}\BibitemShut {NoStop}%
\bibitem [{\citenamefont {K{\"o}hler}\ and\ \citenamefont
  {Schmelcher}(2021)}]{kohler2021}%
  \BibitemOpen
  \bibfield  {author} {\bibinfo {author} {\bibfnamefont {F.}~\bibnamefont
  {K{\"o}hler}}\ and\ \bibinfo {author} {\bibfnamefont {P.}~\bibnamefont
  {Schmelcher}},\ }\href {\doibase 10.1103/PhysRevA.103.043326} {\bibfield
  {journal} {\bibinfo  {journal} {Phys. Rev. A}\ }\textbf {\bibinfo {volume}
  {103}},\ \bibinfo {pages} {043326} (\bibinfo {year} {2021})}\BibitemShut
  {NoStop}%
\bibitem [{\citenamefont {Thomas}\ \emph {et~al.}(2017)\citenamefont {Thomas},
  \citenamefont {Chilcott}, \citenamefont {Chisholm}, \citenamefont {Deb},
  \citenamefont {Horvath}, \citenamefont {Sawyer},\ and\ \citenamefont
  {Kj{\ae}rgaard}}]{thomas2017}%
  \BibitemOpen
  \bibfield  {author} {\bibinfo {author} {\bibfnamefont {R.}~\bibnamefont
  {Thomas}}, \bibinfo {author} {\bibfnamefont {M.}~\bibnamefont {Chilcott}},
  \bibinfo {author} {\bibfnamefont {C.}~\bibnamefont {Chisholm}}, \bibinfo
  {author} {\bibfnamefont {A.~B.}\ \bibnamefont {Deb}}, \bibinfo {author}
  {\bibfnamefont {M.}~\bibnamefont {Horvath}}, \bibinfo {author} {\bibfnamefont
  {B.~J.}\ \bibnamefont {Sawyer}}, \ and\ \bibinfo {author} {\bibfnamefont
  {N.}~\bibnamefont {Kj{\ae}rgaard}},\ }\href {\doibase
  10.1088/1742-6596/875/2/012007} {\bibfield  {journal} {\bibinfo  {journal}
  {J. Phys.: Conf. Ser.}\ }\textbf {\bibinfo {volume} {875}},\ \bibinfo {pages}
  {012007} (\bibinfo {year} {2017})}\BibitemShut {NoStop}%
\bibitem [{\citenamefont {Thomas}\ \emph {et~al.}(2018)\citenamefont {Thomas},
  \citenamefont {Chilcott}, \citenamefont {Tiesinga}, \citenamefont {Deb},\
  and\ \citenamefont {Kj{\ae}rgaard}}]{thomas2018}%
  \BibitemOpen
  \bibfield  {author} {\bibinfo {author} {\bibfnamefont {R.}~\bibnamefont
  {Thomas}}, \bibinfo {author} {\bibfnamefont {M.}~\bibnamefont {Chilcott}},
  \bibinfo {author} {\bibfnamefont {E.}~\bibnamefont {Tiesinga}}, \bibinfo
  {author} {\bibfnamefont {A.~B.}\ \bibnamefont {Deb}}, \ and\ \bibinfo
  {author} {\bibfnamefont {N.}~\bibnamefont {Kj{\ae}rgaard}},\ }\href {\doibase
  10.1038/s41467-018-07375-8} {\bibfield  {journal} {\bibinfo  {journal} {Nat.
  Commun.}\ }\textbf {\bibinfo {volume} {9}},\ \bibinfo {pages} {4895}
  (\bibinfo {year} {2018})}\BibitemShut {NoStop}%
\bibitem [{\citenamefont {Tajima}\ \emph {et~al.}(2020)\citenamefont {Tajima},
  \citenamefont {Takahashi}, \citenamefont {Nakano},\ and\ \citenamefont
  {Iida}}]{tajima2020collisional}%
  \BibitemOpen
  \bibfield  {author} {\bibinfo {author} {\bibfnamefont {H.}~\bibnamefont
  {Tajima}}, \bibinfo {author} {\bibfnamefont {J.}~\bibnamefont {Takahashi}},
  \bibinfo {author} {\bibfnamefont {E.}~\bibnamefont {Nakano}}, \ and\ \bibinfo
  {author} {\bibfnamefont {K.}~\bibnamefont {Iida}},\ }\href {\doibase
  10.1103/PhysRevA.102.051302} {\bibfield  {journal} {\bibinfo  {journal}
  {Phys. Rev. A}\ }\textbf {\bibinfo {volume} {102}},\ \bibinfo {pages}
  {051302(R)} (\bibinfo {year} {2020})}\BibitemShut {NoStop}%
\bibitem [{\citenamefont {Kwasniok}\ \emph {et~al.}(2020)\citenamefont
  {Kwasniok}, \citenamefont {Mistakidis},\ and\ \citenamefont
  {Schmelcher}}]{kwasniok2020}%
  \BibitemOpen
  \bibfield  {author} {\bibinfo {author} {\bibfnamefont {J.}~\bibnamefont
  {Kwasniok}}, \bibinfo {author} {\bibfnamefont {S.~I.}\ \bibnamefont
  {Mistakidis}}, \ and\ \bibinfo {author} {\bibfnamefont {P.}~\bibnamefont
  {Schmelcher}},\ }\href {\doibase 10.1103/PhysRevA.101.053619} {\bibfield
  {journal} {\bibinfo  {journal} {Phys. Rev. A}\ }\textbf {\bibinfo {volume}
  {101}},\ \bibinfo {pages} {053619} (\bibinfo {year} {2020})}\BibitemShut
  {NoStop}%
\bibitem [{\citenamefont {Theel}\ \emph {et~al.}(2021)\citenamefont {Theel},
  \citenamefont {Keiler}, \citenamefont {Mistakidis},\ and\ \citenamefont
  {Schmelcher}}]{theel2021}%
  \BibitemOpen
  \bibfield  {author} {\bibinfo {author} {\bibfnamefont {F.}~\bibnamefont
  {Theel}}, \bibinfo {author} {\bibfnamefont {K.}~\bibnamefont {Keiler}},
  \bibinfo {author} {\bibfnamefont {S.~I.}\ \bibnamefont {Mistakidis}}, \ and\
  \bibinfo {author} {\bibfnamefont {P.}~\bibnamefont {Schmelcher}},\ }\href
  {\doibase 10.1103/PhysRevResearch.3.023068} {\bibfield  {journal} {\bibinfo
  {journal} {Phys. Rev. Research}\ }\textbf {\bibinfo {volume} {3}},\ \bibinfo
  {pages} {023068} (\bibinfo {year} {2021})}\BibitemShut {NoStop}%
\bibitem [{\citenamefont {Goold}\ \emph {et~al.}(2011)\citenamefont {Goold},
  \citenamefont {Fogarty}, \citenamefont {Lo~Gullo}, \citenamefont
  {Paternostro},\ and\ \citenamefont {Busch}}]{goold2011}%
  \BibitemOpen
  \bibfield  {author} {\bibinfo {author} {\bibfnamefont {J.}~\bibnamefont
  {Goold}}, \bibinfo {author} {\bibfnamefont {T.}~\bibnamefont {Fogarty}},
  \bibinfo {author} {\bibfnamefont {N.}~\bibnamefont {Lo~Gullo}}, \bibinfo
  {author} {\bibfnamefont {M.}~\bibnamefont {Paternostro}}, \ and\ \bibinfo
  {author} {\bibfnamefont {T.}~\bibnamefont {Busch}},\ }\href {\doibase
  10.1103/PhysRevA.84.063632} {\bibfield  {journal} {\bibinfo  {journal} {Phys.
  Rev. A}\ }\textbf {\bibinfo {volume} {84}},\ \bibinfo {pages} {063632}
  (\bibinfo {year} {2011})}\BibitemShut {NoStop}%
\bibitem [{\citenamefont {Mistakidis}\ \emph
  {et~al.}(2019{\natexlab{b}})\citenamefont {Mistakidis}, \citenamefont
  {Katsimiga}, \citenamefont {Koutentakis},\ and\ \citenamefont
  {Schmelcher}}]{mistakidis2019a}%
  \BibitemOpen
  \bibfield  {author} {\bibinfo {author} {\bibfnamefont {S.~I.}\ \bibnamefont
  {Mistakidis}}, \bibinfo {author} {\bibfnamefont {G.~C.}\ \bibnamefont
  {Katsimiga}}, \bibinfo {author} {\bibfnamefont {G.~M.}\ \bibnamefont
  {Koutentakis}}, \ and\ \bibinfo {author} {\bibfnamefont {P.}~\bibnamefont
  {Schmelcher}},\ }\href {\doibase 10.1088/1367-2630/ab1045} {\bibfield
  {journal} {\bibinfo  {journal} {New J. Phys.}\ }\textbf {\bibinfo {volume}
  {21}},\ \bibinfo {pages} {043032} (\bibinfo {year}
  {2019}{\natexlab{b}})}\BibitemShut {NoStop}%
\bibitem [{\citenamefont {Kr{\"o}nke}\ \emph {et~al.}(2013)\citenamefont
  {Kr{\"o}nke}, \citenamefont {Cao}, \citenamefont {Vendrell},\ and\
  \citenamefont {Schmelcher}}]{kronke2013}%
  \BibitemOpen
  \bibfield  {author} {\bibinfo {author} {\bibfnamefont {S.}~\bibnamefont
  {Kr{\"o}nke}}, \bibinfo {author} {\bibfnamefont {L.}~\bibnamefont {Cao}},
  \bibinfo {author} {\bibfnamefont {O.}~\bibnamefont {Vendrell}}, \ and\
  \bibinfo {author} {\bibfnamefont {P.}~\bibnamefont {Schmelcher}},\ }\href
  {\doibase 10.1088/1367-2630/15/6/063018} {\bibfield  {journal} {\bibinfo
  {journal} {New J. Phys.}\ }\textbf {\bibinfo {volume} {15}},\ \bibinfo
  {pages} {063018} (\bibinfo {year} {2013})}\BibitemShut {NoStop}%
\bibitem [{\citenamefont {Cao}\ \emph {et~al.}(2013)\citenamefont {Cao},
  \citenamefont {Kr{\"o}nke}, \citenamefont {Vendrell},\ and\ \citenamefont
  {Schmelcher}}]{cao2013}%
  \BibitemOpen
  \bibfield  {author} {\bibinfo {author} {\bibfnamefont {L.}~\bibnamefont
  {Cao}}, \bibinfo {author} {\bibfnamefont {S.}~\bibnamefont {Kr{\"o}nke}},
  \bibinfo {author} {\bibfnamefont {O.}~\bibnamefont {Vendrell}}, \ and\
  \bibinfo {author} {\bibfnamefont {P.}~\bibnamefont {Schmelcher}},\ }\href
  {\doibase 10.1063/1.4821350} {\bibfield  {journal} {\bibinfo  {journal} {J.
  Chem. Phys.}\ }\textbf {\bibinfo {volume} {139}},\ \bibinfo {pages} {134103}
  (\bibinfo {year} {2013})}\BibitemShut {NoStop}%
\bibitem [{\citenamefont {Cao}\ \emph {et~al.}(2017)\citenamefont {Cao},
  \citenamefont {Bolsinger}, \citenamefont {Mistakidis}, \citenamefont
  {Koutentakis}, \citenamefont {Kr{\"o}nke}, \citenamefont {Schurer},\ and\
  \citenamefont {Schmelcher}}]{cao2017}%
  \BibitemOpen
  \bibfield  {author} {\bibinfo {author} {\bibfnamefont {L.}~\bibnamefont
  {Cao}}, \bibinfo {author} {\bibfnamefont {V.}~\bibnamefont {Bolsinger}},
  \bibinfo {author} {\bibfnamefont {S.~I.}\ \bibnamefont {Mistakidis}},
  \bibinfo {author} {\bibfnamefont {G.~M.}\ \bibnamefont {Koutentakis}},
  \bibinfo {author} {\bibfnamefont {S.}~\bibnamefont {Kr{\"o}nke}}, \bibinfo
  {author} {\bibfnamefont {J.~M.}\ \bibnamefont {Schurer}}, \ and\ \bibinfo
  {author} {\bibfnamefont {P.}~\bibnamefont {Schmelcher}},\ }\href {\doibase
  10.1063/1.4993512} {\bibfield  {journal} {\bibinfo  {journal} {J. Chem.
  Phys.}\ }\textbf {\bibinfo {volume} {147}},\ \bibinfo {pages} {044106}
  (\bibinfo {year} {2017})}\BibitemShut {NoStop}%
\bibitem [{\citenamefont {K{\"o}hler}\ \emph {et~al.}(2019)\citenamefont
  {K{\"o}hler}, \citenamefont {Keiler}, \citenamefont {Mistakidis},
  \citenamefont {Meyer},\ and\ \citenamefont {Schmelcher}}]{kohler2019}%
  \BibitemOpen
  \bibfield  {author} {\bibinfo {author} {\bibfnamefont {F.}~\bibnamefont
  {K{\"o}hler}}, \bibinfo {author} {\bibfnamefont {K.}~\bibnamefont {Keiler}},
  \bibinfo {author} {\bibfnamefont {S.~I.}\ \bibnamefont {Mistakidis}},
  \bibinfo {author} {\bibfnamefont {H.-D.}\ \bibnamefont {Meyer}}, \ and\
  \bibinfo {author} {\bibfnamefont {P.}~\bibnamefont {Schmelcher}},\ }\href
  {\doibase 10.1063/1.5104344} {\bibfield  {journal} {\bibinfo  {journal} {J.
  Chem. Phys.}\ }\textbf {\bibinfo {volume} {151}},\ \bibinfo {pages} {054108}
  (\bibinfo {year} {2019})}\BibitemShut {NoStop}%
\bibitem [{\citenamefont {LeBlanc}\ and\ \citenamefont
  {Thywissen}(2007)}]{leblanc2007}%
  \BibitemOpen
  \bibfield  {author} {\bibinfo {author} {\bibfnamefont {L.~J.}\ \bibnamefont
  {LeBlanc}}\ and\ \bibinfo {author} {\bibfnamefont {J.~H.}\ \bibnamefont
  {Thywissen}},\ }\href {\doibase 10.1103/PhysRevA.75.053612} {\bibfield
  {journal} {\bibinfo  {journal} {Phys. Rev. A}\ }\textbf {\bibinfo {volume}
  {75}},\ \bibinfo {pages} {053612} (\bibinfo {year} {2007})}\BibitemShut
  {NoStop}%
\bibitem [{\citenamefont {Lercher}\ \emph {et~al.}(2011)\citenamefont
  {Lercher}, \citenamefont {Takekoshi}, \citenamefont {Debatin}, \citenamefont
  {Schuster}, \citenamefont {Rameshan}, \citenamefont {Ferlaino}, \citenamefont
  {Grimm},\ and\ \citenamefont {N{\"a}gerl}}]{lercher2011}%
  \BibitemOpen
  \bibfield  {author} {\bibinfo {author} {\bibfnamefont {A.~D.}\ \bibnamefont
  {Lercher}}, \bibinfo {author} {\bibfnamefont {T.}~\bibnamefont {Takekoshi}},
  \bibinfo {author} {\bibfnamefont {M.}~\bibnamefont {Debatin}}, \bibinfo
  {author} {\bibfnamefont {B.}~\bibnamefont {Schuster}}, \bibinfo {author}
  {\bibfnamefont {R.}~\bibnamefont {Rameshan}}, \bibinfo {author}
  {\bibfnamefont {F.}~\bibnamefont {Ferlaino}}, \bibinfo {author}
  {\bibfnamefont {R.}~\bibnamefont {Grimm}}, \ and\ \bibinfo {author}
  {\bibfnamefont {H.~C.}\ \bibnamefont {N{\"a}gerl}},\ }\href {\doibase
  10.1140/epjd/e2011-20015-6} {\bibfield  {journal} {\bibinfo  {journal} {Eur.
  Phys. J. D}\ }\textbf {\bibinfo {volume} {65}},\ \bibinfo {pages} {3}
  (\bibinfo {year} {2011})}\BibitemShut {NoStop}%
\bibitem [{\citenamefont {Cazalilla}\ \emph {et~al.}(2011)\citenamefont
  {Cazalilla}, \citenamefont {Citro}, \citenamefont {Giamarchi}, \citenamefont
  {Orignac},\ and\ \citenamefont {Rigol}}]{cazalilla2011}%
  \BibitemOpen
  \bibfield  {author} {\bibinfo {author} {\bibfnamefont {M.~A.}\ \bibnamefont
  {Cazalilla}}, \bibinfo {author} {\bibfnamefont {R.}~\bibnamefont {Citro}},
  \bibinfo {author} {\bibfnamefont {T.}~\bibnamefont {Giamarchi}}, \bibinfo
  {author} {\bibfnamefont {E.}~\bibnamefont {Orignac}}, \ and\ \bibinfo
  {author} {\bibfnamefont {M.}~\bibnamefont {Rigol}},\ }\href {\doibase
  10.1103/RevModPhys.83.1405} {\bibfield  {journal} {\bibinfo  {journal} {Rev.
  Mod. Phys.}\ }\textbf {\bibinfo {volume} {83}},\ \bibinfo {pages} {1405}
  (\bibinfo {year} {2011})}\BibitemShut {NoStop}%
\bibitem [{\citenamefont {Ketterle}\ and\ \citenamefont {{van
  Druten}}(1996)}]{ketterle1996}%
  \BibitemOpen
  \bibfield  {author} {\bibinfo {author} {\bibfnamefont {W.}~\bibnamefont
  {Ketterle}}\ and\ \bibinfo {author} {\bibfnamefont {N.~J.}\ \bibnamefont
  {{van Druten}}},\ }\href {\doibase 10.1103/PhysRevA.54.656} {\bibfield
  {journal} {\bibinfo  {journal} {Phys. Rev. A}\ }\textbf {\bibinfo {volume}
  {54}},\ \bibinfo {pages} {656} (\bibinfo {year} {1996})}\BibitemShut
  {NoStop}%
\bibitem [{\citenamefont {G{\"o}rlitz}\ \emph {et~al.}(2001)\citenamefont
  {G{\"o}rlitz}, \citenamefont {Vogels}, \citenamefont {Leanhardt},
  \citenamefont {Raman}, \citenamefont {Gustavson}, \citenamefont
  {{Abo-Shaeer}}, \citenamefont {Chikkatur}, \citenamefont {Gupta},
  \citenamefont {Inouye}, \citenamefont {Rosenband},\ and\ \citenamefont
  {Ketterle}}]{gorlitz2001}%
  \BibitemOpen
  \bibfield  {author} {\bibinfo {author} {\bibfnamefont {A.}~\bibnamefont
  {G{\"o}rlitz}}, \bibinfo {author} {\bibfnamefont {J.~M.}\ \bibnamefont
  {Vogels}}, \bibinfo {author} {\bibfnamefont {A.~E.}\ \bibnamefont
  {Leanhardt}}, \bibinfo {author} {\bibfnamefont {C.}~\bibnamefont {Raman}},
  \bibinfo {author} {\bibfnamefont {T.~L.}\ \bibnamefont {Gustavson}}, \bibinfo
  {author} {\bibfnamefont {J.~R.}\ \bibnamefont {{Abo-Shaeer}}}, \bibinfo
  {author} {\bibfnamefont {A.~P.}\ \bibnamefont {Chikkatur}}, \bibinfo {author}
  {\bibfnamefont {S.}~\bibnamefont {Gupta}}, \bibinfo {author} {\bibfnamefont
  {S.}~\bibnamefont {Inouye}}, \bibinfo {author} {\bibfnamefont
  {T.}~\bibnamefont {Rosenband}}, \ and\ \bibinfo {author} {\bibfnamefont
  {W.}~\bibnamefont {Ketterle}},\ }\href {\doibase
  10.1103/PhysRevLett.87.130402} {\bibfield  {journal} {\bibinfo  {journal}
  {Phys. Rev. Lett.}\ }\textbf {\bibinfo {volume} {87}},\ \bibinfo {pages}
  {130402} (\bibinfo {year} {2001})}\BibitemShut {NoStop}%
\bibitem [{\citenamefont {Inouye}\ \emph {et~al.}(1998)\citenamefont {Inouye},
  \citenamefont {Andrews}, \citenamefont {Stenger}, \citenamefont {Miesner},
  \citenamefont {{Stamper-Kurn}},\ and\ \citenamefont {Ketterle}}]{inouye1998}%
  \BibitemOpen
  \bibfield  {author} {\bibinfo {author} {\bibfnamefont {S.}~\bibnamefont
  {Inouye}}, \bibinfo {author} {\bibfnamefont {M.~R.}\ \bibnamefont {Andrews}},
  \bibinfo {author} {\bibfnamefont {J.}~\bibnamefont {Stenger}}, \bibinfo
  {author} {\bibfnamefont {H.-J.}\ \bibnamefont {Miesner}}, \bibinfo {author}
  {\bibfnamefont {D.~M.}\ \bibnamefont {{Stamper-Kurn}}}, \ and\ \bibinfo
  {author} {\bibfnamefont {W.}~\bibnamefont {Ketterle}},\ }\href {\doibase
  10.1038/32354} {\bibfield  {journal} {\bibinfo  {journal} {Nature}\ }\textbf
  {\bibinfo {volume} {392}},\ \bibinfo {pages} {151} (\bibinfo {year}
  {1998})}\BibitemShut {NoStop}%
\bibitem [{\citenamefont {Fedichev}\ \emph {et~al.}(1996)\citenamefont
  {Fedichev}, \citenamefont {Kagan}, \citenamefont {Shlyapnikov},\ and\
  \citenamefont {Walraven}}]{fedichev1996}%
  \BibitemOpen
  \bibfield  {author} {\bibinfo {author} {\bibfnamefont {P.~O.}\ \bibnamefont
  {Fedichev}}, \bibinfo {author} {\bibfnamefont {Y.}~\bibnamefont {Kagan}},
  \bibinfo {author} {\bibfnamefont {G.~V.}\ \bibnamefont {Shlyapnikov}}, \ and\
  \bibinfo {author} {\bibfnamefont {J.~T.~M.}\ \bibnamefont {Walraven}},\
  }\href {\doibase 10.1103/PhysRevLett.77.2913} {\bibfield  {journal} {\bibinfo
   {journal} {Phys. Rev. Lett.}\ }\textbf {\bibinfo {volume} {77}},\ \bibinfo
  {pages} {2913} (\bibinfo {year} {1996})}\BibitemShut {NoStop}%
\bibitem [{\citenamefont {Bouton}\ \emph {et~al.}(2020)\citenamefont {Bouton},
  \citenamefont {Nettersheim}, \citenamefont {Adam}, \citenamefont {Schmidt},
  \citenamefont {Mayer}, \citenamefont {Lausch}, \citenamefont {Tiemann},\ and\
  \citenamefont {Widera}}]{bouton2020}%
  \BibitemOpen
  \bibfield  {author} {\bibinfo {author} {\bibfnamefont {Q.}~\bibnamefont
  {Bouton}}, \bibinfo {author} {\bibfnamefont {J.}~\bibnamefont {Nettersheim}},
  \bibinfo {author} {\bibfnamefont {D.}~\bibnamefont {Adam}}, \bibinfo {author}
  {\bibfnamefont {F.}~\bibnamefont {Schmidt}}, \bibinfo {author} {\bibfnamefont
  {D.}~\bibnamefont {Mayer}}, \bibinfo {author} {\bibfnamefont
  {T.}~\bibnamefont {Lausch}}, \bibinfo {author} {\bibfnamefont
  {E.}~\bibnamefont {Tiemann}}, \ and\ \bibinfo {author} {\bibfnamefont
  {A.}~\bibnamefont {Widera}},\ }\href {\doibase 10.1103/PhysRevX.10.011018}
  {\bibfield  {journal} {\bibinfo  {journal} {Phys. Rev. X}\ }\textbf {\bibinfo
  {volume} {10}},\ \bibinfo {pages} {011018} (\bibinfo {year}
  {2020})}\BibitemShut {NoStop}%
\bibitem [{\citenamefont {Adam}\ \emph {et~al.}(2021)\citenamefont {Adam},
  \citenamefont {Bouton}, \citenamefont {Nettersheim}, \citenamefont
  {Burgardt},\ and\ \citenamefont {Widera}}]{adam2021}%
  \BibitemOpen
  \bibfield  {author} {\bibinfo {author} {\bibfnamefont {D.}~\bibnamefont
  {Adam}}, \bibinfo {author} {\bibfnamefont {Q.}~\bibnamefont {Bouton}},
  \bibinfo {author} {\bibfnamefont {J.}~\bibnamefont {Nettersheim}}, \bibinfo
  {author} {\bibfnamefont {S.}~\bibnamefont {Burgardt}}, \ and\ \bibinfo
  {author} {\bibfnamefont {A.}~\bibnamefont {Widera}},\ }\href@noop {} {\
  (\bibinfo {year} {2021})},\ \Eprint {http://arxiv.org/abs/2105.03331}
  {arXiv:2105.03331} \BibitemShut {NoStop}%
\bibitem [{\citenamefont {Albiez}\ \emph {et~al.}(2005)\citenamefont {Albiez},
  \citenamefont {Gati}, \citenamefont {F{\"o}lling}, \citenamefont {Hunsmann},
  \citenamefont {Cristiani},\ and\ \citenamefont {Oberthaler}}]{albiez2005}%
  \BibitemOpen
  \bibfield  {author} {\bibinfo {author} {\bibfnamefont {M.}~\bibnamefont
  {Albiez}}, \bibinfo {author} {\bibfnamefont {R.}~\bibnamefont {Gati}},
  \bibinfo {author} {\bibfnamefont {J.}~\bibnamefont {F{\"o}lling}}, \bibinfo
  {author} {\bibfnamefont {S.}~\bibnamefont {Hunsmann}}, \bibinfo {author}
  {\bibfnamefont {M.}~\bibnamefont {Cristiani}}, \ and\ \bibinfo {author}
  {\bibfnamefont {M.~K.}\ \bibnamefont {Oberthaler}},\ }\href {\doibase
  10.1103/PhysRevLett.95.010402} {\bibfield  {journal} {\bibinfo  {journal}
  {Phys. Rev. Lett.}\ }\textbf {\bibinfo {volume} {95}},\ \bibinfo {pages}
  {010402} (\bibinfo {year} {2005})}\BibitemShut {NoStop}%
\bibitem [{\citenamefont {Z{\"o}llner}\ \emph {et~al.}(2008)\citenamefont
  {Z{\"o}llner}, \citenamefont {Meyer},\ and\ \citenamefont
  {Schmelcher}}]{zollner2008}%
  \BibitemOpen
  \bibfield  {author} {\bibinfo {author} {\bibfnamefont {S.}~\bibnamefont
  {Z{\"o}llner}}, \bibinfo {author} {\bibfnamefont {H.-D.}\ \bibnamefont
  {Meyer}}, \ and\ \bibinfo {author} {\bibfnamefont {P.}~\bibnamefont
  {Schmelcher}},\ }\href {\doibase 10.1103/PhysRevLett.100.040401} {\bibfield
  {journal} {\bibinfo  {journal} {Phys. Rev. Lett.}\ }\textbf {\bibinfo
  {volume} {100}},\ \bibinfo {pages} {040401} (\bibinfo {year}
  {2008})}\BibitemShut {NoStop}%
\bibitem [{Note1()}]{Note1}%
  \BibitemOpen
  \bibinfo {note} {The characteristic length of the box potential is set such
  that the resulting interaction scales are comparable with the ones used in
  the case of a harmonically trapped medium.}\BibitemShut {Stop}%
\bibitem [{\citenamefont {Raab}(2000)}]{raab2000}%
  \BibitemOpen
  \bibfield  {author} {\bibinfo {author} {\bibfnamefont {A.}~\bibnamefont
  {Raab}},\ }\href {\doibase 10.1016/S0009-2614(00)00200-1} {\bibfield
  {journal} {\bibinfo  {journal} {Chem. Phys. Lett.}\ }\textbf {\bibinfo
  {volume} {319}},\ \bibinfo {pages} {674} (\bibinfo {year}
  {2000})}\BibitemShut {NoStop}%
\bibitem [{Note2()}]{Note2}%
  \BibitemOpen
  \bibinfo {note} {Throughout this work the characterization}\BibitemShut
  {NoStop}%
\bibitem [{\citenamefont {Schmidt}(1907)}]{schmidt1907}%
  \BibitemOpen
  \bibfield  {author} {\bibinfo {author} {\bibfnamefont {E.}~\bibnamefont
  {Schmidt}},\ }\href {\doibase 10.1007/BF01449770} {\bibfield  {journal}
  {\bibinfo  {journal} {Math. Ann.}\ }\textbf {\bibinfo {volume} {63}},\
  \bibinfo {pages} {433} (\bibinfo {year} {1907})}\BibitemShut {NoStop}%
\bibitem [{\citenamefont {Ekert}\ and\ \citenamefont
  {Knight}(1995)}]{ekert1995}%
  \BibitemOpen
  \bibfield  {author} {\bibinfo {author} {\bibfnamefont {A.}~\bibnamefont
  {Ekert}}\ and\ \bibinfo {author} {\bibfnamefont {P.~L.}\ \bibnamefont
  {Knight}},\ }\href {\doibase 10.1119/1.17904} {\bibfield  {journal} {\bibinfo
   {journal} {Am. J. Phys.}\ }\textbf {\bibinfo {volume} {63}},\ \bibinfo
  {pages} {415} (\bibinfo {year} {1995})}\BibitemShut {NoStop}%
\bibitem [{\citenamefont {Horodecki}\ \emph {et~al.}(2009)\citenamefont
  {Horodecki}, \citenamefont {Horodecki}, \citenamefont {Horodecki},\ and\
  \citenamefont {Horodecki}}]{horodecki2009}%
  \BibitemOpen
  \bibfield  {author} {\bibinfo {author} {\bibfnamefont {R.}~\bibnamefont
  {Horodecki}}, \bibinfo {author} {\bibfnamefont {P.}~\bibnamefont
  {Horodecki}}, \bibinfo {author} {\bibfnamefont {M.}~\bibnamefont
  {Horodecki}}, \ and\ \bibinfo {author} {\bibfnamefont {K.}~\bibnamefont
  {Horodecki}},\ }\href {\doibase 10.1103/RevModPhys.81.865} {\bibfield
  {journal} {\bibinfo  {journal} {Rev. Mod. Phys.}\ }\textbf {\bibinfo {volume}
  {81}},\ \bibinfo {pages} {865} (\bibinfo {year} {2009})}\BibitemShut
  {NoStop}%
\bibitem [{\citenamefont {Li}\ and\ \citenamefont {Haldane}(2008)}]{li2008}%
  \BibitemOpen
  \bibfield  {author} {\bibinfo {author} {\bibfnamefont {H.}~\bibnamefont
  {Li}}\ and\ \bibinfo {author} {\bibfnamefont {F.~D.~M.}\ \bibnamefont
  {Haldane}},\ }\href {\doibase 10.1103/PhysRevLett.101.010504} {\bibfield
  {journal} {\bibinfo  {journal} {Phys. Rev. Lett.}\ }\textbf {\bibinfo
  {volume} {101}},\ \bibinfo {pages} {010504} (\bibinfo {year}
  {2008})}\BibitemShut {NoStop}%
\bibitem [{\citenamefont {Light}\ \emph {et~al.}(1985)\citenamefont {Light},
  \citenamefont {Hamilton},\ and\ \citenamefont {Lill}}]{light1985}%
  \BibitemOpen
  \bibfield  {author} {\bibinfo {author} {\bibfnamefont {J.~C.}\ \bibnamefont
  {Light}}, \bibinfo {author} {\bibfnamefont {I.~P.}\ \bibnamefont {Hamilton}},
  \ and\ \bibinfo {author} {\bibfnamefont {J.~V.}\ \bibnamefont {Lill}},\
  }\href {\doibase 10.1063/1.448462} {\bibfield  {journal} {\bibinfo  {journal}
  {J. Chem. Phys.}\ }\textbf {\bibinfo {volume} {82}},\ \bibinfo {pages} {1400}
  (\bibinfo {year} {1985})}\BibitemShut {NoStop}%
\bibitem [{\citenamefont {{Garc{\'i}a-March}}\ \emph
  {et~al.}(2014)\citenamefont {{Garc{\'i}a-March}}, \citenamefont
  {{Juli{\'a}-D{\'i}az}}, \citenamefont {Astrakharchik}, \citenamefont {Busch},
  \citenamefont {Boronat},\ and\ \citenamefont {Polls}}]{garcia-march2014}%
  \BibitemOpen
  \bibfield  {author} {\bibinfo {author} {\bibfnamefont {M.~A.}\ \bibnamefont
  {{Garc{\'i}a-March}}}, \bibinfo {author} {\bibfnamefont {B.}~\bibnamefont
  {{Juli{\'a}-D{\'i}az}}}, \bibinfo {author} {\bibfnamefont {G.~E.}\
  \bibnamefont {Astrakharchik}}, \bibinfo {author} {\bibfnamefont
  {T.}~\bibnamefont {Busch}}, \bibinfo {author} {\bibfnamefont
  {J.}~\bibnamefont {Boronat}}, \ and\ \bibinfo {author} {\bibfnamefont
  {A.}~\bibnamefont {Polls}},\ }\href {\doibase 10.1088/1367-2630/16/10/103004}
  {\bibfield  {journal} {\bibinfo  {journal} {New J. Phys.}\ }\textbf {\bibinfo
  {volume} {16}},\ \bibinfo {pages} {103004} (\bibinfo {year}
  {2014})}\BibitemShut {NoStop}%
\bibitem [{\citenamefont {Pa{\v s}kauskas}\ and\ \citenamefont
  {You}(2001)}]{paskauskas2001}%
  \BibitemOpen
  \bibfield  {author} {\bibinfo {author} {\bibfnamefont {R.}~\bibnamefont
  {Pa{\v s}kauskas}}\ and\ \bibinfo {author} {\bibfnamefont {L.}~\bibnamefont
  {You}},\ }\href {\doibase 10.1103/PhysRevA.64.042310} {\bibfield  {journal}
  {\bibinfo  {journal} {Phys. Rev. A}\ }\textbf {\bibinfo {volume} {64}},\
  \bibinfo {pages} {042310} (\bibinfo {year} {2001})}\BibitemShut {NoStop}%
\bibitem [{\citenamefont {Jain}\ and\ \citenamefont
  {Boninsegni}(2011)}]{jain2011}%
  \BibitemOpen
  \bibfield  {author} {\bibinfo {author} {\bibfnamefont {P.}~\bibnamefont
  {Jain}}\ and\ \bibinfo {author} {\bibfnamefont {M.}~\bibnamefont
  {Boninsegni}},\ }\href {\doibase 10.1103/PhysRevA.83.023602} {\bibfield
  {journal} {\bibinfo  {journal} {Phys. Rev. A}\ }\textbf {\bibinfo {volume}
  {83}},\ \bibinfo {pages} {023602} (\bibinfo {year} {2011})}\BibitemShut
  {NoStop}%
\bibitem [{\citenamefont {Bandyopadhyay}\ \emph {et~al.}(2017)\citenamefont
  {Bandyopadhyay}, \citenamefont {Roy},\ and\ \citenamefont
  {Angom}}]{bandyopadhyay2017}%
  \BibitemOpen
  \bibfield  {author} {\bibinfo {author} {\bibfnamefont {S.}~\bibnamefont
  {Bandyopadhyay}}, \bibinfo {author} {\bibfnamefont {A.}~\bibnamefont {Roy}},
  \ and\ \bibinfo {author} {\bibfnamefont {D.}~\bibnamefont {Angom}},\ }\href
  {\doibase 10.1103/PhysRevA.96.043603} {\bibfield  {journal} {\bibinfo
  {journal} {Phys. Rev. A}\ }\textbf {\bibinfo {volume} {96}},\ \bibinfo
  {pages} {043603} (\bibinfo {year} {2017})}\BibitemShut {NoStop}%
\bibitem [{\citenamefont {Mueller}\ \emph {et~al.}(2006)\citenamefont
  {Mueller}, \citenamefont {Ho}, \citenamefont {Ueda},\ and\ \citenamefont
  {Baym}}]{mueller2006}%
  \BibitemOpen
  \bibfield  {author} {\bibinfo {author} {\bibfnamefont {E.~J.}\ \bibnamefont
  {Mueller}}, \bibinfo {author} {\bibfnamefont {T.-L.}\ \bibnamefont {Ho}},
  \bibinfo {author} {\bibfnamefont {M.}~\bibnamefont {Ueda}}, \ and\ \bibinfo
  {author} {\bibfnamefont {G.}~\bibnamefont {Baym}},\ }\href {\doibase
  10.1103/PhysRevA.74.033612} {\bibfield  {journal} {\bibinfo  {journal} {Phys.
  Rev. A}\ }\textbf {\bibinfo {volume} {74}},\ \bibinfo {pages} {033612}
  (\bibinfo {year} {2006})}\BibitemShut {NoStop}%
\bibitem [{\citenamefont {Z{\"o}llner}\ \emph
  {et~al.}(2006{\natexlab{a}})\citenamefont {Z{\"o}llner}, \citenamefont
  {Meyer},\ and\ \citenamefont {Schmelcher}}]{zollner2006}%
  \BibitemOpen
  \bibfield  {author} {\bibinfo {author} {\bibfnamefont {S.}~\bibnamefont
  {Z{\"o}llner}}, \bibinfo {author} {\bibfnamefont {H.-D.}\ \bibnamefont
  {Meyer}}, \ and\ \bibinfo {author} {\bibfnamefont {P.}~\bibnamefont
  {Schmelcher}},\ }\href {\doibase 10.1103/PhysRevA.74.053612} {\bibfield
  {journal} {\bibinfo  {journal} {Phys. Rev. A}\ }\textbf {\bibinfo {volume}
  {74}},\ \bibinfo {pages} {053612} (\bibinfo {year}
  {2006}{\natexlab{a}})}\BibitemShut {NoStop}%
\bibitem [{\citenamefont {Z{\"o}llner}\ \emph
  {et~al.}(2006{\natexlab{b}})\citenamefont {Z{\"o}llner}, \citenamefont
  {Meyer},\ and\ \citenamefont {Schmelcher}}]{zollner2006a}%
  \BibitemOpen
  \bibfield  {author} {\bibinfo {author} {\bibfnamefont {S.}~\bibnamefont
  {Z{\"o}llner}}, \bibinfo {author} {\bibfnamefont {H.-D.}\ \bibnamefont
  {Meyer}}, \ and\ \bibinfo {author} {\bibfnamefont {P.}~\bibnamefont
  {Schmelcher}},\ }\href {\doibase 10.1103/PhysRevA.74.063611} {\bibfield
  {journal} {\bibinfo  {journal} {Phys. Rev. A}\ }\textbf {\bibinfo {volume}
  {74}},\ \bibinfo {pages} {063611} (\bibinfo {year}
  {2006}{\natexlab{b}})}\BibitemShut {NoStop}%
\bibitem [{\citenamefont {Murphy}\ \emph {et~al.}(2007)\citenamefont {Murphy},
  \citenamefont {McCann}, \citenamefont {Goold},\ and\ \citenamefont
  {Busch}}]{murphy2007}%
  \BibitemOpen
  \bibfield  {author} {\bibinfo {author} {\bibfnamefont {D.~S.}\ \bibnamefont
  {Murphy}}, \bibinfo {author} {\bibfnamefont {J.~F.}\ \bibnamefont {McCann}},
  \bibinfo {author} {\bibfnamefont {J.}~\bibnamefont {Goold}}, \ and\ \bibinfo
  {author} {\bibfnamefont {T.}~\bibnamefont {Busch}},\ }\href {\doibase
  10.1103/PhysRevA.76.053616} {\bibfield  {journal} {\bibinfo  {journal} {Phys.
  Rev. A}\ }\textbf {\bibinfo {volume} {76}},\ \bibinfo {pages} {053616}
  (\bibinfo {year} {2007})}\BibitemShut {NoStop}%
\bibitem [{Note3()}]{Note3}%
  \BibitemOpen
  \bibinfo {note} {To discern among the different phases we rely on the
  integrated two-body density of the impurities $\DOTSI \intop \ilimits@
  _{-\infty }^{0} dx_{1}^Idx_{2}^I\rho _{II}^{(2)}(x_1^I, x_2^I)$. It
  immediately distinguishes the cases of two separated (coalesced) impurities
  where it acquires the value 0.0 (0.5). Furthermore, it allows to also
  identify intermediate regimes where both the diagonal and the off-diagonal
  elements of $\rho _{II}^{(2)}(x_1^I, x_2^I)$ are populated. Notice that for a
  box confined medium we additionally weight the integrated density by
  $\protect \mathrm {sgn}(g_{BI})$ in order to distinguish between regimes (I)
  and (III) appearing at strongly repulsive and attractive $g_{BI}$,
  respectively.}\BibitemShut {Stop}%
\bibitem [{Note4()}]{Note4}%
  \BibitemOpen
  \bibinfo {note} {Throughout this work we refer to weakly (strongly)
  interacting impurities when $g_{II}<g_{BB}$ ($g_{II}>g_{BB}$). Similarly, we
  use the term weak (intermediate, strong) impurity-medium interaction strength
  if $g_{BI}<g_{BB}$ ($g_{BI}>g_{BB}$, $g_{BI}\gg g_{BB}$).}\BibitemShut
  {Stop}%
\bibitem [{\citenamefont {Pflanzer}\ \emph {et~al.}(2009)\citenamefont
  {Pflanzer}, \citenamefont {Z{\"o}llner},\ and\ \citenamefont
  {Schmelcher}}]{pflanzer2009}%
  \BibitemOpen
  \bibfield  {author} {\bibinfo {author} {\bibfnamefont {A.~C.}\ \bibnamefont
  {Pflanzer}}, \bibinfo {author} {\bibfnamefont {S.}~\bibnamefont
  {Z{\"o}llner}}, \ and\ \bibinfo {author} {\bibfnamefont {P.}~\bibnamefont
  {Schmelcher}},\ }\href {\doibase 10.1088/0953-4075/42/23/231002} {\bibfield
  {journal} {\bibinfo  {journal} {J. Phys. B: At. Mol. Opt. Phys.}\ }\textbf
  {\bibinfo {volume} {42}},\ \bibinfo {pages} {231002} (\bibinfo {year}
  {2009})}\BibitemShut {NoStop}%
\bibitem [{\citenamefont {Pflanzer}\ \emph {et~al.}(2010)\citenamefont
  {Pflanzer}, \citenamefont {Z{\"o}llner},\ and\ \citenamefont
  {Schmelcher}}]{pflanzer2010}%
  \BibitemOpen
  \bibfield  {author} {\bibinfo {author} {\bibfnamefont {A.~C.}\ \bibnamefont
  {Pflanzer}}, \bibinfo {author} {\bibfnamefont {S.}~\bibnamefont
  {Z{\"o}llner}}, \ and\ \bibinfo {author} {\bibfnamefont {P.}~\bibnamefont
  {Schmelcher}},\ }\href {\doibase 10.1103/PhysRevA.81.023612} {\bibfield
  {journal} {\bibinfo  {journal} {Phys. Rev. A}\ }\textbf {\bibinfo {volume}
  {81}},\ \bibinfo {pages} {023612} (\bibinfo {year} {2010})}\BibitemShut
  {NoStop}%
\bibitem [{\citenamefont {Liu}\ and\ \citenamefont {Zhang}(2015)}]{liu2015}%
  \BibitemOpen
  \bibfield  {author} {\bibinfo {author} {\bibfnamefont {Y.}~\bibnamefont
  {Liu}}\ and\ \bibinfo {author} {\bibfnamefont {Y.}~\bibnamefont {Zhang}},\
  }\href {\doibase 10.1103/PhysRevA.91.053610} {\bibfield  {journal} {\bibinfo
  {journal} {Phys. Rev. A}\ }\textbf {\bibinfo {volume} {91}},\ \bibinfo
  {pages} {053610} (\bibinfo {year} {2015})}\BibitemShut {NoStop}%
\bibitem [{\citenamefont {Rath}\ and\ \citenamefont
  {Schmidt}(2013)}]{rath2013}%
  \BibitemOpen
  \bibfield  {author} {\bibinfo {author} {\bibfnamefont {S.~P.}\ \bibnamefont
  {Rath}}\ and\ \bibinfo {author} {\bibfnamefont {R.}~\bibnamefont {Schmidt}},\
  }\href {\doibase 10.1103/PhysRevA.88.053632} {\bibfield  {journal} {\bibinfo
  {journal} {Phys. Rev. A}\ }\textbf {\bibinfo {volume} {88}},\ \bibinfo
  {pages} {053632} (\bibinfo {year} {2013})}\BibitemShut {NoStop}%
\bibitem [{Note5()}]{Note5}%
  \BibitemOpen
  \bibinfo {note} {We have verified that varying from $g_{II}/\protect
  \mathaccentV {tilde}07E{g}_{\protect \rm {box}}=0.0$ to 2.0 the overall shape
  of the one-body density of both species does not qualitatively alter for
  fixed $g_{BI}$. However, quantitative deviation are in play, for example, for
  strong $g_{II}$ for which the periodic motion consisting of the impurities
  collision and expansion is more pronounced than for weak
  $g_{II}$.}\BibitemShut {Stop}%
\bibitem [{Note6()}]{Note6}%
  \BibitemOpen
  \bibinfo {note} {A similar localization behavior of the impurities is also
  observed for the dynamical evolution at intermediate and strong attractive
  $g_{BI}$ (not shown here). However, in this case also the medium localizes
  around the trap center.}\BibitemShut {Stop}%
\bibitem [{Note7()}]{Note7}%
  \BibitemOpen
  \bibinfo {note} {However, we remark that for an adequate description of the
  quasi-particle notion more elaborated measures need to be considered such as
  the residue, which goes beyond the present scope.}\BibitemShut {Stop}%
\bibitem [{\citenamefont {Ronzheimer}\ \emph {et~al.}(2013)\citenamefont
  {Ronzheimer}, \citenamefont {Schreiber}, \citenamefont {Braun}, \citenamefont
  {Hodgman}, \citenamefont {Langer}, \citenamefont {McCulloch}, \citenamefont
  {{Heidrich-Meisner}}, \citenamefont {Bloch},\ and\ \citenamefont
  {Schneider}}]{ronzheimer2013}%
  \BibitemOpen
  \bibfield  {author} {\bibinfo {author} {\bibfnamefont {J.~P.}\ \bibnamefont
  {Ronzheimer}}, \bibinfo {author} {\bibfnamefont {M.}~\bibnamefont
  {Schreiber}}, \bibinfo {author} {\bibfnamefont {S.}~\bibnamefont {Braun}},
  \bibinfo {author} {\bibfnamefont {S.~S.}\ \bibnamefont {Hodgman}}, \bibinfo
  {author} {\bibfnamefont {S.}~\bibnamefont {Langer}}, \bibinfo {author}
  {\bibfnamefont {I.~P.}\ \bibnamefont {McCulloch}}, \bibinfo {author}
  {\bibfnamefont {F.}~\bibnamefont {{Heidrich-Meisner}}}, \bibinfo {author}
  {\bibfnamefont {I.}~\bibnamefont {Bloch}}, \ and\ \bibinfo {author}
  {\bibfnamefont {U.}~\bibnamefont {Schneider}},\ }\href {\doibase
  10.1103/PhysRevLett.110.205301} {\bibfield  {journal} {\bibinfo  {journal}
  {Phys. Rev. Lett.}\ }\textbf {\bibinfo {volume} {110}},\ \bibinfo {pages}
  {205301} (\bibinfo {year} {2013})}\BibitemShut {NoStop}%
\bibitem [{\citenamefont {Theel}\ \emph {et~al.}(2020)\citenamefont {Theel},
  \citenamefont {Keiler}, \citenamefont {Mistakidis},\ and\ \citenamefont
  {Schmelcher}}]{theel2020}%
  \BibitemOpen
  \bibfield  {author} {\bibinfo {author} {\bibfnamefont {F.}~\bibnamefont
  {Theel}}, \bibinfo {author} {\bibfnamefont {K.}~\bibnamefont {Keiler}},
  \bibinfo {author} {\bibfnamefont {S.~I.}\ \bibnamefont {Mistakidis}}, \ and\
  \bibinfo {author} {\bibfnamefont {P.}~\bibnamefont {Schmelcher}},\ }\href
  {\doibase 10.1088/1367-2630/ab6eab} {\bibfield  {journal} {\bibinfo
  {journal} {New J. Phys.}\ }\textbf {\bibinfo {volume} {22}},\ \bibinfo
  {pages} {023027} (\bibinfo {year} {2020})}\BibitemShut {NoStop}%
\bibitem [{Note8()}]{Note8}%
  \BibitemOpen
  \bibinfo {note} {Since in this scenario the size of the impurities
  double-well potential is not altered, the absence of the dynamical response
  corresponding to Figure~\ref {fig:box_gpop_dm2}(c1) provides further evidence
  for the involvement of finite size effects in this latter case.}\BibitemShut
  {Stop}%
\bibitem [{Note9()}]{Note9}%
  \BibitemOpen
  \bibinfo {note} {We remark that such a diminishing response for these values
  of $g_{BI}$ is observed besides the one-body density level also for the
  corresponding two-body densities which preserve their initial diagonal or
  off-diagonal shape depending on the species (not shown here).}\BibitemShut
  {Stop}%
\bibitem [{\citenamefont {Mistakidis}\ \emph
  {et~al.}(2020{\natexlab{b}})\citenamefont {Mistakidis}, \citenamefont
  {Volosniev},\ and\ \citenamefont {Schmelcher}}]{mistakidis2020b}%
  \BibitemOpen
  \bibfield  {author} {\bibinfo {author} {\bibfnamefont {S.~I.}\ \bibnamefont
  {Mistakidis}}, \bibinfo {author} {\bibfnamefont {A.~G.}\ \bibnamefont
  {Volosniev}}, \ and\ \bibinfo {author} {\bibfnamefont {P.}~\bibnamefont
  {Schmelcher}},\ }\href {\doibase 10.1103/PhysRevResearch.2.023154} {\bibfield
   {journal} {\bibinfo  {journal} {Phys. Rev. Research}\ }\textbf {\bibinfo
  {volume} {2}},\ \bibinfo {pages} {023154} (\bibinfo {year}
  {2020}{\natexlab{b}})}\BibitemShut {NoStop}%
\bibitem [{\citenamefont {Kivelson}(1982)}]{kivelson1982}%
  \BibitemOpen
  \bibfield  {author} {\bibinfo {author} {\bibfnamefont {S.}~\bibnamefont
  {Kivelson}},\ }\href {\doibase 10.1103/PhysRevB.26.4269} {\bibfield
  {journal} {\bibinfo  {journal} {Phys. Rev. B}\ }\textbf {\bibinfo {volume}
  {26}},\ \bibinfo {pages} {4269} (\bibinfo {year} {1982})}\BibitemShut
  {NoStop}%
\bibitem [{\citenamefont {Kivelson}\ and\ \citenamefont
  {Schrieffer}(1982)}]{kivelson1982a}%
  \BibitemOpen
  \bibfield  {author} {\bibinfo {author} {\bibfnamefont {S.}~\bibnamefont
  {Kivelson}}\ and\ \bibinfo {author} {\bibfnamefont {J.~R.}\ \bibnamefont
  {Schrieffer}},\ }\href {\doibase 10.1103/PhysRevB.25.6447} {\bibfield
  {journal} {\bibinfo  {journal} {Phys. Rev. B}\ }\textbf {\bibinfo {volume}
  {25}},\ \bibinfo {pages} {6447} (\bibinfo {year} {1982})}\BibitemShut
  {NoStop}%
\bibitem [{\citenamefont {Mistakidis}\ \emph
  {et~al.}(2019{\natexlab{c}})\citenamefont {Mistakidis}, \citenamefont
  {Hilbig},\ and\ \citenamefont {Schmelcher}}]{mistakidis2019b}%
  \BibitemOpen
  \bibfield  {author} {\bibinfo {author} {\bibfnamefont {S.~I.}\ \bibnamefont
  {Mistakidis}}, \bibinfo {author} {\bibfnamefont {L.}~\bibnamefont {Hilbig}},
  \ and\ \bibinfo {author} {\bibfnamefont {P.}~\bibnamefont {Schmelcher}},\
  }\href {\doibase 10.1103/PhysRevA.100.023620} {\bibfield  {journal} {\bibinfo
   {journal} {Phys. Rev. A}\ }\textbf {\bibinfo {volume} {100}},\ \bibinfo
  {pages} {023620} (\bibinfo {year} {2019}{\natexlab{c}})}\BibitemShut
  {NoStop}%
\bibitem [{\citenamefont {Bergschneider}\ \emph {et~al.}(2018)\citenamefont
  {Bergschneider}, \citenamefont {Klinkhamer}, \citenamefont {Becher},
  \citenamefont {Klemt}, \citenamefont {Z{\"u}rn}, \citenamefont {Preiss},\
  and\ \citenamefont {Jochim}}]{bergschneider2018spin}%
  \BibitemOpen
  \bibfield  {author} {\bibinfo {author} {\bibfnamefont {A.}~\bibnamefont
  {Bergschneider}}, \bibinfo {author} {\bibfnamefont {V.~M.}\ \bibnamefont
  {Klinkhamer}}, \bibinfo {author} {\bibfnamefont {J.~H.}\ \bibnamefont
  {Becher}}, \bibinfo {author} {\bibfnamefont {R.}~\bibnamefont {Klemt}},
  \bibinfo {author} {\bibfnamefont {G.}~\bibnamefont {Z{\"u}rn}}, \bibinfo
  {author} {\bibfnamefont {P.~M.}\ \bibnamefont {Preiss}}, \ and\ \bibinfo
  {author} {\bibfnamefont {S.}~\bibnamefont {Jochim}},\ }\href {\doibase
  10.1103/PhysRevA.97.063613} {\bibfield  {journal} {\bibinfo  {journal} {Phys.
  Rev. A}\ }\textbf {\bibinfo {volume} {97}},\ \bibinfo {pages} {063613}
  (\bibinfo {year} {2018})}\BibitemShut {NoStop}%
\bibitem [{\citenamefont {Chen}\ \emph {et~al.}(2018)\citenamefont {Chen},
  \citenamefont {Schurer},\ and\ \citenamefont {Schmelcher}}]{chen2018}%
  \BibitemOpen
  \bibfield  {author} {\bibinfo {author} {\bibfnamefont {J.}~\bibnamefont
  {Chen}}, \bibinfo {author} {\bibfnamefont {J.~M.}\ \bibnamefont {Schurer}}, \
  and\ \bibinfo {author} {\bibfnamefont {P.}~\bibnamefont {Schmelcher}},\
  }\href {\doibase 10.1103/PhysRevLett.121.043401} {\bibfield  {journal}
  {\bibinfo  {journal} {Phys. Rev. Lett.}\ }\textbf {\bibinfo {volume} {121}},\
  \bibinfo {pages} {043401} (\bibinfo {year} {2018})}\BibitemShut {NoStop}%
\bibitem [{Note10()}]{Note10}%
  \BibitemOpen
  \bibinfo {note} {Additionally, the trap frequency of the harmonic confinement
  has been adjusted according to $\omega _I'=\omega _I\protect \sqrt {\protect
  \frac {m_I}{m_I'}}$ in order to preserve the initial shape of the double-well
  potential.}\BibitemShut {Stop}%
\end{thebibliography}%

\end{document}